\documentclass[longauth]{aa}

\usepackage{txfonts}
\usepackage{newtxtext,newtxmath}
\usepackage{amssymb}
\usepackage{adjustbox}
\usepackage{graphicx,color,booktabs,tabularx}
\usepackage{multirow}
\usepackage{epsfig}
\usepackage{setspace}
\usepackage{lscape}
\usepackage{latexsym}
\usepackage{mathrsfs}
\usepackage{tikz}
\usepackage{placeins}
\usepackage{rotating}
\usepackage{multicol}
\usepackage{acronym}
\usepackage{pgfplots}
\usepackage{csquotes}
\usepackage{times}
\usepackage{amsmath} 
\usepackage{caption, makecell, siunitx}

\usepackage{natbib}
\usepackage[colorlinks=true,
            linkcolor=blue,
            urlcolor=blue,
            citecolor=blue]{hyperref}
\usepackage[flushleft]{threeparttable} 
            
\begin{document} 
   \title{A new discovery space opened by eROSITA:  }
   \subtitle{ Ionised AGN outflows from X--ray selected samples}
   
   \author{B. Musiimenta
          \inst{1,2}
          \and
          M. Brusa\inst{1,2}
          \and
          T. Liu\inst{3}
          \and
          M. Salvato\inst{3}
          \and
          J. Buchner\inst{3}
          \and
          Z. Igo\inst{3}
          \and
          S. Waddell\inst{3}
          \and
          Y. Toba\inst{6,7,8}
          \and
          R. Arcodia\inst{3}
          \and
          J. Comparat\inst{3}
          \and
          D. Alexander \inst{13}
          \and
          F. Shankar \inst{14}
          \and
          A. Lapi \inst{10}
          \and
          C. Ramos Almeida \inst{11,12}
          \and A. Georgakakis   \inst{4}
          \and
          A. Merloni\inst{3}
          \and
          T. Urrutia\inst{5}
          \and
            J. Li \inst{15,16,17}
            \and 
            Y. Terashima \inst{9}
            \and 
            Y. Shen \inst{18,19}
            \and 
            Q. Wu \inst{18,20}
        \and T. Dwelly   \inst{3}
          \and K. Nandra   \inst{3}
        \and J. Wolf \inst{3}
          }
          
   \institute{Dipartimento di Fisica e Astronomia 'Augusto Righi', Alma Mater Studiorum - Università di Bologna, Via Gobetti 93/2, 40129 Bologna, Italy
         \and
             INAF-Osservatorio di Astrofisica e Scienza dello Spazio via Gobetti 93/3, 40129 Bologna, Italy
             \and 
             Max-Planck-Institut f\"ur extraterrestrische Physik, Giessenbachstra{\ss}e 1, D-85748 Garching bei M\"unchen, Germany
             \and Institute for Astronomy and Astrophysics, National Observatory of Athens, V. Paulou and I. Metaxa, 11532, Greece
             \and
             Leibniz-Institut für Astrophysik Potsdam, An der Sternwarte 16, 14482 Potsdam, Germany
             \and Department of Astronomy, Kyoto University, Kitashirakawa-Oiwake-cho, Sakyo-ku, Kyoto 606-8502, Japan
             \and Academia Sinica Institute of Astronomy and Astrophysics, 11F of Astronomy-Mathematics Building, AS/NTU, No. 1, Section 4, Roosevelt Road, Taipei 10617, Taiwan
             \and Research Center for Space and Cosmic Evolution, Ehime University, 2-5 Bunkyo-cho, Matsuyama, Ehime 790-8577, Japan\label{in:Ehime}
             \and   Graduate School of Science and Engineering, Ehime University, 2-5 Bunkyo-cho, Matsuyama, Ehime 790-8577, Japan
             \and SISSA, Via Bonomea 265, 34136 Trieste, Italy
             \and 
             Instituto de Astrofísica de Canarias, Calle Vía Láctea, s/n, 38205 La Laguna, Tenerife, Spain 
             \and 
             Departamento de Astrofísica, Universidad de La Laguna, 38206 La Laguna, Tenerife, Spain
             \and
             Centre for Extragalactic Astronomy, Department of Physics, Durham University, South Road, Durham, DH1 3LE, UK
             \and
             School of Physics and Astronomy, University of Southampton, Highfield, SO17 1BJ, UK
             \and
             Kavli Institute for the Physics and Mathematics of the Universe, The University of Tokyo, Kashiwa 277-8583 (Kavli IPMU, WPI), Japan 
             \and
             School of Astronomy and Space Science, University of Science and Technology of China, Hefei 230026, People's Republic of China
             \and
             Department of Astronomy, University of Illinois at Urbana-Champaign, Urbana, IL 61801, USA
             \and 
             National Center for Supercomputing Applications, University of Illinois at Urbana-Champaign, Urbana, IL 61801, USA
             \and
             Center for AstroPhysical Surveys, National Center for Supercomputing Applications, University of Illinois at Urbana-Champaign, Urbana, IL 61801, USA
             \and
             National Center for Supercomputing Applications, University of Illinois at Urbana-Champaign, Urbana, IL 61801, USA
     }

   \date{Received 28 November 2022; accepted 12 September 2023}

  \abstract
   { In the context of an evolutionary model, the outflow phase of an active galactic nucleus (AGN) occurs at the peak of its activity, once the central supermassive black hole (SMBH) is massive enough to generate sufficient power to counterbalance the potential well of the host galaxy. This outflow feedback phase plays a vital role in galaxy evolution.  }
   {Our aim in this paper is to apply various selection methods to isolate powerful AGNs in the feedback phase, trace and characterise outflows in these AGNs, and explore the link between AGN luminosity and outflow properties.}
   {We applied a combination of methods to the Spectrum Roentgen Gamma (SRG) eROSITA Final Equatorial Depth survey (eFEDS) catalogue and isolated $\sim$1400 candidates at z$>$0.5 out of $\sim11750$ AGNs ($\sim12$\%). Furthermore, we narrowed down our selection to 427 sources that have 0.5$<$z$<$1. We tested the robustness of our selection on the small subsample of 50 sources with available good quality SDSS spectra at 0.5$<$z$<$1 and, for which we fitted the [OIII] emission line complex and searched for the presence of ionised gas outflow signatures.}
   { Out of the 50 good quality SDSS spectra, we identified 23 quasars ($\sim45$\%) with evidence of ionised outflows based on the presence of significant broad and/or shifted components in [OIII]$\lambda5007\text{\AA}$. They are on average more luminous ($\mathrm{log L_{bol}\sim ~45.2~erg~s^{-1}}$) and more obscured ($\mathrm{N_{H}}$ $\mathrm{\sim10^{22}~cm^{-2}}$) than the parent sample of $\sim$427 candidates, although this may be ascribed to selection effects affecting the good quality SDSS spectra sample. By adding 118 quasars at 0.5$<$z$<$3.5 with evidence of outflows reported in the literature, we find a weak correlation between the maximum outflow velocity and the AGN bolometric luminosity. On the contrary, we recovered strong correlations between the mass outflow rate and outflow kinetic power with the AGN bolometric luminosity.}
   {About 30\% of our sample have kinetic coupling efficiencies, $\dot{E}/L_{bol}\!>\!1$\%, suggesting that the outflows could have a significant effect on their host galaxies.
   We find that the majority of the outflows have momentum flux ratios lower than 20 which rules out an energy-conserving nature. Our present work points to the unequivocal existence of a rather short AGN outflow phase, paving the way towards a new avenue to dissect AGN outflows in large samples within eROSITA and beyond.}

 \maketitle  
 
\section{Introduction}

\label{intro}
Active galactic nuclei (AGNs) are central nuclei of massive galaxies that are powered by the accretion of matter towards the central supermassive black holes (SMBHs) \citep{soltan1982,rees1984}.
The energy produced during the accretion episodes by AGNs in the form of winds, radiation or jets is argued to have a great impact on the interstellar medium (ISM) \citep[]{alexanderDM2012,Fabian2012,harrison2017}.   
The link between the energy produced by the AGNs and the surrounding ISM is called AGN feedback \citep{silk1998,dimatteo2005,hopkins2009}, and it is considered a key element in galaxy evolution models and simulations (e.g. \citealt{schaye2015}).

To better understand galaxy evolution, it is crucial to determine how SMBHs form and evolve. Various processes have been proposed to trigger the formation and evolution of SMBHs, including internal processes such as disc or bar instabilities and external processes such as mergers, among others \citep{hopkins2008}. 
In the merger scenario, the black hole grows within a dust-enshrouded and star-forming environment, followed by high accretion close to the Eddington limit. This in turn releases a tremendous amount of energy (the blow-out phase) through radiation pressure-driven winds or outflows to the surrounding environment.

These AGN-driven outflows may provide the mechanism needed to remove the gas and thus quench star formation (SF) \citep{costa2018c}, limiting the further growth of the galaxy and the SMBHs as well \citep[e.g.][]{dimatteo2005}. As discussed in \cite{harrison2018}, 
an alternative mechanism is the injection of energy in the ISM or intergalactic medium (IGM), which indirectly heats the halo and prevents future in-fall of matter onto the galaxy. Referred to as the maintenance mode of feedback, this phenomenon is primarily caused by jet-driven outflows at large scales. When these outflows collide, they introduce turbulence and generate shocks in their vicinity, effectively preventing cooling and impeding star formation \citep{croton2006,ciotti2010,nelson2019}. The outflows influence further fueling the AGN activity affects the fueling of star formation and regulates its duty cycle.

According to hydrodynamical simulations, during the key 'blow-out phase', the AGN is characterised by high nuclear obscuration with column densities ($\mathrm{N_{H}}$) $\mathrm{>10^{22} cm^{-2}}$, high AGN bolometric luminosities, and accretion close to the Eddington limit \citep{hopkins2005,dimatteo2005,blecha2018}. As discussed in the recent study by \cite{blecha2018} with simulations of AGN evolution through galaxy mergers, the AGN luminosity and column densities are enhanced during the late stages of galaxy mergers where the blow-out phase is also expected to occur. Surveys that are robust to obscuration are needed to study these kinds of objects. 

In addition, this phase is expected to be short, and possibly even shorter than the AGN lifetime of $\sim10^{5}$ years \citep{Schawinski2015,kingnixon2015}, hence the need for large-area surveys. AGN outflows may occur during the early phases of SMBH and galaxy evolution according to some models \citep[e.g.][]{lapi2014,lapi2018}, and could occur on relatively short timescales depending on the exact modelling of the light curve.

AGN outflows at all gas phases (ionised or neutral, molecular or atomic) and all scales (from the launching region to host galaxy scale) have been revealed and studied in sources detected both in the local \citep{tombesi2010,gofford2013,feruglio2013a,feruglio2013b,rupke2013,harrison2014,aalto2015,morganti2016,toba2017a,toba201b,igo2020,ramosalmeida2022,speranza2022} and distant universe \citep{canodiaz2012,maiolino2012,cresci2015b,cresci2015a,cicone2015,perna2015a,carniani2016,Brusa2015,brusa2016,harrison2016,kakkad2016,bischetti2017,perrotta2019,Scholtz2020,kakkad2020,ishikawa2021,vayner2021,brusa2022} through integrated or spatially resolved spectroscopy \citep[see also][for a review]{Cicone2018}.  

We note that to quantify and characterise AGN outflows, assumptions on different parameters such as electron density, velocity, radius, geometry, etc., are made when it is not possible to measure them. This makes it challenging to determine the correlations between the outflow properties, the AGN properties and the host galaxy properties \citep[e.g.][]{fiore2017}. Moreover, due to their multi-phase nature, we can only access and characterise the gas motions by emission line tracers that highlight different and incomplete gas phases. This complexity makes it more difficult to compare observations with theoretical predictions \citep{harrison2018}. Whether these AGN outflows affect their host galaxies and how is still a puzzling question.

Although exploring individual systems with outflows is important, from the observational point of view it is also paramount to study large samples and explore efficient selection techniques to 1) better provide constraints for models and simulations; and 2) isolate the best targets for multiwavelength follow-up with current and future facilities. As discussed in detail in the next sections, several studies have investigated the properties of AGN outflows. They have pre-selected these AGNs with outflows using different observational techniques in the X-ray, UV, optical, infrared and radio \citep[e.g.][ among others]{zakamska2016,perrotta2019,brusa2022}. The different selection methods used in these studies, which are still prone to incompleteness, are an indication that there is no unique way of selecting AGNs in the feedback phase. This calls for the need for more studies to explore different multi-wavelength approaches to pre-select these kinds of objects. Moreover,
AGNs in this key evolutionary phase are more sensitive to obscuration in the optical and soft X-rays than IR and hard X-rays \citep{alexanderDM2012,blecha2018}.

This study aims to apply a combination of different methods that have been used to select this class of AGNs to sources detected in a new large X--ray survey field, imaged by the extended Roentgen Survey with an Imaging Telescope Array (eROSITA), and explore the scaling relations between AGN properties and outflow properties combining results from this work with literature samples published so far.
This paper is organised as follows. Section \ref{sampleselection} describes the different selection criteria of AGN outflows from eFEDS. Section \ref{results} describes the properties of our selected candidates, spectral analysis and outflow properties. Discussions of the results and our conclusions are presented in Sect. \ref{discussion} and \ref{conclusions}, respectively. Throughout the paper, we adopt the cosmological parameters $\rm{H_{0}=70 kms^{-1}Mpc^{-1}}$, $\rm{\Omega_{m}=0.3}$ and $\rm{\Omega_{\Lambda}=0.7 }$ \citep{spergel2003}. We use the AB magnitudes system unless otherwise stated. 

\section{Selection of AGNs in the feedback phase}
\label{sampleselection}
\subsection{The eFEDS sample}
eROSITA is an X--ray imaging telescope aboard the Spectrum Roentgen Gamma (SRG, \citealt{sunyaev2021}) observing within 0.2 -- 10 keV band and it is expected to detect millions of AGNs over the all-sky at the end of a 4 years survey \citep{meloni2012,predehl2021}. A large ($\sim$140 deg²) extragalactic area, the eROSITA Final Equatorial Depth survey (eFEDS), rich in photometric and spectroscopic multi-wavelength coverage, was observed for four days during eROSITA's performance and verification phase. One of the main purposes of the eFEDS survey is to allow eROSITA scientists to test their techniques on this smaller sample to prepare for the future all-sky larger samples and explore the possible science that can emerge from the eROSITA X--ray sources population. 

The eFEDS X-ray catalogue consists of 27369 point-like X--ray sources detected in the 0.2-2.3 keV band \citep[Main catalogue;][]{brunner2022}. 
The optical and IR counterparts are presented in \cite{salvato2021}, along with multi-band photometric (from Galex FUV to Wise W4) and redshift (spectroscopic redshifts when available, and photometric redshifts otherwise) information. 
From this catalogue, \cite{liuT2021} presents, the X--ray spectral properties along with monochromatic luminosities at 5100$\text{\AA}$ and 2500$\text{\AA}$, for the AGN subsample (22097 objects). 
While only 1\% of these sources are significantly detected
in the 2.3-5 keV band (Hard catalogue, Nandra et al. in prep), \cite{liuT2021} analysed the 0.2-8 keV X-ray spectra of all AGNs in the main catalogue, and estimated 2-10 keV X-ray fluxes (absorption corrected in the rest-frame band). The next step is to isolate AGNs with a high chance of being in the feedback phase. To do so, we have adopted the criteria discussed in detail in Sects.  \ref{ColorselectionineFEDS} and \ref{opticalspectralproperties}, as follows.

\subsection{Colour selection in eFEDS (Sample A)}
\label{ColorselectionineFEDS}
During the blow-out phase, enhanced obscuration is expected, evidenced by red colours in the optical to infrared bands. Previous studies \citep[e.g.][]{urrutia2008,banerji2012} have studied colour-based pre-selection of red quasars at z$>$0.5 based on the colour cut
\begin{equation}
    \label{rmw1}
    \mathrm{r - W1}> 4 
\end{equation} 
and
\begin{equation}
    \label{imw3}
    \mathrm{i - W3} > 4.6,
\end{equation}
 where i is the magnitude in the i band, r is the magnitude in the r band, and W1 and W3 are the magnitudes in the WISE W1 and W3 bands, respectively.

 These criteria have also been utilised in prior investigations \citep[e.g.][]{ross2015,hamann2016,zakamska2016,perrotta2019} to select targets for spectroscopic follow-up. Specifically, Eq. \ref{imw3} has been employed to identify extreme red quasars \citep[ERQs; e.g.][]{hamann2016,perrotta2019,vayner2021} characterised by their reddish hues, indicative of outflow phenomena. 

In other investigations instead, the red colours have been coupled with a selection involving also a  high hard X--ray to optical flux ratio given by
\begin{equation}
    \label{hardXrayfluxtoOpticalflux}
    \rm{log\left ( \frac{F_{2-10 keV}}{F_{opt}}\right)} > 1,
\end{equation}
where $\mathrm{F_{2-10 keV}}$ is the X--ray flux in 2-10 keV band and $\rm{F_{opt}}$ is the flux in the optical r-band \citep{brusa2010,Brusa2015,perna2015a,lamassa2016}. As a pilot study, \cite{brusa2022} applied the combination of Eq.\ref{rmw1} and Eq.\ref{hardXrayfluxtoOpticalflux} 
to the eFEDS Hard sample ($\sim$250 sources). This method isolated three sources. The only one with an available optical spectrum from SDSS (XID 439, hereafter ID 608 from the main sample) turned out to be an obscured source with N$_{H}>10^{22}$ cm$^{-2}$, and X--ray luminous, with L$_{\rm X}>10^{44}$ erg s$^{-1}$. The analysis of the optical spectrum revealed the presence of a broad and red-shifted [OIII] line, with associated mass outflow rates of $\sim$1.4 $\mathrm{M_{\odot} yr^{-1} }$ \citep{brusa2022}. This is the first outflowing quasar isolated from eROSITA data. 

We now extend this approach to the larger eFEDS AGN sample \citep{liuT2021}. From a sample of 22,097 extragalactic sources, 14930 have reliable spectroscopic or photometric redshift (redshift grade 4 or 5 -- see \citealt{salvato2021}). We select AGNs with redshift z$>$0.5, and we isolate 11754 AGNs as our starting sub-sample. The redshift of 0.5 is chosen to encompass the range in which high luminous AGNs peak and the fact that AGN radiative feedback is relevant or common at these high redshift ranges.

\begin{figure*}[!htpb]
\centering
\begin{tabular}{cc}
    \includegraphics[width=0.45\textwidth]{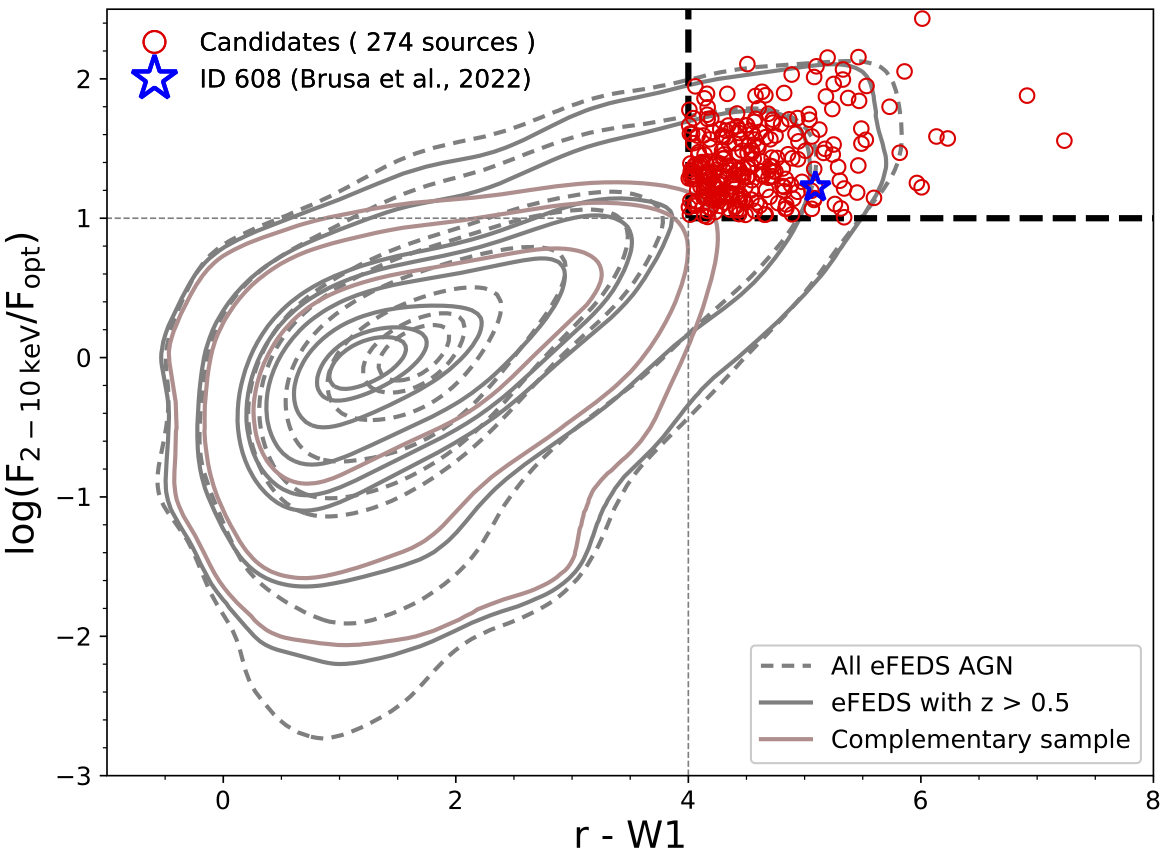} &
    \includegraphics[width=0.45\textwidth]{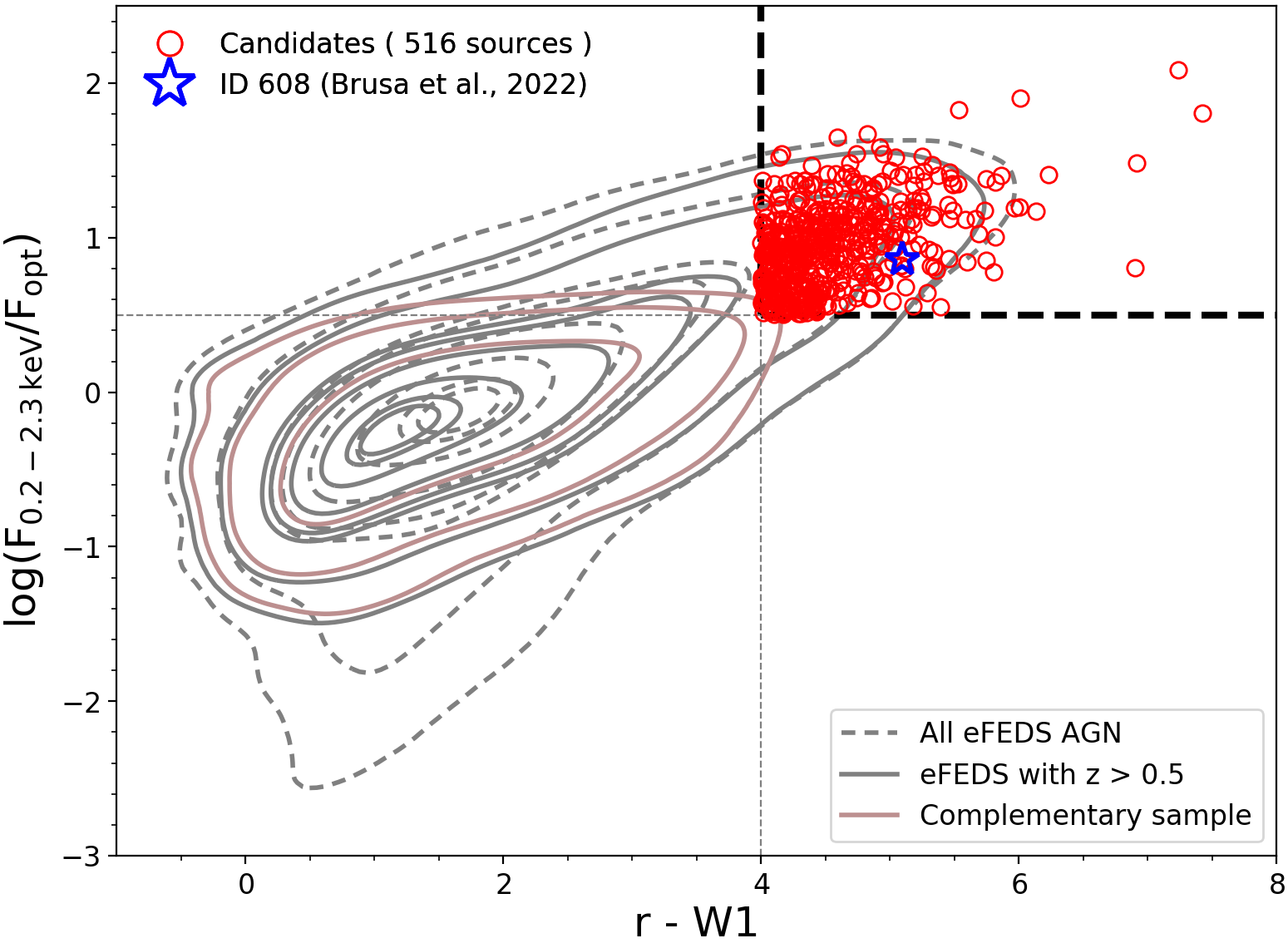} \\
    \includegraphics[width=0.45\textwidth]{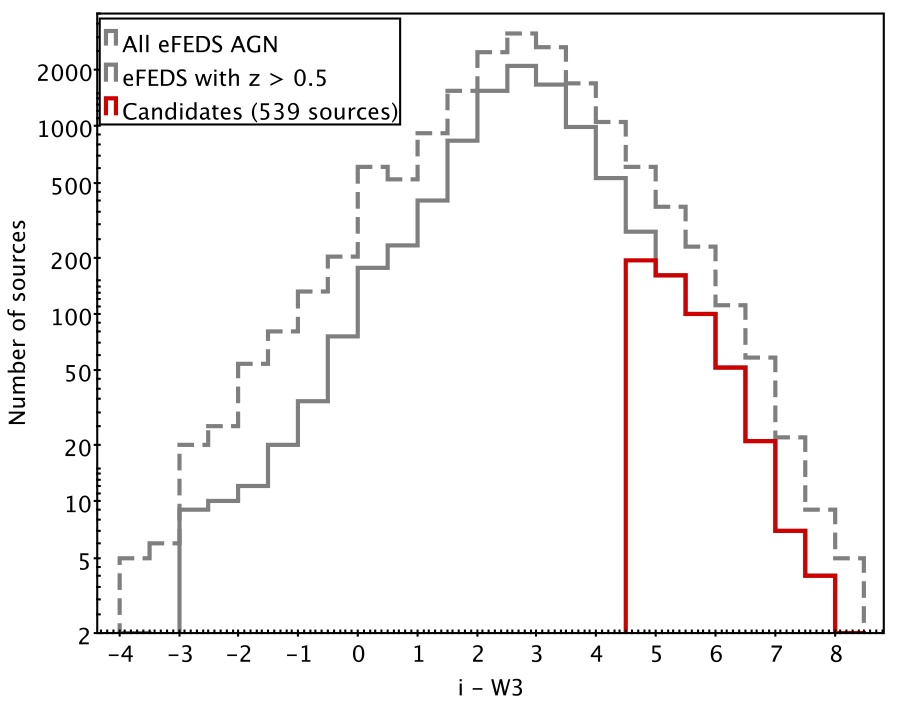} &
    \includegraphics[width=0.45\textwidth]{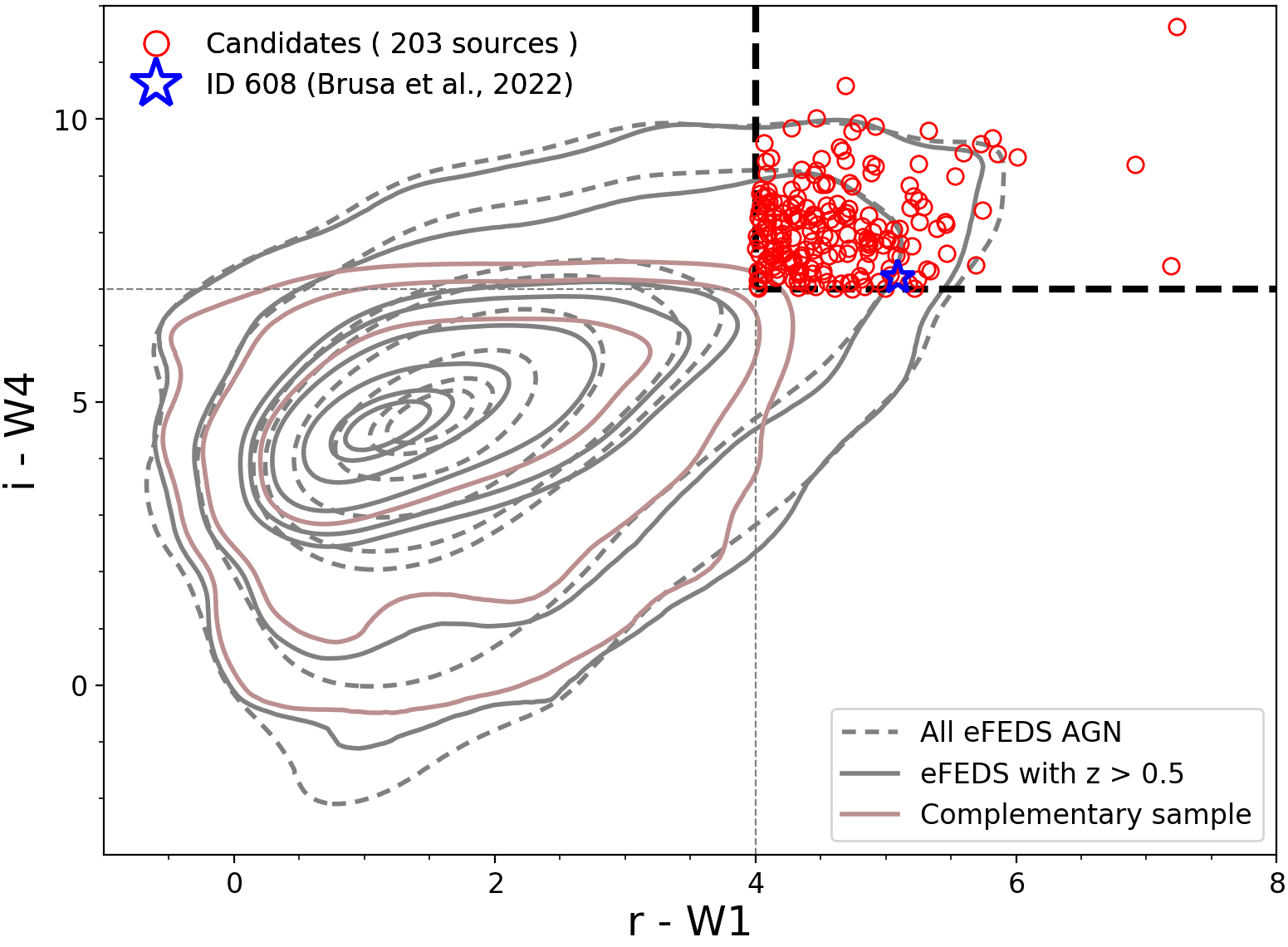} 
\end{tabular}
\caption{
Plots showing the colour selections to isolate AGNs in the feedback phase. Top left panel: The $2-10$ keV to optical flux ratio versus r-W1 colour. The black thick dashed lines represent Eq. \ref{hardXrayfluxtoOpticalflux} and \ref{rmw1} selection locus. Top right panel: The $0.2-2.3$ keV to optical flux ratio versus r-W1 colour. The black thick dashed lines represent Eq. \ref{softXrayfluxtoOpticalflux} and \ref{rmw1} selection locus.  Bottom left panel: Distribution of i-W3 selected sample. Red distribution are selected candidates with Eq. \ref{imw3}. Bottom right panel: i-W4 versus r-W1 selection method. The black thick dashed lines represent the Equations \ref{imw4} and \ref{rmw1} selection locus. In all panels, the red circles are the selected candidates. Brown contours are sources in the complementary locus. Solid grey contours or distribution indicate the eFEDS AGN sample after applying the cut on redshift (z$>$0.5) and dashed grey contours or distribution show the full eFEDS AGN sample. XID 439 or ID 608 \cite{brusa2022} is shown as a blue star. 
}
\label{fig:eFEDSColorSelectionMethods}
\end{figure*}

 We now apply the different criteria to select AGNs in the feedback phase by colour and flux ratios described above to the eFEDS AGN sample. First, we apply a combination of Eq. \ref{rmw1} and Eq.\ref{hardXrayfluxtoOpticalflux} (Eq. \ref{rmw1} $\land$ Eq.\ref{hardXrayfluxtoOpticalflux}) to the eFEDS main sample as illustrated in Figure~\ref{fig:eFEDSColorSelectionMethods}. This selection retrieved 274 sources. Figure~\ref{fig:eFEDSColorSelectionMethods} shows, on the upper left panel, the eFEDS Main catalogue sources populated in this selection plane (contours) and as empty circles the selected candidates. 

Because at z$>$1 the soft X-ray band partially samples the rest-frame hard X-ray emission, allowing us to use the large Main sample, we consider also the soft X-ray to optical flux ratio. The upper right panel of Fig. \ref{fig:eFEDSColorSelectionMethods} shows the soft X--ray to optical flux ratio coupled with the r-W1 selection, that is,
\begin{equation}
    \label{softXrayfluxtoOpticalflux}
    \rm{\left (r - W1> 4 \right) \land \left ( log \frac{F_{0.2-2.3 keV}}{F_{opt}} > 0.5 \right)},
\end{equation}
where $\mathrm{F_{0.2-2.3 keV}}$ is the X--ray flux in 0.2-2.3 keV band and $\mathrm{F_{opt }}$ is the optical flux in r band. From this method, 516 AGNs are isolated. 

Finally, because IR brightness is also an indicator of obscuration, the MIR to optical flux ratio (see also \citealt{Brusa2015,perna2015b}) is combined with the optical to NIR red colour as
\begin{equation}
    \label{imw4}
    \mathrm{\left (r - W1> 4 \right) \land (i - W4} > 7).
\end{equation}
 We present this plane, in the bottom right panel of Fig. \ref{fig:eFEDSColorSelectionMethods}. This criterion selects 203 candidates.

Unlike the other colour selection criteria, used as combinations, Eq.~\ref{imw3} criterion is applied independently as in the previous studies and because AGNs that satisfy this criterion have already demonstrated a correlation between colour and [O III] line properties \citep{perrotta2019}. As stated in \citet{perrotta2019}, a redder colour corresponds to a broader [O III] line profile. Additionally, previous studies by \cite{perrotta2019,vayner2021} have identified outflows in 100\% of their samples that were previously selected with Eq. \ref{imw3}. With the i-W3 colour cut (Eq.~\ref{imw3}, \citet{hamann2016,perrotta2019}), 539 candidates are selected. These are shown in the lower left panel of Fig. \ref{fig:eFEDSColorSelectionMethods}. The number of candidates is comparable to the r-W1 selection.

 Altogether, the above colour selection methods result in a sample of 853 unique candidates which are most likely luminous and obscured and hence, with high chances of containing AGNs in the feedback phase. We refer to these 853 candidates as 'sample A', hereafter. Of these, 352 have redshift of 0.5$<$z$<$1. 
 
In Fig. \ref{fig:eFEDSAllColorSelectionMethods}, all the colour selection methods that resulted in sample A are plotted in the selection diagnostics together.  In this Figure, the selected candidates are scattered in almost all the regions of the planes defined by the main eFEDS AGN sample. This shows that we are selecting these candidates AGNs with strong winds homogeneously, which means minimal bias in our selection methods. It tells us how complicated it is to select these candidates since they do not fall into a defined region. This implies that the complete and pure way to pre-select AGNs with strong winds is by applying a combination of methods as we have done in this study. 

\subsection{Selection based on X--ray and optical spectral properties (Sample B)} 
\label{opticalspectralproperties}
To exert powerful radiative feedback, AGNs in the feedback phase are expected to experience high mass accretion rates and high Eddington ratios~\citep[e.g.][]{hopkins2005,dimatteo2005}. This may drive outflows of gas.

The gas along the line-of-sight is unstable to (Thompson) radiation pressure if the column density is low and the Eddington ratio is high. The wedge is computed by \citet{fabian2008,fabian2009} considering dusty media, with an enhanced cross-section.  \citet{fabian2008,fabian2009} defined a plane of column density ($\mathrm{N_{H}}$) vs. Eddington ratio ($\mathrm{\lambda_{Edd}}$) in which such kinds of sources are expected to appear that is, the region where $\mathrm{N_{H}>10^{21.5} cm^{-2}}$ and $\mathrm{\lambda_{Edd}}$ is greater than the effective Eddington limit for different values of $\mathrm{N_{H}}$. This position in the $\mathrm{N_{H}}$-$\mathrm{\lambda_{Edd}}$ plane has been used to select AGNs in the feedback phase in previous studies \citep[e.g.][]{kakkad2016,Ricci2017,lansbury2020,alonsoherrero2021}. We applied the same selection to the eFEDS sample.

We derive $\mathrm{\lambda_{Edd}}$ for a sub-sample of eFEDS sources with optical spectroscopic observations. It is possible to derive virial black hole masses (MBH) using broad line region (BLR) and the velocity dispersion of the BLR gas from single epoch spectra and line widths \citep[e.g.][]{peterson2004,vestergaard2006,shen2011,woo2015}. 
Following the approach by \cite{WuandShen2022}, BH masses are available for all the sources included on the SDSS DR16 and SDSS DR17 quasar catalogue (DR16Q and DR17Q), by applying \texttt{PyQSOFit}  \citep{guo2018,Shen_2019} to their optical spectra. 4931 sources in eFEDS have BH masses estimated in this way, with 3813 at z$>$0.5.
  
\begin{figure}[!t]
\centering
    \includegraphics[width=0.45\textwidth]{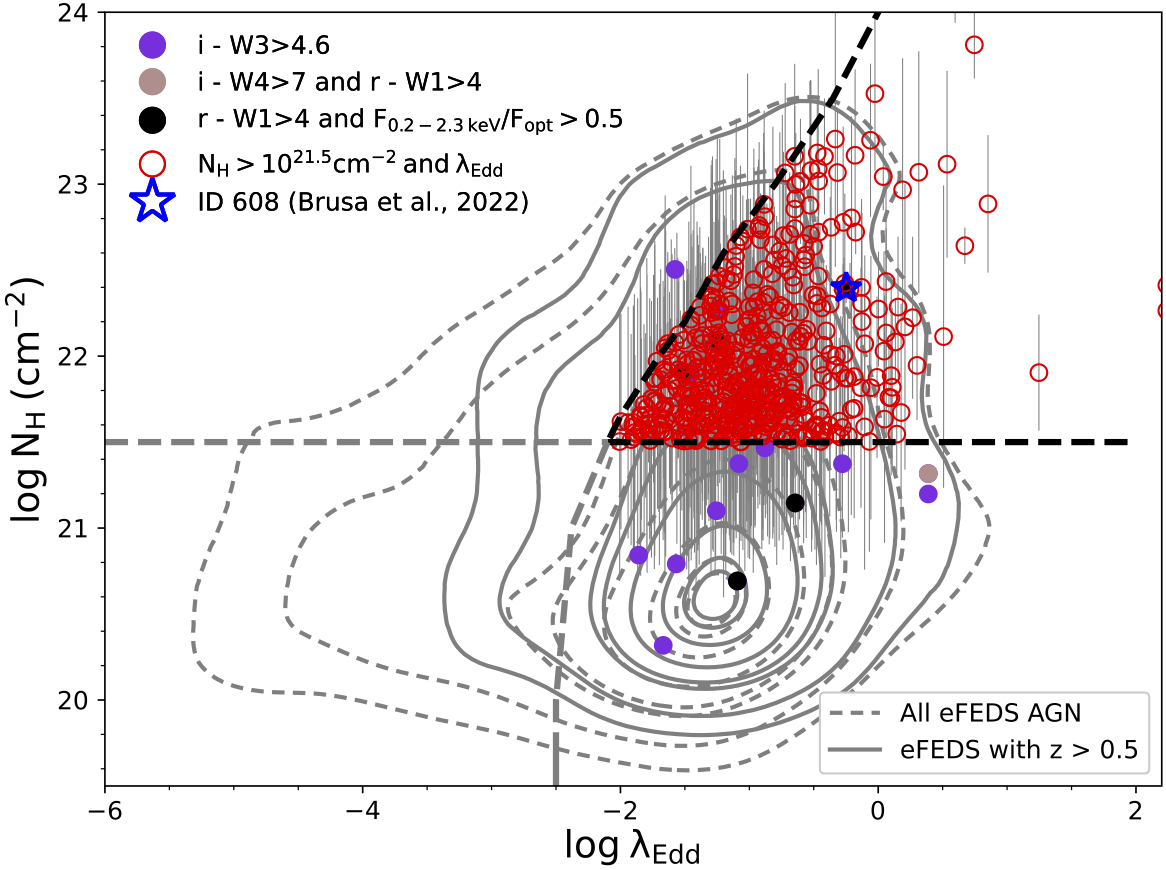}
\caption{ $\mathrm{N_{H}}$ plotted against Eddington ratio. Sources marked in red are the selected candidates. Solid grey contours are eFEDS sample after applying the cut on redshift (z$>$0.5) and dashed grey contours show the full eFEDS sample. ID 608 \citep{brusa2022} (blue star) and the sources isolated from Sect. \ref{ColorselectionineFEDS} are shown with violet circles. The black thick dashed line and curve represent the region $\mathrm{N_{H}>10^{21.5} cm^{-2}}$ and $\mathrm{\lambda_{Edd}>}$ effective Eddington limit. The dotted curve (thick and thin) is the effective Eddington limit for different values of $\mathrm{N_{H}}$ \citep{fabian2009}.}
    \label{fig:eFEDS_NH_Eddratio}
\end{figure}

To obtain $\mathrm{\lambda_{Edd}}$ which is defined as $\mathrm{L_{bol}/L_{Edd}(MBH)}$, we computed the Eddington luminosity from the BH masses. The AGN bolometric luminosity was obtained from the catalogued X--ray luminosity \citep{liuT2021}, applying the bolometric corrections using Eq. 3 in \citet{duras2020}. We populated our sources in the $\mathrm{N_{H}}$-$\mathrm{\lambda_{Edd}}$ plane and we isolated a sample of 528 candidates which show $\mathrm{N_{H}}>10^{21.5}$ cm $^{-2}$ and $\mathrm{\lambda_{Edd}> \lambda_{eff}^{limit}}$. We use the column density values in \citep{liuT2021} that were obtained using a single power law model. The selected sources are shown as red circles in Fig. \ref{fig:eFEDS_NH_Eddratio}. We refer to these 528 candidates as 'sample B', hereafter. Of these, 78 have a redshift of 0.5$<$z$<$1. 

In Fig. \ref{fig:eFEDSAllColorSelectionMethods}, our sample B selected by $\mathrm{N_H}$ and $\mathrm{\lambda_{Edd}}$ appear in the blue region of the colour selection planes used to construct sample A (see previous subsection).
This is due to the fact that these obscured and highly accreting sources are likely being observed immediately before the blow-out phase, or to the fact that the obscuration is inhomogeneous (see Appendix \ref{sectionAPX} for details).

\section{Results}
\label{results}

\begin{figure}[!ht]
\centering
    \includegraphics[width=0.45\textwidth]{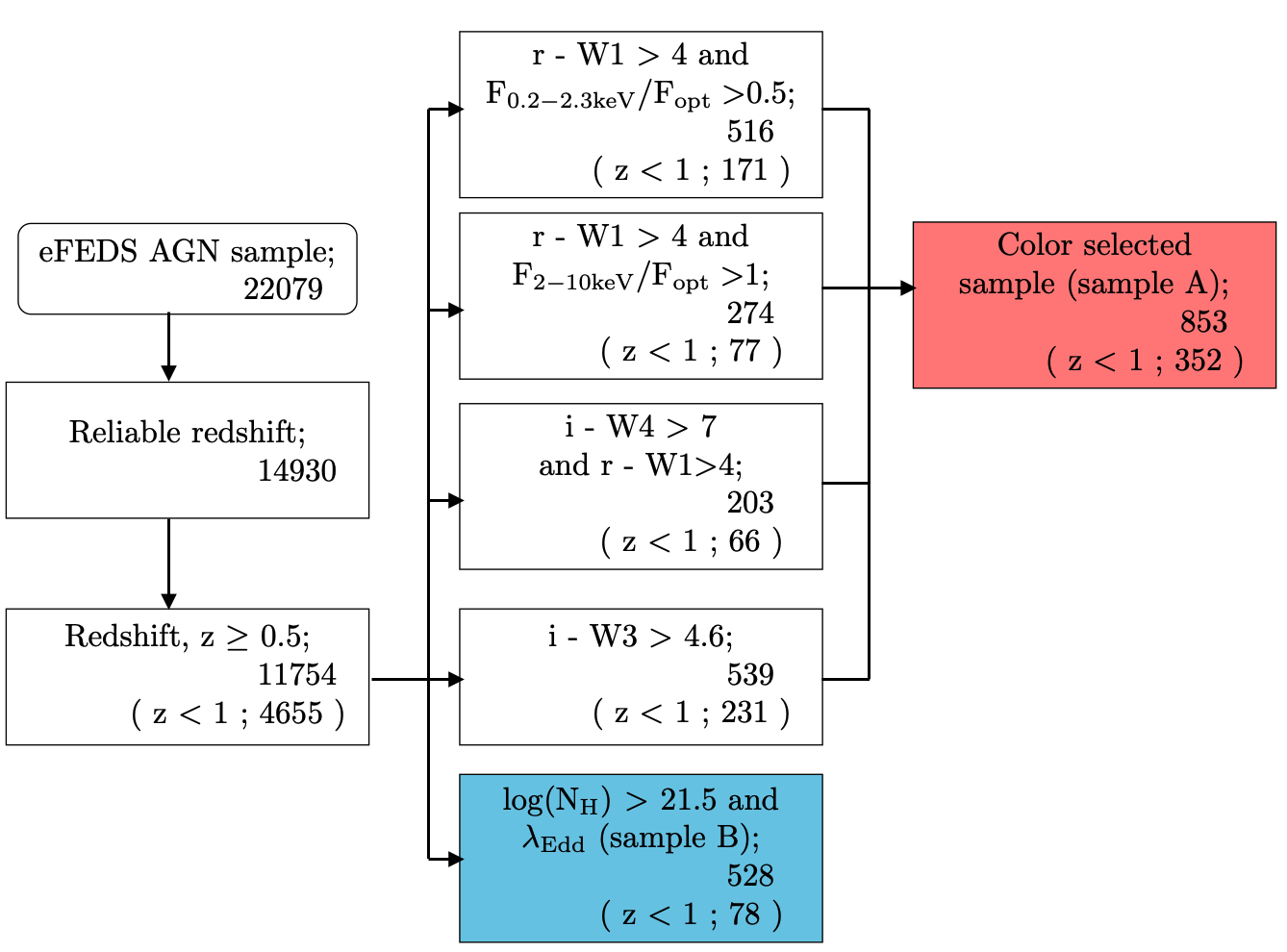} 
\caption{ Flow chart summarising our selection criteria used to isolate candidates from the eFEDS AGN sample. Each box contains a selection method with the cuts applied and the number of sources obtained in the bottom right corner of each box. In brackets, we indicate the number of sources for z less than one (0.5$<$z$<$1) which we apply later on for SDSS spectra analysis. By reliable spectroscopic or photometric redshift, we only consider redshift grade 4 or 5 --see \citealt{salvato2021}. $\rm{F_{opt}}$ refers to flux in the optical r-band. The box highlighted in red is sample A obtained after applying the colour selections in the top right of Fig. \ref{fig:eFEDSColorSelectionMethods}. The box highlighted in blue is the $\mathrm{N_{H} >10^{21.5} cm^{-2}}$ and $\mathrm{\lambda_{Edd}> \lambda_{eff}^{limit}}$ selection or sample B in Fig. \ref{fig:eFEDS_NH_Eddratio} }
\label{fig:flow}
\end{figure}

\begin{figure}[!ht]
\centering
    \includegraphics[width=0.42\textwidth]{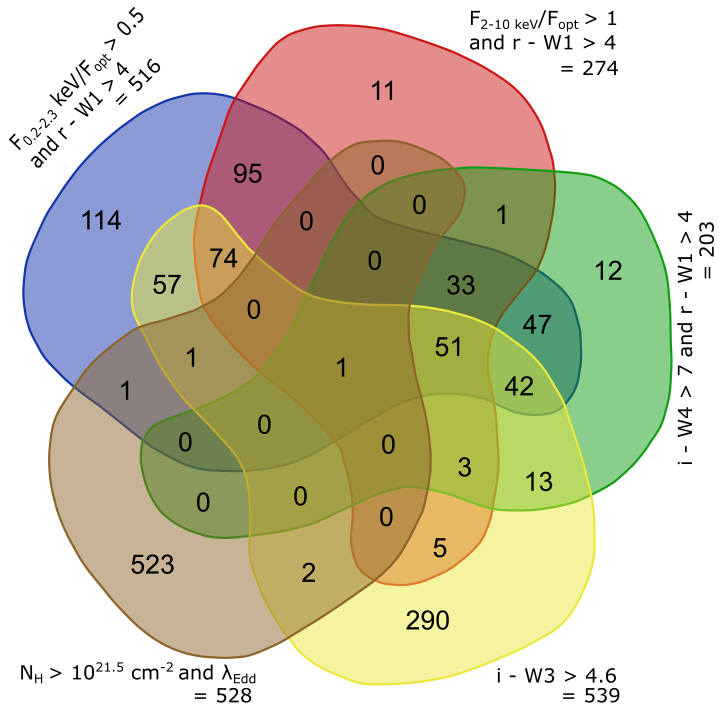} 
\caption{ Venn diagram showing how many sources overlap from our selection criteria. The source that is common in all the four selection method is ID 608 \citep{brusa2022}. $\rm{F_{opt}}$ refers to flux in the optical r-band. The selection methods are indicated by different colours. The Venn diagram was drawn using bioinformatics and evolutionary genomics webtool at
\tiny{\texttt{https://bioinformatics.psb.ugent.be/webtools/Venn/}}
}
\label{fig:venn}
\end{figure}

 The results from the selections presented in Sect.~\ref{ColorselectionineFEDS} and \ref{opticalspectralproperties}
 are summarised in the flow chart shown in Fig. \ref{fig:flow}, with the final Sample A and B and the relevant numbers highlighted in the red and blue boxes, respectively. In total, we isolated 1376 ($\sim$1400) unique candidates ($\sim12$\% of the z$>0.5$ eFEDS AGN population). 
 To clearly visualise how many of our selected candidates from different selection methods overlap, we represent all our samples on the Venn diagram as shown in Fig. \ref{fig:venn}. 950 (69\% of our selected sources) candidates are uniquely selected in an individual selection method. This shows how unique each method is and why using this strategy of applying different methods to isolate candidate AGNs in the feedback phase minimises the chances of missing potential candidates. 
Only one source satisfies all the selection criteria, and this is ID 608, the archetypal outflowing quasar already presented in \citet{brusa2022}.

\subsection{Properties of eFEDS outflowing quasar candidates}
\begin{figure}[!t]
\centering
    \includegraphics[width=0.45\textwidth]{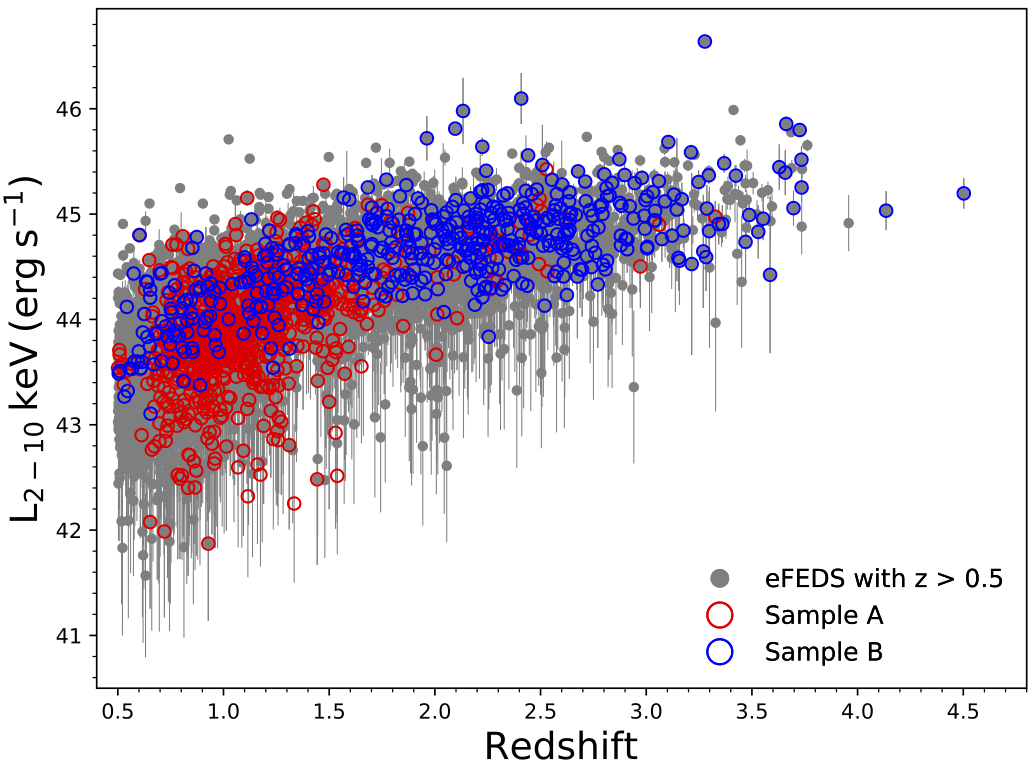} 
\caption{X--ray luminosity and redshift distributions of our two main candidate samples. The grey points indicate the eFEDS AGN sample at z$>$0.5. The red circles indicate sample A obtained after applying the colour selections described in Sect.~\ref{ColorselectionineFEDS} while blue circles indicate Sample B selected as detailed in \ref{opticalspectralproperties}.}
\label{fig:lxz}
\end{figure}

We present the X--ray luminosity and redshift distribution for Sample A and Sample B in Fig. \ref{fig:lxz}, compared with the full eFEDS sample at z$>0.5$. Sample A (in red) populates mainly the low redshift and with lower X--ray luminosities range, while Sample B spans a wider range of both luminosities and redshift. 

\begin{figure}[tpb]
\centering
    \includegraphics[width=0.45\textwidth]{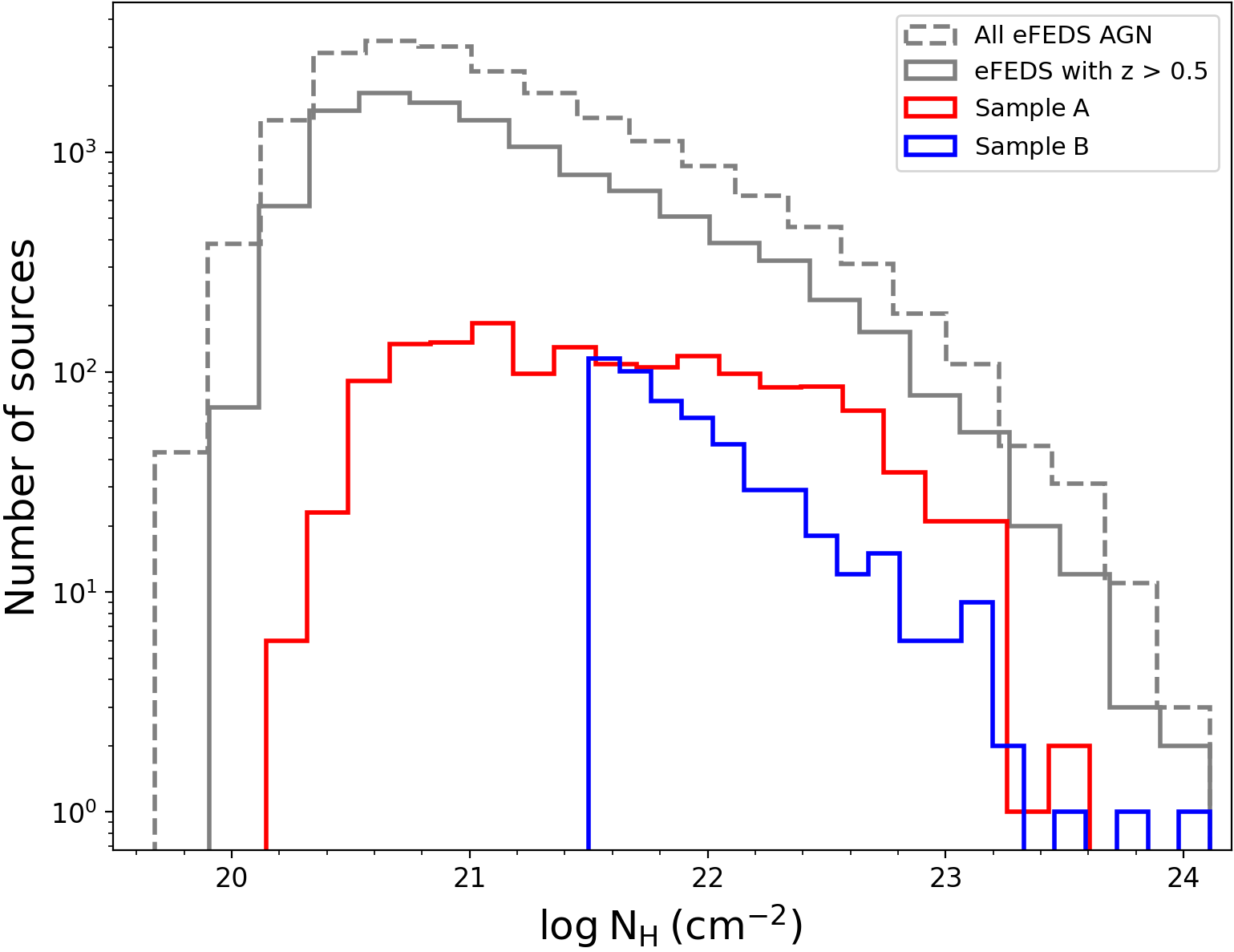} 
\caption{$\mathrm{N_{H}}$ distribution of our selected candidates from the different methods compared with the overall eFEDS AGN sample and subsample at z$>$0.5. Our isolated candidates appear more obscured than the overall eFEDS AGN population and subsample at z$>$0.5. }
\label{fig:NHeFEDSSelectionMethods}
\end{figure}

Figure \ref{fig:NHeFEDSSelectionMethods} presents the $\mathrm{N_{H}}$ distributions of all the candidates outflowing quasars from sample A and sample B. Given that objects in sample B have been selected on the basis of N$_{\rm H}>10^{21.5}$ cm$^{-2}$, their histogram is reported for completeness. 
The objects in sample A appear more obscured than the general eFEDS AGN population. In particular, about 11\% of the eFEDS z$>0.5$ AGN sub-sample have $\mathrm{N_{H}}\mathrm{>10^{22}~cm^{-2}}$ while 
our candidate AGNs in the feedback phase from Sample A show a 3 times higher obscured fraction (30\% have $\mathrm{N_{H}}\mathrm{>10^{22}~cm^{-2}}$). This confirms that the colour selection methods are efficient in isolating more obscured sources than the overall AGN population. 

\begin{figure}[!htpb]
    \centering
    \includegraphics[width=0.45\textwidth]{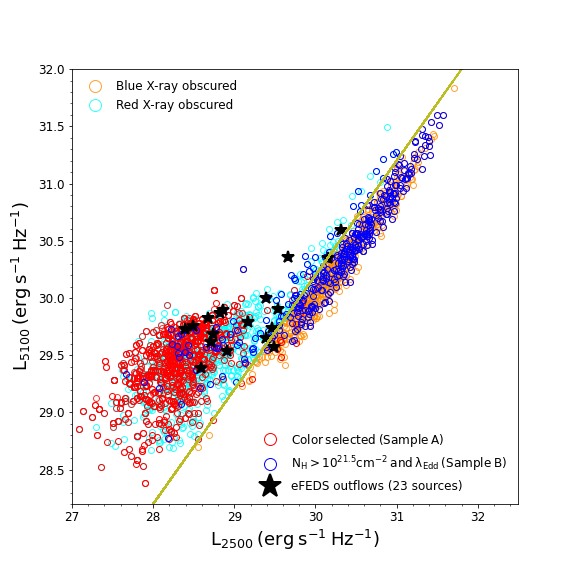}
\caption{Luminosity at 5100 \AA ($\rm{L_{5100\text{\AA}}}$) versus Luminosity at 2500 \AA ($\rm{L_{2500\text{\AA}}}$) for sources in Sample A and B. Representation of the selected samples with the X--ray obscured sources ($\mathrm{N_{H} >10^{21.5} cm^{-2}}$: red obscured and blue obscured based on the line $\rm{L_{5100\text{\AA}}}$-$\rm{L_{2500\text{\AA}}}$=0.2; see Text for details). Candidates in sample B appear in the blue side (selects mostly type 1 AGN). It also correlates well with $\rm{L_{5100\text{\AA}}}$-$\rm{L_{2500\text{\AA}}}$=0.2 relation which indicates their type 1 nature. Sample A indeed appear in the red side (selects mostly type 2 AGN).  
We indicate in black stars the 23 ionised outflows from the eFEDS sample (see Sect. \ref{efeds_spectra} for the discussion of these 23 sources )}.
\label{fig:obscured}
\end{figure}

 Figure \ref{fig:obscured} shows the position of the sources in Sample A and B in the $\mathrm{\rm{L_{5100\text{\AA}}}}$-$\rm{L_{2500\text{\AA}}}$ plane\footnote{The rest frame luminosities at 2500\AA \: and  5100\AA \: have been derived and catalogued in \citep{liuT2021}}. 
 The dividing line at $\mathrm{\rm{L_{5100\text{\AA}}}}$-$\rm{L_{2500\text{\AA}}}$=0.2 can be used  
 to divide the sample in 'blue' and 'red' on the basis of their relative fluxes in the rest frame optical and UV band  \citep{liuT2021}: sources above the dividing line are expected to be reddened. 
  In this figure, we also show all the AGNs in eFEDS with significant X--ray obscuration (N$_{\rm H}>10^{21.5}$ cm$^{-2}$), colour coded according to their position with respect to the dividing line. We refer to sources above the dividing line as 'red X-ray obscured' and those below the dividing line as 'blue X-ray obscured'.
  
  The vast majority of the sources in Sample A lie above the dividing line (see Fig. \ref{fig:obscured}), overlapping the locus of the 'red obscured' sources, as expected. Our Sample B instead correlates well with the orange points ('blue obscured'), in agreement with the fact that they show broad lines in the optical spectra and therefore exhibit less attenuation in the $\rm{L_{2500\text{\AA}}}$. A few sources in our Sample B, however, appear in the region of red obscured sources.

\subsection{ $\rm{[OIII]}$ line SDSS spectral analysis of eFEDS outflowing quasar candidates at 0.5$<$z$<$1 and outflow properties}
\label{efeds_spectra}

Outflowing gas can give rise to asymmetric and/or broad optical line profiles. The presence of ionised outflowing gas is assessed by spectral fitting of the [OIII]4959,5007 forbidden line complex, in the rest-frame optical spectra. This line has been extensively used in the literature for ionised outflow search and characterisation (see e.g. \citealt{fiore2017} and \citealt{leung2019} for recent compilations)

For the spectral analysis, we exploit the Sloan Digital Sky Survey (SDSS I-II-III and IV;  \cite{york2000,gunn2006,smee2013,ahumada2020}). In particular, we use the SDSS public spectra mainly from SDSS-IV \citep{blanton2017} from SDSS data release 16 (DR16) and 17 (DR17) \citep{lyke2020,Abdurro2022}. The SDSS spectral range is such that the [OIII]4959,5007 line complex is sampled only up to z$\sim$1. Therefore, in order to confirm the nature of our candidates, we isolated from Samples A and B all AGNs with spectroscopic redshift of 0.5$<$z$<$1.

Only 78 AGNs from Sample B (out of 528) have 0.5$<$z$<$1, all with spectra. On the other hand, from sample A, only 352 out of the 853 candidates have 0.5$<$z$<$1 and only 6 out of 352 have their spectra available.Two of these 6 sources also appear in sample B (ID608 and ID13349). Therefore, given that 2 sources appear in both samples A and B, in total, we retrieved 82 spectra at 0.5$<$z$<$1 for further spectral analysis.

We then fit our spectra with the python fitting code (PyQSOFit)\footnote{This is a python fitting code used to measure spectral properties of quasars which is also used to decompose different components of the quasar spectra \citep{guo2018,Shen_2019}}. PyQSOFit takes the spectrum, redshift, and line fitting parameters specifying the range and fitting constraints as inputs parameter, performs a continuum modelling and host galaxy subtraction and a multi-component fit of the emission lines. The line best-fit parameters are then returned as outputs \citep{guo2018,Shen_2019}.

We fit the $\mathrm{[OIII]\lambda5007}$ line complex with two Gaussian functions: one is included to model the systemic component, tracing the NLR and the second one is included to model any additional broad component, possibly indicating the presence of outflowing gas. In both cases, the flux ratios between $\mathrm{[OIII]\lambda{4959}}$ and $\mathrm{[OIII]\lambda{5007}}$ were fixed at 1:2.99 \citep{osterbrock1981}. The FWHM of the narrow component was fixed to be less than 550 km s$^{-1}$. The velocity shifts were obtained from the velocity peak of either the broad or narrow emission lines and the systemic velocity.

We ran \texttt{PyQSOFit} and visually inspected all 82 spectra. We discarded 32 of them either because of a very noisy continuum around the [OIII] region or bad residuals after the continuum and host galaxy subtraction. Two out of 32 sources are from sample A.

We then considered the remaining 50 sources. From the spectral fitting of the $\mathrm{[OIII]\lambda{5007}}$ and $\mathrm{[OIII]\lambda{4959}}$ line complex, we found that 21 sources are significantly best fit with two Gaussian components (e.g. the additional broad component has a flux to flux error ratio $>$2.5), showing a blue or red shifted broad $\mathrm{[OIII]\lambda{5007}}$ with FWHM $\mathrm{\sim 600 - 2800 ~km~ s^{-1}}$. Another 17 spectra are best fitted with two Gaussian components. Here the broad component is less significant, with a flux to flux error ratio $<$2.5. The remaining 12 sources are instead fitted with a single Gaussian component. Two of them have FWHM$\mathrm{>800~km~ s^{-1}} $. Our outflow sample is composed of 21 sources with significant broad component detection and the 2 sources with a single component fit, but FWHM$\mathrm{>800~km~ s^{-1}} $. This approach follows previous works \citep[e.g.][]{bischetti2017,perrotta2019,brusa2022}.

In total, we retrieved 23 sources with outflows. Only one of the [OIII] spectral fit of these candidates is shown in the main text in Fig. \ref{fig:efedsSDSSspectra}, while all the other fits (22 sources) are presented in the Appendix (see Appendix \ref{rest of the spectra} for details). We started with a sample of $\sim11750$ AGN, from which we selected $\sim$1400 which are AGN feedback candidates. Out of these, 427 sources have 0.5$<$z$<$1. To test the robustness of our selection, we selected the 50 AGN feedback candidates with best quality SDSS spectra. We found evidence for outflowing gas in $\sim$45\% of the sample (23/50), with a higher fraction if we consider sample A only (3/4). 
\begin{figure}[!t]
\centering
    \includegraphics[width=0.45\textwidth]{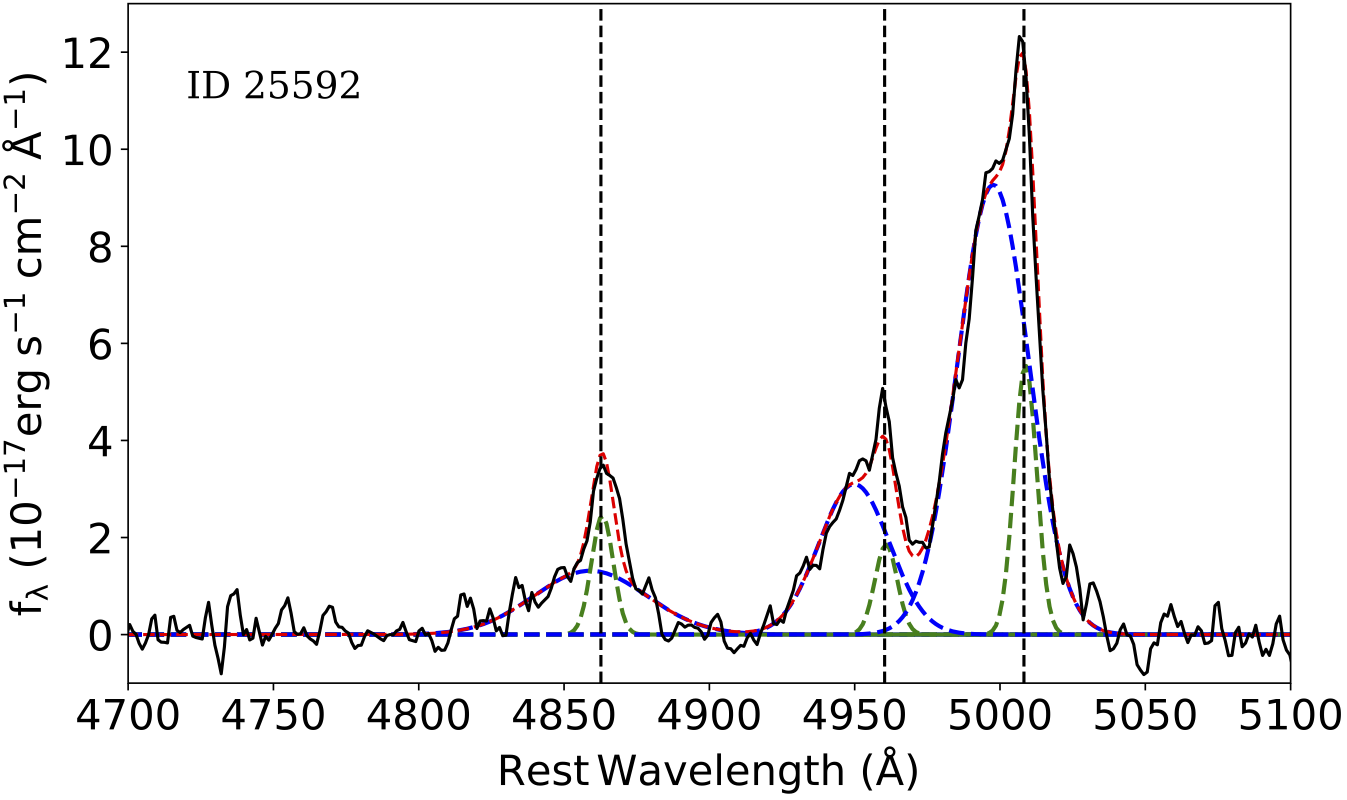} 
\caption{SDSS emission line profiles fit of the H$\beta$+[OIII] line continuum subtracted complex for one of our candidates. The blue dashed line indicated the broad component, the green fit indicates the narrow component. The red dashed line indicates the total fit. The vertical dotted lines indicate the peaks at 4862.68, 4960.30, and 5008.24 for H$_{\beta}$ and [OIII] rest-frame wavelength.}
    \label{fig:efedsSDSSspectra}
\end{figure}

\begin{figure*}[!htpb]
\centering
    \begin{tabular}{cc}
    \includegraphics[width=0.45\textwidth]{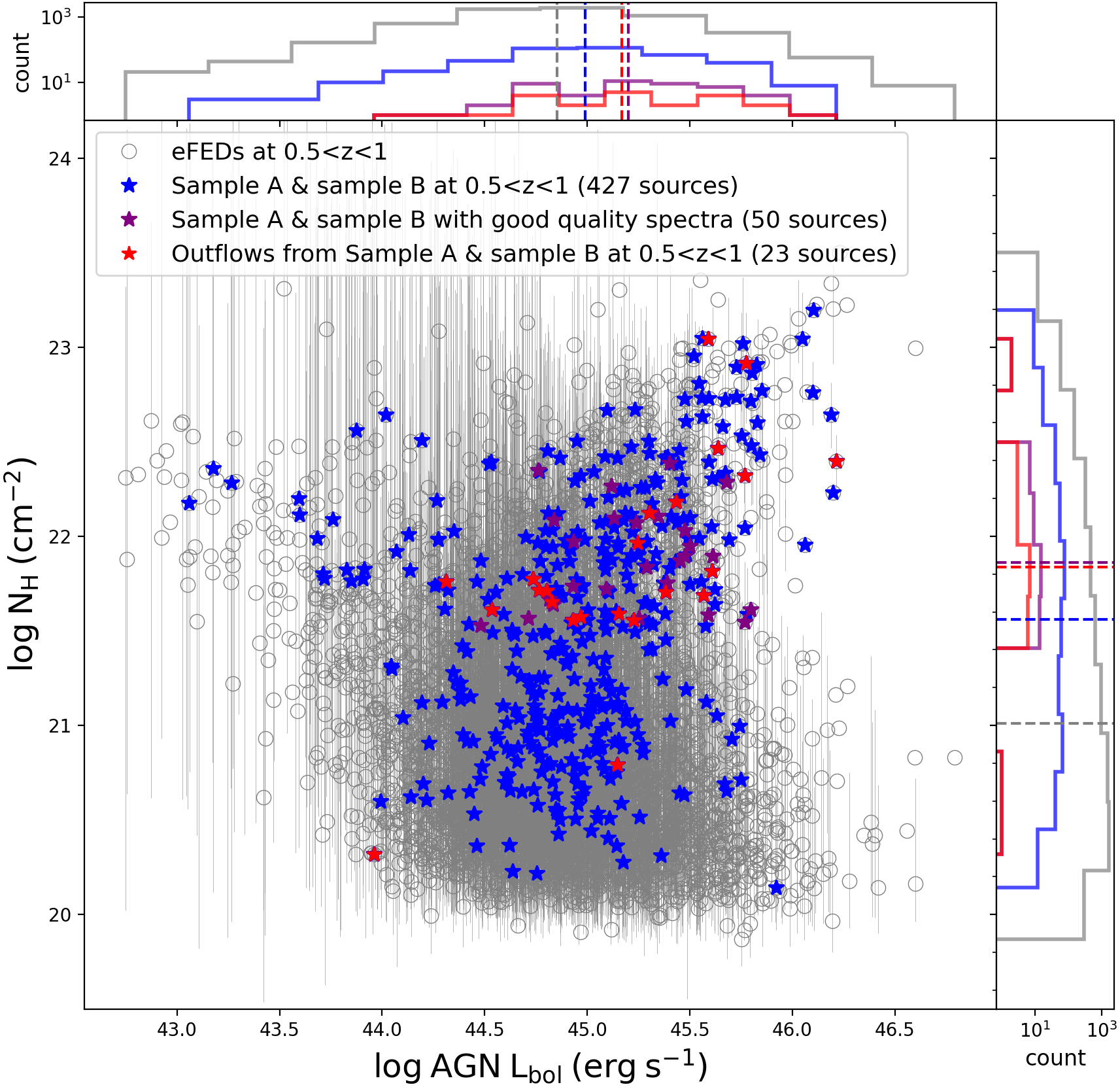} &
    \includegraphics[width=0.45\textwidth]{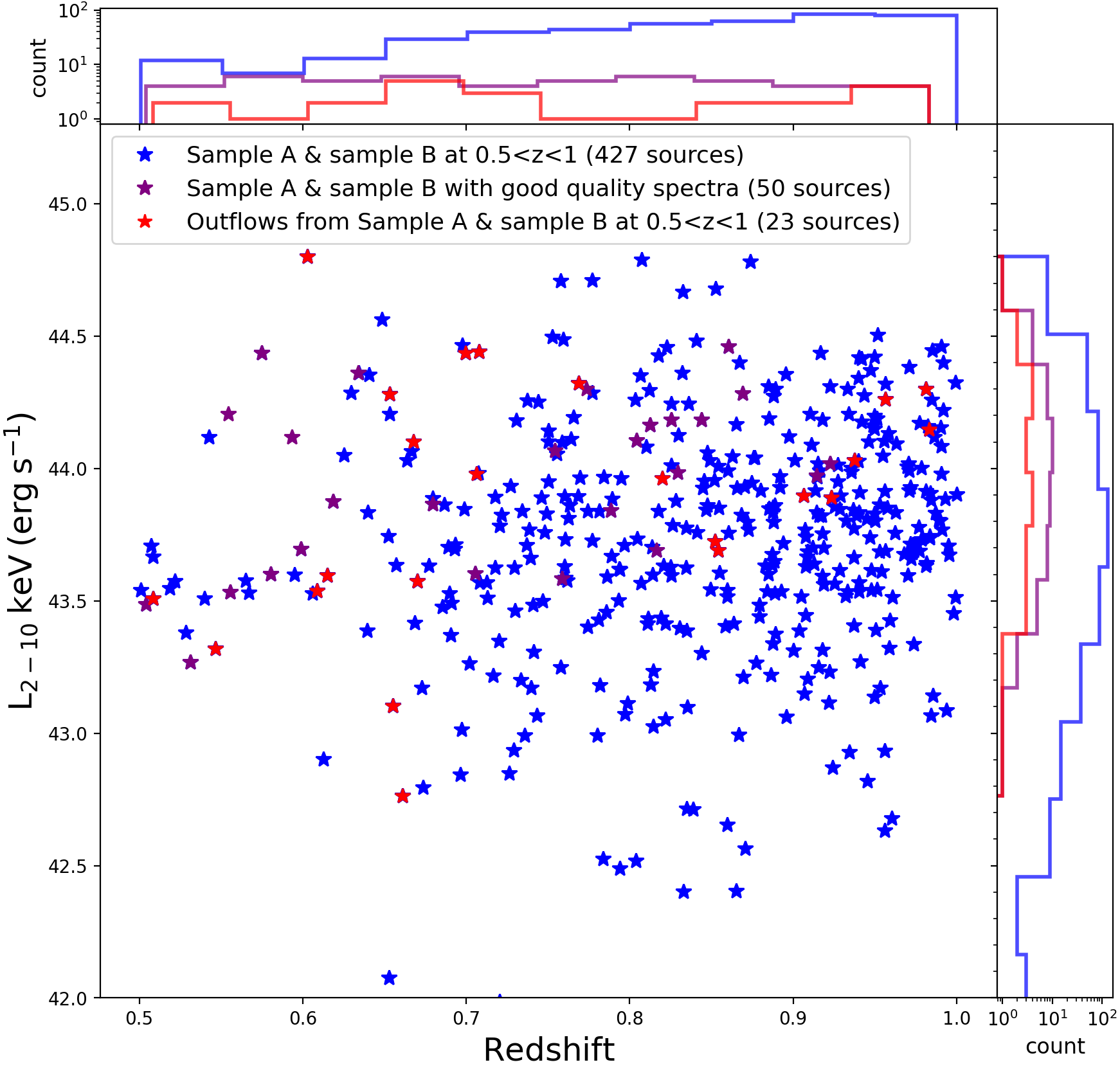} 
    \end{tabular}
\caption{Comparisons of the different samples. Left panel: $\mathrm{N_{H}}$ and AGN bolometric representation of our isolated candidates. Open grey circles represent eFEDS sources at 0.5$<$z$<$1, blue-filled stars represent sample A and sample B at 0.5$<$z$<$1, purple-filled stars represent sample A and sample B with good quality SDSS spectra and the red-filled stars represent the 23 candidates confirmed with outflows. The colours used in the histograms are identical to the colours used in the legend of the scatter plot for the same samples. The dotted lines represent the means of the respective distributions. Right panel: The X--ray luminosity and redshift distributions of three samples; sample A and sample B at 0.5$<$z$<$1, sample A and sample B with good quality SDSS spectra and 23 candidates confirmed with outflows. The colours used in the histograms are identical to the colours used in the legend of the scatter plot for the same samples. }
\label{fig:nh_lbol}
\end{figure*}

 In order to investigate the limitations and biases arising from our small sample sizes, we check how they accurately represent the original samples. We focus on parameters such as column density (N${\rm H}$), bolometric luminosity (L${\rm bol}$), X-ray luminosity (L$_{\rm x}$), and redshift (z). Figure \ref{fig:nh_lbol} illustrates this representation using various samples which include: The eFEDS sources at 0.5$<$z$<$1, Sample A and Sample B at 0.5$<$z$<$1 (consisting of 427 sources; 'parent sample'), Sample A and Sample B with good quality spectra (comprising 50 sources), and 23 sources identified with outflows.

From the left panel of this Figure, we observe that the sources with good-quality SDSS spectra (indicated by purple stars) are clustered towards high column densities and slightly high bolometric luminosities with respect to their parent sample (indicated by blue stars).  On the other hand, the sources with clear outflow signatures (indicated by red stars) cover the same parameter space as those with good-quality SDSS spectra. The fact that the sources with outflows have on average N$_{\rm H}\sim10^{22}$ cm$^{-2}$ and high bolometric luminosities $\mathrm{log L_{bol}\sim ~45.2~erg~s^{-1}}$ than the parent sample may be ascribed to a selection bias affecting the sample with good quality SDSS spectra, which is dominated by Sample B. Since N$_{\rm H}$ is a direct parameter in one of our selection methods, we plot the X--ray luminosity and redshift distributions again (right panel) but only for three samples to asses their representation. We find that our samples are generally consistent with each other in terms of redshift, but we do tend to pick up sources with slightly higher X--ray luminosities. 

We refer the reader to Appendix \ref{selection_summary} for comprehensive details regarding the number of sources identified with outflows from individual selection methods. Having in mind that the sources with detected outflows may be a genuine representation only of the most obscured and luminous sources in the parent sample, we use our small samples to discuss the reliability and validity of the statistical analyses conducted in the study.

The number of sources with outflows has narrowed down to less than half of the candidates with the best quality SDSS spectra. This study primarily relies on SDSS spectra, which are prone to poor signal-to-noise conditions, consequently imposing limitations on the accuracy of our results. Furthermore, our ability to trace outflows is constrained by the wavelength range available, thereby restricting our study to a single phase (ionised). Our relatively small sample size can be attributed to these inherent limitations. However, we note that the identification of these outflows, despite their limited numbers, could potentially be linked to the duty cycle of outflows and the blow-out phase being rather short \citep{Schawinski2015,kingnixon2015} compared to the average AGN timescales in evolutionary models. This means that only a few fractions of AGNs can be detected in the blow-out phase. 

We list in Table \ref{tab:fitresults1} the fitting results for these 23 sources, that is the narrow and broad FWHM, the broad [OIII] flux (and its significance) and the corresponding luminosity with their uncertainties as estimated using a Monte Carlo approach that is already included in PyQSOFit package \citep{guo2018,Shen_2019}. 
For each source, we defined a maximum outflow velocity (V$_{\rm max}$) as 
\begin{equation}
    \label{vmax}
    \rm{V_{max}= |\Delta{V}|+2\sigma_{[OIII]}^{broad}},
\end{equation}
where $\Delta{V}$ is the velocity shift between the velocity peak of the broad emission line and the systemic velocity and $\sigma$ is the velocity dispersion (see Appendix~\ref{assumptions} for a detailed discussion on the computation of outflow velocities in the literature). This quantity is also listed in Table \ref{tab:fitresults1}, along with $\Delta{V}$ and $\sigma$. Other properties of these 23 sources (N$_{\rm H}$, L$_{\rm X}$, L$_{\rm bol}$ and M$_{*}$) are reported in the first part of Table~\ref{tab:fitresults2}.

\begin{table*}
    \centering
   \caption{ X-ray properties, stellar mass and outflow properties of our candidates with outflows} 
   \begin{threeparttable}
    \begin{tabular}{ |c|c|c|c|c|c|c|c|c|c|}
\hline
 &  &  &  &  &  &  &  &  &  \\
EROID & $\mathrm{log \:N_{H}}$ & log  $\rm{L_{x}}$ & log $\mathrm{L_{bol}}$ & log ${M_{*}}$ & $\rm{R_{out}}$ & $\rm{M_{out}}$ & ${\dot{M}}$ & ${\dot{E}}$ & ${\dot{P}_{OF}}$  \\
 & $\mathrm{cm^{-2}}$ & $\mathrm{erg~s^{-1}}$ & $\mathrm{erg~s^{-1}}$ & $\mathrm{{M}_\odot}$ & kpc & $\mathrm{{M}_\odot}$ ($10^6$) & $\mathrm{{M}_\odot {yr}^{-1}}$ & $\mathrm{erg~s^{-1}}$ ($10^{41}$) & $\mathrm{dyne}$ ($10^{34}$)  \\
 \hline
25592 & 20.3 & 42.7 & 43.9 & 11.2 & 5.4 & 10.7$\pm$0.1 & 17.1$\pm$0.3 & 241.9$\pm$8.4E-6 & 22.9 \\
6743 & 23.0 & 44.2 & 45.5 & 10.8 & 6.5 & 0.8$\pm$0.03 & 0.6$\pm$0.06 & 2.6$\pm$9.4E-5 & 0.4 \\
608 & 22.3 & 44.8 & 46.2 & -99.0 & 10.0 & 4.4$\pm$0.2 & 2.6$\pm$0.1 & 17.2$\pm$8.4E-5 & 2.4 \\
19520 & 22.1 & 44.0 & 45.3 & 9.8 & 4.5 & 13.0$\pm$0.9 & 22.6$\pm$2.2 & 260.6$\pm$0.005 & 27.3 \\
18777 & 21.5 & 43.7 & 44.9 & 10.9 & 9.0 & 1.8$\pm$0.1 & 2.7$\pm$0.3 & 94.7$\pm$0.005 & 5.7 \\
21835$^*$ & 21.8 & 44.3 & 45.6 & 10.9 & 9.3 & 6.3$\pm$0.5 & 2.1$\pm$0.3 & 4.2$\pm$6.2E-4 & 1.1 \\
12686 & 22.4 & 44.3 & 45.6 & 10.9 & 6.0 & 4.1$\pm$0.3 & 3.1$\pm$0.6 & 12.0$\pm$0.006 & 2.2 \\
31136 & 21.7 & 43.1 & 44.3 & -99.0 & 10.0 & 4.5$\pm$0.4 & 2.9$\pm$0.3 & 24.0$\pm$8.7E-4 & 3.0 \\
5896$^*$ & 22.3 & 44.4 & 45.7 & -99.0 & 10.0 & 9.1$\pm$0.9 & 3.4$\pm$0.5 & 8.9$\pm$0.001 & 2.0 \\
16271 & 21.7 & 43.5 & 44.7 & 10.3 & 2.0 & 0.6$\pm$0.09 & 1.5$\pm$0.3 & 7.3$\pm$0.007 & 1.2 \\
14896 & 21.5 & 43.9 & 45.2 & 11.2 & 3.5 & 1.7$\pm$0.2 & 2.5$\pm$0.5 & 13.9$\pm$0.008 & 2.1 \\
10152 & 20.7 & 43.8 & 45.1 & -99.0 & 10.0 & 2.2$\pm$0.3 & 1.9$\pm$0.9 & 27.4$\pm$0.4 & 2.6\\
11596 & 21.6 & 44.2 & 45.5 & 10.9 & 3.9 & 7.3$\pm$1.3 & 20.2$\pm$4.0 & 452.4$\pm$0.04 & 34.0\\
14390 & 21.7 & 43.5 & 44.7 & 10.2 & 3.0 & 5.7$\pm$1.1 & 6.4$\pm$1.3 & 13.7$\pm$5.4E-4 & 3.3 \\
16050 & 21.6 & 43.5 & 44.8 & 11.1 & 7.7 & 2.8$\pm$0.6 & 1.4$\pm$0.6 & 3.9$\pm$0.04 & 0.8 \\
19964 & 21.7 & 43.5 & 44.8 & 10.5 & 5.7 & 0.4$\pm$0.1 & 0.1$\pm$0.06 & 0.1$\pm$4.7E-4 & 0.1 \\
14541 & 21.5 & 43.8 & 45.1 & 10.0 & 4.6 & 2.1$\pm$0.6 & 6.8$\pm$2.9 & 274.6$\pm$2.5 & 15.4 \\
19174 & 22.9 & 44.4 & 45.7 & 11.0 & 5.7 & 12.3$\pm$3.9 & 11.9$\pm$8.7 & 70.1$\pm$6.3 & 10.3 \\
28337 & 21.9 & 43.9 & 45.2 & -99.0 & 2.9 & 1.0$\pm$0.3 & 0.7$\pm$0.5 & 0.7$\pm$0.04 & 0.3 \\
16799 & 21.6 & 43.3 & 44.5 & 10.8 & 7.8 & 0.6$\pm$0.2 & 0.1$\pm$0.2 & 0.1$\pm$0.1 
 & 0.04\\
19841 & 21.5 & 43.6 & 44.9 & -99.0 & 10.0 & 1.5$\pm$0.5 & 0.4$\pm$0.1 & 0.9$\pm$7.1E-4 & 0.2\\
17754 & 21.7 & 44.1 & 45.3 & 10.9 & 6.3 & 0.5$\pm$0.1 & 0.2$\pm$0.1 & 0.6$\pm$0.004 & 0.1 \\
15417 & 22.1 & 44.1 & 45.4 & 10.3 & 7.8 & 9.9$\pm$3.5 & 6.2$\pm$2.3 & 28.1$\pm$0.02 & 4.7 \\

\hline
\end{tabular}
    \begin{tablenotes}
        
        \item Notes: The columns (left-right) are eROSITA ID (EROID ) of our candidates with outflows, column density ($\mathrm{log \:N_{H}}$), X--ray luminosity ( $\rm{L_{x}}$; 2-10 keV), AGN bolometric luminosity (AGN $\mathrm{L_{bol}}$) and stellar mass (${M_{*}}$). Column density and X--ray luminosity are obtained from \citep{liuT2021}, AGN bolometric luminosity calculated from X--ray luminosity and the bolometric corrections using Eq. 3 in \cite{duras2020} relation. Followed by outflow properties; radius ($\rm{R_{out}}$), outflow mass ($\rm{M_{out}}$), mass outflow rate (${\dot{M}}$), outflow kinetic power (${\dot{E}}$) and momentum flux of the outflow(${\dot{P}_{OF}}$) as estimated in this study. -99.0 is used to indicate missing values. Sources marked with $^*$ have single Gaussian components.
      \end{tablenotes}
    \end{threeparttable}
\label{tab:fitresults2}
\end{table*}

\subsection{Ionised AGN outflow properties}
\label{outflowproperties} 
From the spectral analysis, quantities such as the measured velocities, and the emission line luminosities can be derived. Outflow physical properties such as mass outflow, mass outflow rate and kinetic power, can then be constrained by applying a set of standardised assumptions.

As far as the $\mathrm{[OIII]}5007$ line is concerned, and adopting the approach proposed by \cite{canodiaz2012} and \citet{carniani2015}, the ionised mass outflow is given by:
\begin{equation}
    \label{mout}
   \mathrm{ M_{out}^{ion}}=\rm{5.33\times{10^7}M_{\odot}\left (\frac{C}{10^{[O/H]}}\right)\left(\frac{L_{[OIII]}}{10^{44}erg s^{-1}}\right)\left(\frac{<n_{e}>}{10^3 cm^{-3}}\right)^{-1}},
\end{equation}
where $\rm{L_{44}([OIII])}$ is the luminosity of the broad component of the  $\mathrm{[OIII]}$ line in units of $\mathrm{10^{44}~erg~s^{-1}}$, $\mathrm{n_{e}}$ is the outflowing gas electron density in units of $\mathrm{10^3  cm^{-3}}$,  $\mathrm{10^{[O/H]}}$ is the oxygen abundance in solar units and C is a factor that encodes the condensation factor of the gas clouds (see \citealt[][for details]{canodiaz2012}. The ionising gas clouds are always assumed to have the same density thus the condensation factor C is approximated to 1, and the metallicity of the outflowing material is always assumed solar.

From Eq. \ref{mout}, the mass outflow rate can be calculated either assuming spherical or shell$-$like geometry. In the case of spherical mass outflow rate, from the fluid field continuity equation, $\mathrm{\rho=\frac{3M_{out}}{\Omega{\pi}R^3}}$ is the mean density of an outflow, while $\mathrm{\Omega{\pi}}$ is the solid angle covered by the outflow.  The mass outflow rate can thus be written as; 
\begin{equation}
    \label{moutrate1}   {\dot{\textrm{M}}_\mathrm{{out}}^{\rm{ion}}}=\rm{\Omega{\pi}R^2\rho{\nu}= 3\frac{M_{out}^{ion}V_{out}}{R_{out}}},
\end{equation}
where $\rm{V_{out}}$ is the outflow velocity, and $\rm{R_{out}}$ (in kpc) is the radius at which the outflow is computed\footnote{For an outflow assumed to propagate in a thin shell of thickness $\rm{\Delta{R}}$, the mass outflow rate is given by
\begin{equation}
    \label{moutrate3}   {\dot{\textrm{M}}_\mathrm{{out}}^{\rm{ion}}}=\rm{\frac{M_{out}^{ion}V_{out}}{\Delta{R}}}. 
\end{equation}
}.  

By substituting Eq. \ref{mout} in \ref{moutrate1}, and assuming solar metallicity and a condensation factor C$=$1, the mass outflow rate in units of $\rm{M_{\odot}yr^{-1}}$ can be rewritten as;
\begin{equation}
\begin{split}
  \label{moutrate2}
    {\dot{\textrm{M}}_\mathrm{{out}}^{\rm{ion}}}=  \rm{164\left(\frac{kpc}{R_{out}}\right)\left(\frac{L_{[OIII]}}{10^{44}erg~s^{-1}}\right)} \\ 
    {\left( \frac{\rm{V_{out}}}{1000\; \rm{km s^{-1}}}\right)\left(\frac{<\rm{n_{e}}>}{10^3\; \rm{cm^{-3}}}\right)^{-1}}.
\end{split}
\end{equation}
The momentum flux of the outflow can then be computed as,
\begin{equation}
    \label{mf}
    {\dot{\textrm{P}}}={\dot{\textrm{M}}_\mathrm{{out}}^{\rm{ion}}}\times{\rm{V_{out}}}.
\end{equation}
Finally, the kinetic luminosity can be deduced as,
\begin{equation}
    \label{ke}
    {\dot{\textrm{E}}_\mathrm{{kin}}}={\frac{1}{2}\times{ {\dot{\textrm{M}}_\mathrm{{out}}^{\rm{ion}}}\times{\rm{V^{2}_{out}}}}}.
\end{equation}

For all the sources in Table \ref{tab:fitresults2}, we calculated the outflow properties using Eq. \ref{moutrate2}, \ref{mf} and \ref{ke}. 
Given that for our eFEDS sources we do not have spatially resolved data, we assumed as outflow radius (R$_{out}$) the galaxy half-light radius as estimated from AGN-host galaxy image decomposition (\cite{li2023} following the method described in \citealt{lij2021}) on deep i-band data obtained in the footprint of the Hyper Suprime Cam (HSC, \citealt{miyazaki2018}) Subaru Strategic Program (HSC-SSP: \citealt{aihara2018a,aihara2018b,aihara2019}). Most likely, the outflow will be contained within the host galaxy therefore the assumed radius will provide conservative estimates of the mass outflow rates. As outflow velocity V$_\mathrm{out}$, we used the maximum velocity calculated from our fit parameters using Eq. \ref{vmax}, while to compute the [OIII] line luminosity we considered the flux associated with the detected broad component.

Finally, our data do not permit a direct estimate of the gas electron density. Therefore, we assumed for n$_e$ a value of 200 $\mathrm{cm^{-3}}$, that is the one used in \cite{fiore2017} and within the range of the measured values of other high-redshift targets \citep[][among others]{nesvadba2006,perna2015a,Brusa2015,brusa2016}. 

We measure mass outflow rates in the range $\mathrm{\sim 0.2 - 23~ M_{\odot}~yr^{-1} }$, kinetic powers in the range ${log(\dot{E})~\sim~40-44}$~erg~s$^{-1}$ and outflow momentum rates in the range $\rm{4.8~\times~10^{32}~dyne}$ - $\rm{3.4~\times~10^{35}~dyne}$. When compared to the radiative momentum flux from the central black hole (${\dot{\textrm{P}}_\mathrm{{AGN}}=\mathrm{L_{bol}/c}}$), we obtain momentum flux ratios in ranges of $\sim$0.02 - 75 with only one source having  ${\dot{\textrm{P}}_\mathrm{{OF}}}/\dot{\textrm{P}}_\mathrm{{AGN}}$ above 20. All these values are listed in the second part of Table~\ref{tab:fitresults2}. The several assumptions on different parameters such as geometry, electron density, temperature, and in some cases, the velocity of the outflowing gas and radius introduce at least 2-3 orders of magnitude uncertainties in the final estimated values of the outflow properties (see Appendix \ref{assumptions} for details). 

\section{Discussion}
\label{discussion}
\subsection{Selection of quasars in the feedback phase}
\label{selection_discussion}
Any single selection criterion may be incomplete and biased for the identification of the blow-out candidates. This is especially problematic when the sub-population of interest constitutes a small minority with uncertain and potentially diverse observational signatures. This is the case for AGNs in the blow-out phase, where we expect different observed and physical properties depending on the exact coupling of the AGN feedback with the ISM and the level of the obscuration, and where everything is further complicated by redshift effects. This work merges several complementary selection criteria to enable a more complete selection and minimise the selection biases of individual selections.

In Sect. \ref{sampleselection}, we applied a variety of methods used in previous studies \citep{Brusa2015,kakkad2016,perrotta2019,vayner2021,brusa2022} to a large, homogenous sample of X--ray selected AGNs in the eFEDS field \citep{liuT2021}. We constructed two samples of candidate AGNs with outflows, highly complementary in terms of luminosity and redshift distribution. The two samples were defined based on the combination of colour-selection methods targeting mostly sources where reddening, extinction or obscuration are important (Sample A) and on the  $\mathrm{N_{H}}$-$\lambda_{\rm Edd}$ diagnostic (Sample B). This resulted in a total of $\sim$1400 unique sources.

We tested the efficiency of the selection via spectroscopic analysis of SDSS optical spectra, looking for unsettled gas motions in the [OIII] line emission. We can conduct this experiment only for the $\sim430$ sources at 0.5$<$z$<$1. All sources in Sample B have available spectra, but only $\sim\!15$\% are at 0.5$<$z$<$1. On the other hand, despite $\sim \!40$\% of sources in Sample A having 0.5$<$z$<$1, optical spectra exist only for 6 of them. This implies that we are able to validate the efficiency of the selection only on limited samples, for example, 20\% of the 0.5$<$z$<$1 sample and $\sim$6\% of the overall candidates. Although incompleteness and selection biases may still affect our subsample of outflowing sources, our work still enables us to increase the overall number of outflowing AGNs so far detected and thus provides additional validation tests for the scaling relations previously published in the literature.

We report that $\sim$45\% of the sources with available good quality spectra from the combined sample (Sample A + B) have clear signatures of outflows (see Sect.~\ref{efeds_spectra}) and this may be as high as 80\% if we consider sources for which a broad component can be accommodated in the fit, but at a lower significance. This is a striking confirmation of the efficiency of the combined selection to recover this rare population, at least at 0.5$<$z$<$1, and among the most obscured and luminous sources.

When only sources from Sample A are considered, the fraction of significant broad line detections rises from 45\% to 75\%. In particular, strong winds in red sources are expected in the very initial stages of models in which AGN feedback is radiatively driven and launched by trapped IR radiation  \citep[e.g.][]{costa2018a, costa2018b}. In these same models, the optical depth in the UV and IR is initially very high along all lines of sight (Costa, private communication). Based on the limited and very small but satisfactory data obtained from the colour selection (3/4 sources), it appears that our combined colour selection method could potentially be the most effective approach for isolating sources in the initial phases of the outflow.

The column density versus Eddington ratio-based selection ('forbidden region' in the $\lambda_{Edd}$-N$_{H}$ plane which defines Sample B) may still trace radiatively driven outflows episodes launched by trapped IR radiation. The adopted selection diagnostic returned, however, mostly AGNs with blue optical/IR colours. This is largely due to a selection bias: BH masses are hardly measurable in optical spectra for red and/or Type 2 AGN, for which instead dedicated deep or NIR follow-up is needed \citep[see e.g.][]{bongiorno2014}. In most hydrodynamical simulations the timescale for the blow-out phase is predicted to be very short ($<$ few$\times 10$-100 Myr;  \cite[see e.g.][]{hopkins2008, yutani2022}) and the average value of the optical depth also decreases very rapidly. After 10 Myr, the average line of sight UV and IR depth may be already diminished by a factor of $\sim100$, facilitating the leaking of blue continuum radiation from the inner regions of the accretion disc.

\subsection{AGN and outflow properties correlations}
\label{correlations}

\begin{figure}[!t]
    \includegraphics[width=0.45\textwidth]{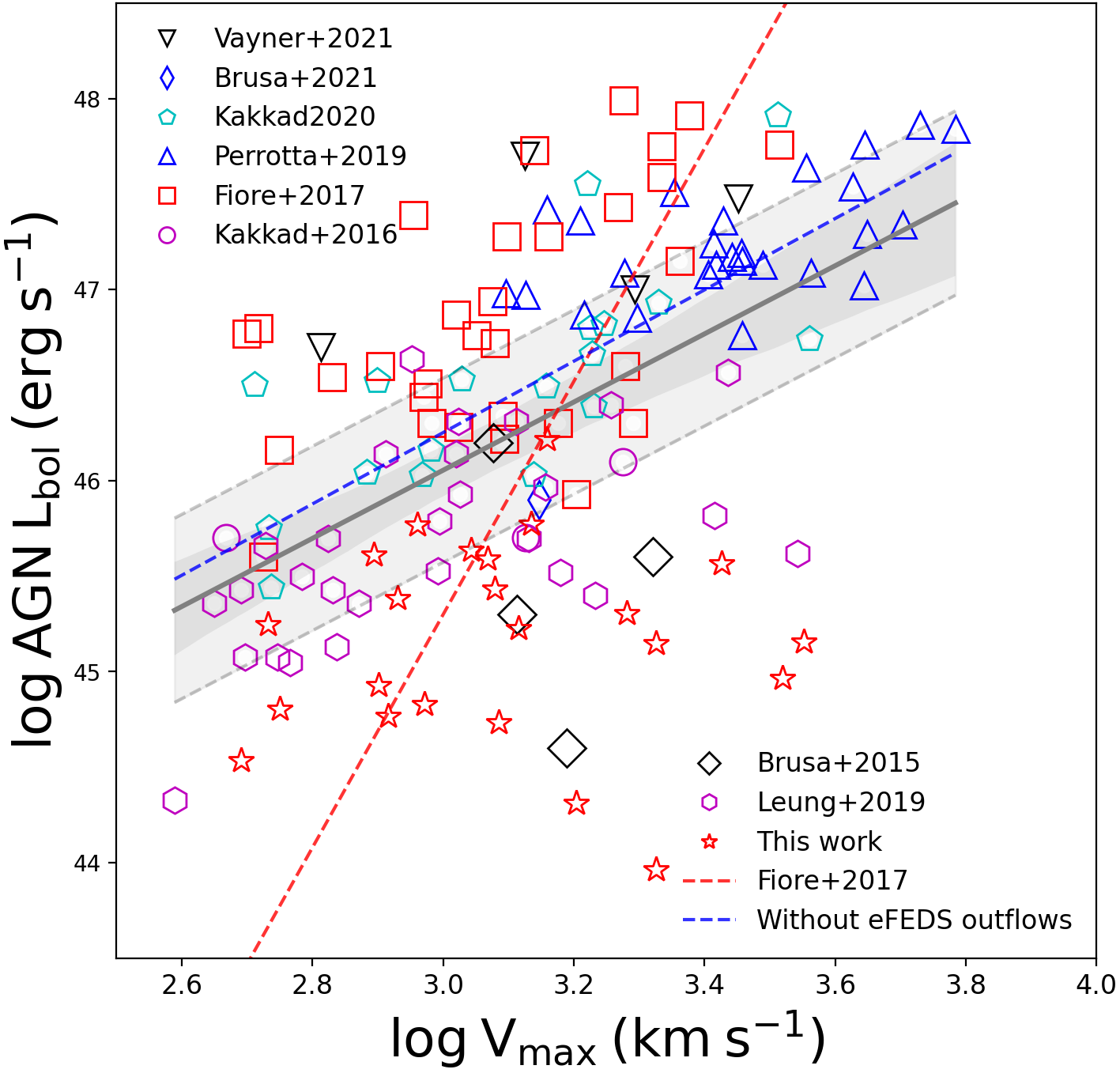} 
\caption{ AGN bolometric luminosity as a function of outflow velocity. Different shapes represent different samples from literature, as labelled.  The solid grey line is the scaling relation obtained from fitting between the two variables including our AGN outflows from eFEDS (see Table 2). The grey region indicates the 95\% confidence interval and the shaded filled region between two dashed black lines is the 1 $\sigma$. The red dashed line is the scaling relation from \cite{fiore2017}, obtained for a sample of $\sim55$ ionised outflows without imposing a redshift cut z$>0.5$. The blue dashed line is the scaling relation obtained from fitting the two variables excluding our eFEDS sources. The relation still appears flattish as compared to \cite{fiore2017}. It is important to note that we maintain the same legend for QWO and eFEDS sources in the following figures. } 
\label{fig:lbol_vmax_qwo}
\end{figure}

\begin{figure}[!t]
    \includegraphics[width=0.45\textwidth]{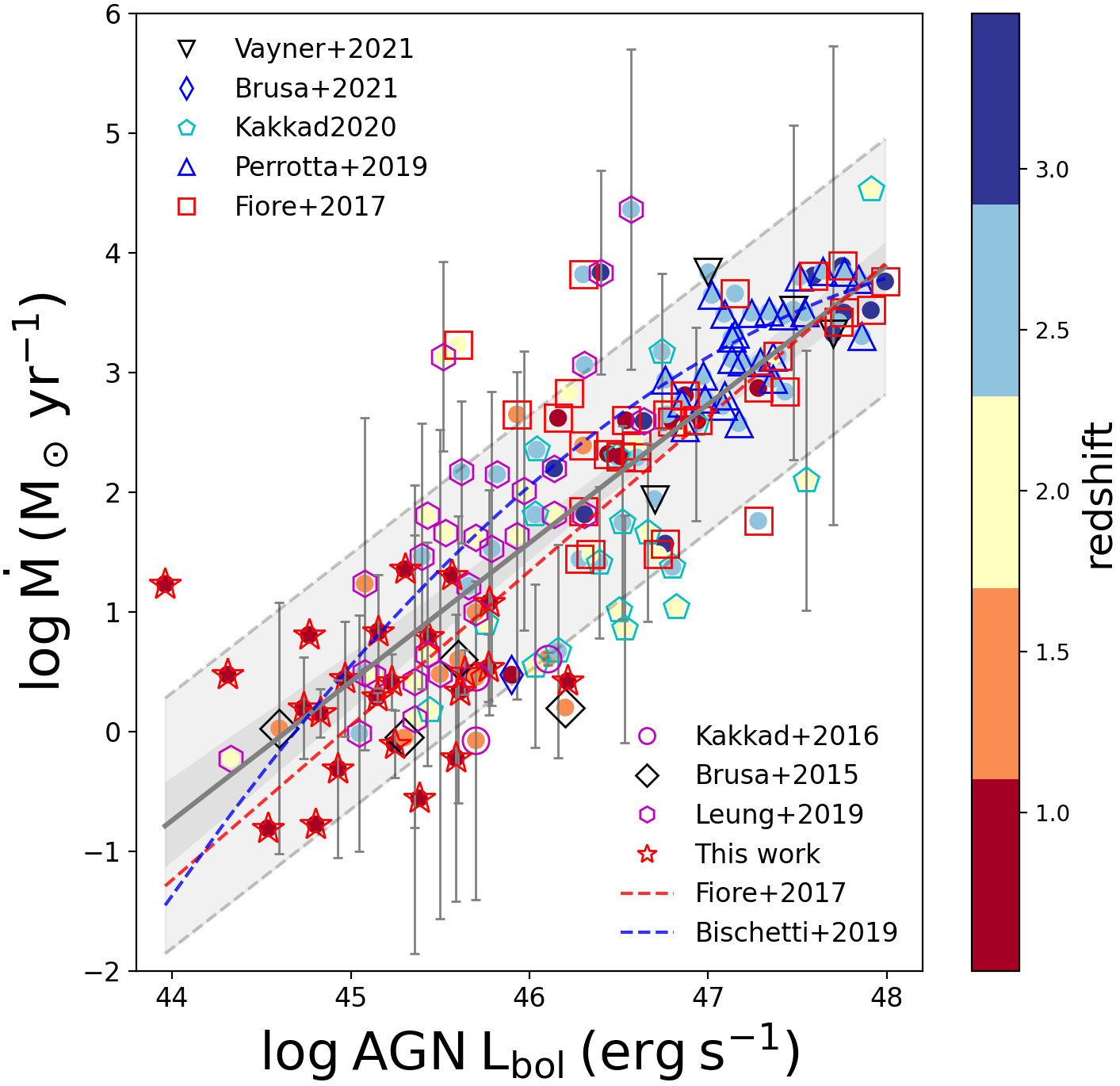} 
\caption{Ionised mass outflow rate as a function of AGN bolometric luminosity. Different shapes represent different literature samples (as labelled). The filled circles are colour-coded with redshift. The blue dashed line is the best-fit relation obtained for ionised outflows in \cite{bischetti2019}. The solid grey line is the scaling relation obtained in this work, from fitting between the two variables (see Table 2). The grey region indicates the 95\% confidence interval and the shaded filled region between two dashed black lines is the 1 $\sigma$. The red dashed line is the scaling relation from \cite{fiore2017}, obtained for a sample of $\sim55$ ionised outflows without imposing a redshift cut z$>0.5$.  The error bars on mass outflow rate values were obtained by extrapolating the errors on FWHM or $\rm{\delta{V}}$ whenever available. These values may have 2-3 orders of magnitude uncertainties due to the different assumptions applied.} 
\label{fig:mdot_lbol_qwo_z}
\end{figure}

 AGN feedback is one of the suggested mechanisms which contributes to SF quenching. If it happens, how it happens, on which scale does it affect the SF (impact on nuclear regions or the whole galaxy) and how much it contributes to the quenching are all still open questions. 
 A way to address this topic, at least statistically, is to look at correlations between outflow properties and properties of AGNs and host galaxies.
 From an observational point of view, several studies in the past reported correlations between the physical properties of outflows at all scales and AGNs and/or 
host galaxies properties  \citep[e.g.][among others]{fiore2017,bischetti2017,matzeu2023}, the most striking being the correlation between mass outflow rate and AGN bolometric luminosity. The mutually inconsistent recipes used to compute the outflow properties in the literature, not considering different scenarios such as multi-phase nature, multi-scale and environment (the wind that may have escaped in the least resistance media or more resistance) can contribute to diluting the correlation between these properties. 

In the following, we explore the correlations between outflow properties and AGN luminosity for a large sample of sources with ionised gas outflows.
 More specifically, our newly detected 23 sources with ionised outflows are presented in Sect.~\ref{outflowproperties}. We add a sample of AGN with outflows compiled from the literature and including previously confirmed AGNs with outflows at z$>$0.5 and for which the physical properties have been derived, with a focus on ionised outflows traced by [OIII], $\rm{H_{\alpha}}$ or $\rm{H_{\beta}}$ emission lines. We refer to this sample as the Quasar With Outflows (QWO) sample hereafter.  
 
 The QWO sample contains 118 sources at z$>$0.5,  compiled from \cite{Brusa2015,kakkad2016,fiore2017,leung2019,perrotta2019,kakkad2020,vayner2021,brusa2022}. \cite{fiore2017} contains ionised outflows detections and an estimate of their main physical parameters from a number of references which include: \cite{nesvadba2008,harrison2012,maiolino2012, rupke2013,liuG2013,harrison2014,genzel2014,cresci2015a,perna2015a,perna2015b,carniani2015,cicone2015, brusa2016,bischetti2017}. For all these sources AGN bolometric luminosity is available, although derived in a heterogeneous way: the bolometric luminosity in \cite{leung2019} is obtained from $\mathrm{[OIII]}$ line luminosity (L[OIII]) by applying a bolometric correction of 600 \citep{kauffmann2009}, \cite{kakkad2020} obtain an estimate of L$_{\rm bol}$ by SED fitting, \cite{fiore2017} and the rest from either SED fitting or bolometric correction from different wavebands. The uncertainty that may arise due to bolometric luminosity obtained from different approaches is assumed negligible. Finally, we note that the several assumptions on different parameters such as geometry, electron density, temperature, and in some cases, the velocity of the outflowing gas and radius introduce at least one order of magnitudes uncertainties in the estimated values of the outflow properties, and this could contribute to increasing the scatter in the correlations. 

\begin{figure*}[!t]
\begin{tabular}{c c}
\includegraphics[width=0.45\textwidth]{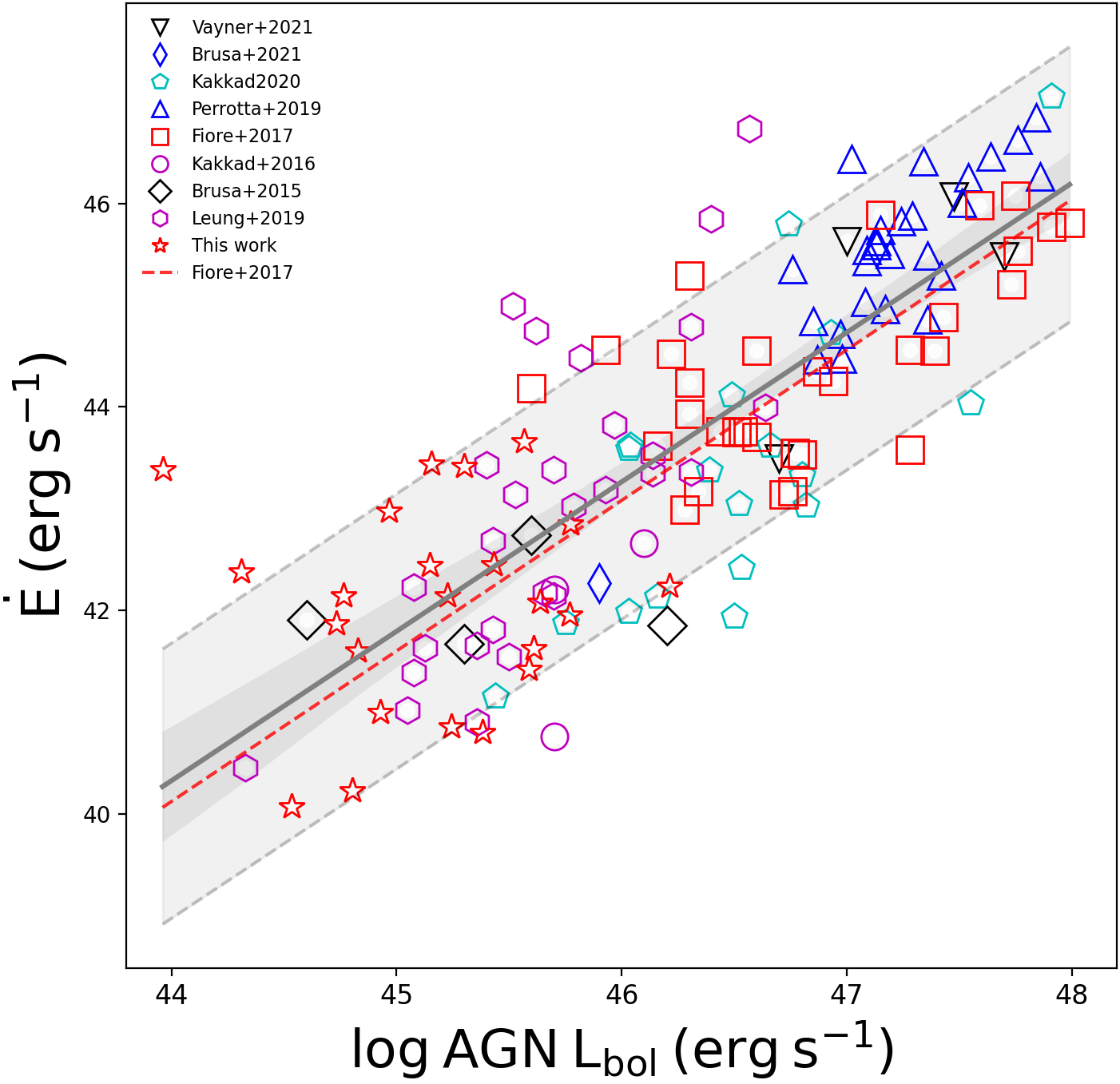} &
\includegraphics[width=0.45\textwidth]{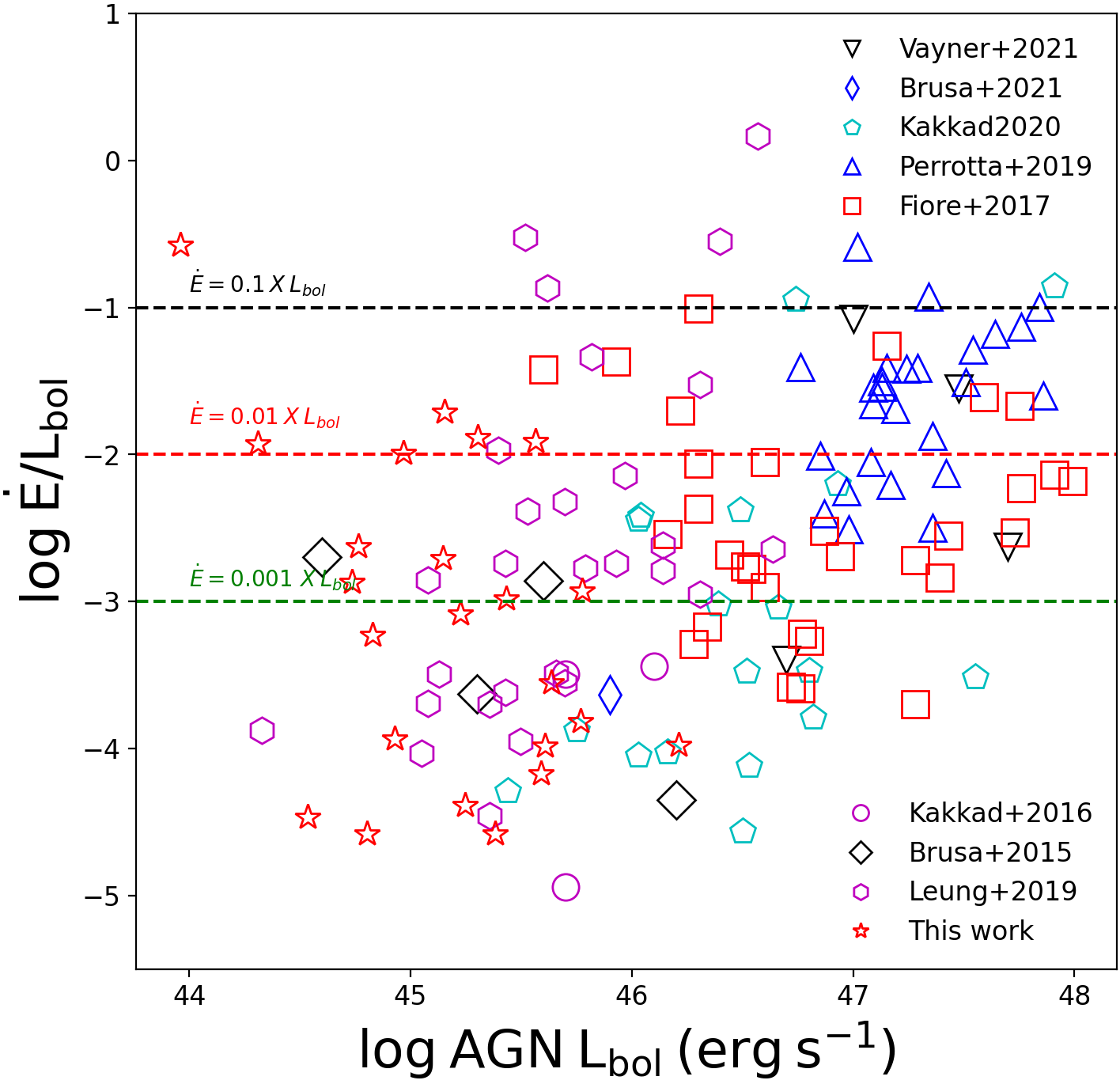} 
\end{tabular}
\caption{Ionised kinetic power and kinetic coupling efficiency as a function of AGN bolometric luminosity. Left panel: Ionised kinetic power as a function of AGN bolometric luminosity. The red dashed line is the best scaling relation obtained in \citet{fiore2017}. The solid grey line in both panels is our scaling relation obtained from OLS fitting between the two variables. The grey region indicates the 95\% confidence interval and the faded filled region between two dashed black lines is the 1 $\sigma$.  Right panel: Kinetic coupling efficiency (${\dot{E}/L_{bol}}$) versus AGN bolometric luminosity. The black dashed line represents ${\dot{E}=0.1L_{bol}}$, red dashed line represents ${\dot{E}=0.01L_{bol}}$ and green dashed line represents ${\dot{E}=0.001L_{bol}}$. 30\% of the candidates appear above ${\dot{E}/L_{bol}}$=0.01 or 1\% . These values have 2-3 orders of magnitude uncertainties due to the different assumptions applied.}
\label{fig:edot_lbol_QWO}
\end{figure*}

For the QWO sample, we collected all the parameters as measured from spectral fitting ($\rm{[O III]}$ or line luminosity, FWHM, $\sigma$ and  $\mathrm{n_{e}}$ if reported, the quoted outflow velocities otherwise) and from observations (outflow radius if available). 
We computed mass outflow, mass outflow rate, and outflow kinetic power following the same receipt used for our eFEDS sources\footnote{In case of parameters derived from H$\alpha$ or H$\beta$ spectral fit, the outflow parameters can be derived following an approach similar to the one presented in Sect.\ref{outflowproperties}, with prescriptions changed accordingly to convert hydrogen flux to mass,  \citep[see e.g.][]{nesvada2017,leung2019,riffel2023} }. We maintain the reported outflow properties for sources in \cite{fiore2017}. The limited spatial resolution of the available observations required several assumptions to estimate outflow properties (see Appendix \ref{assumptions} for the details of the properties we re-computed and the assumptions applied).

We investigate several scaling relations between outflow properties and AGN bolometric luminosities, for a total of 141 sources in the distant Universe (z$>0.5$; QWO+eFEDS sample). 
This is the largest sample of AGNs with ionised outflows at z$>$0.5 and it is $\sim$3 times larger than the previous compilations (e.g. \citealt{fiore2017,leung2019}). Adopting the linear regression fitting procedure that uses the ordinary least square (OLS) model between $\mathrm{V_{max}}$ and bolometric luminosity, we obtain a best-fit scaling relation as $\mathrm{L_{bol}\propto{V_{max}}^{ 1.78\pm0.2} }$   
which significantly differs from $\mathrm{L_{bol}\propto{V_{max}}^{6.1\pm4.4} }$ reported in \cite{fiore2017}. The QWO+eFEDS sample in the L$_{\rm bol}$-V$_{\rm max}$ plane is shown in Fig. \ref{fig:lbol_vmax_qwo} along with the best-fit relation (grey shaded region) and the \citet{fiore2017} relation (red dashed line).  
There is still a significant difference with the \cite{fiore2017} correlation when the fitting is done on the QWO sample, excluding our confirmed outflows from eFEDS. In this case we obtain a scaling relation of $\mathrm{L_{bol}\propto{V_{max}}^{ 1.87\pm0.21} }$, shown as a blue dashed line in Fig. \ref{fig:lbol_vmax_qwo}\footnote{When fitting our sample in the $\mathrm{V_{max}}\propto{L_{bol}}$ plane, we find 
$\mathrm{V_{max}}\propto{L_{bol}^{0.15\pm0.02} }$, and $\mathrm{V_{max}}\propto{L_{bol}^{0.22\pm0.03} }$ when eFEDS sources are excluded. This value is consistent with the slope of 0.27 reported by \citet{leung2019}.}.
We note that the much steeper \citet{fiore2017} correlation has been derived on a factor of $\sim3$ smaller sample, and more heterogeneous in terms of redshift.
Indeed, the \citet{fiore2017} sample includes 51 sources detected at all redshifts, with about 35\% at z$<$0.5. This is also reflected in the relatively large error associated with the measured slope reported in their work. 
 The main reason for a flatter trend is due to the fact that sources in the QWO+eEFEDS sample have on average lower bolometric luminosities ($\sim 1 dex$)  than the original \citet{fiore2017} sample, despite having high outflow velocities. This points once again to the importance of the sample selections for correlation studies.  
 In particular, the primary selection of our eFEDS sample is the presence of bright X-ray emission, whereas this is not the case for the majority of the sources from the QWO (and in the \citealt{fiore2017} sample). Our thesis is that the X-ray active, obscured phase may be the best tracer of the fastest phase of the winds, and their velocity may not necessarily depend on the AGN bolometric luminosity (see e.g. \citealt{Brusa2015}). Even though 23 sources from eFEDS may be a relatively small sample detected with outflows, they occupy a region of parameter space never explored before in the $\mathrm{L_{bol}}$-$\mathrm{V_{max}}$ plane. This points, when looking at the data at face value in their entirety, to a non-existent correlation between these two variables.
 
In Fig. \ref{fig:mdot_lbol_qwo_z}, we show that for our combined sample the ionised mass outflow rate, ${\dot{\textrm{M}}_\textrm{{ion}}}$ (computed on the basis of standardised and homogenised assumptions) correlates well with the AGN bolometric luminosity.  We obtain the scaling relation as ${\dot{\textrm{M}}_\textrm{{ion}}\propto \mathrm{{L_{bol}}}^{1.16\pm0.07} }$ with Spearman rank correlation coefficient of 0.64. The more luminous the AGN, the more prevalent the outflows \citep[see also][]{mullaney2013,zakamskaandgreen2014}. 
Our scaling relations of mass outflow rate with AGN bolometric luminosity are only slightly flatter than what has been published in previous studies on smaller samples, which reported scaling relations of ${\dot{\textrm{M}}_\textrm{{ion}}\propto \mathrm{{L_{bol}}}^{1.29\pm0.38}} $  \citep{fiore2017} and ${\dot{\textrm{M}}_\textrm{{ion}}\propto \mathrm{{L_{bol}}}^{1.34\pm0.37} }$ \citep{leung2019}. We have increased the degrees of freedom threefold than those in \cite{fiore2017}. Our correlation is also consistent with the best-fit relation obtained for ionised outflows in \cite{bischetti2019}, shown with a blue curve in Fig. \ref{fig:mdot_lbol_qwo_z}. 

The colour bar in Fig. \ref{fig:mdot_lbol_qwo_z} highlights that there is a slight trend in redshift with AGN bolometric luminosity (and therefore mass outflow rate). 
We also found correlation trends between the mass outflow rates, black hole mass, and Eddington ratio. These trends are reported in the Appendix \ref{x-raypropertiesofoutflows}, contrary to the weak correlation between these properties found in \cite{kakkad2022} for low redshift quasars. This may imply that there is a single mechanism dominant in driving the outflows for the sources in our compilation. The scatter observed in the AGN bolometric luminosity and mass outflow rate scaling relation indicates that outflow power depends not only on $\mathrm{L_{bol}}$ but also on other factors such as the coupling between outflows and host and amount or geometry of dense gas in the nuclear regions (as discussed in e.g. \citealt{ramosalmeida2022})

We computed the kinetic power from the above mass outflow rates, using standardised assumptions for both the QWO sample and our eFEDS targets. As shown in the left panel of Fig. \ref{fig:edot_lbol_QWO}, by applying linear regression fitting procedure that uses the ordinary least square (OLS) model, we obtained a best-fit scaling relation of ${\dot{\textrm{E}}_\textrm{{ion}}\propto \mathrm{{L_{bol}}}^{1.47\pm0.09} }$.  
\citet{fiore2017} and \citet{leung2019} reported ${\dot{\textrm{E}}_\textrm{{ion}}\propto \mathrm{{L_{bol}}}^{1.48 \pm0.37} }$ and ${\dot{\textrm{E}}_\textrm{{ion}}\propto \mathrm{{L_{bol}}}^{1.87 \pm0.51} }$, respectively with slopes that are comparably similar to our scaling relation obtained from our sample. The results from all our fittings are represented in Table \ref{tab:qwofits} which shows the slope, Spearman rank correlation coefficient and the constant with their respective errors. 

Mass outflow rates and kinetic powers are the quantities that, on the basis of theoretical models, are expected to correlate with the AGN bolometric luminosity and depends both on the outflow velocity and the total entrained mass. For a given bolometric luminosity the same mass outflow rate can be observed in sources with fast winds and low entrained mass (for instance, winds caught at an early stage of the feedback phase), or with massive but slower winds (for instance, winds caught at a later stage of feedback phase). Overall, combining our results shown in Fig.\ref{fig:lbol_vmax_qwo}, \ref{fig:mdot_lbol_qwo_z} and \ref{fig:edot_lbol_QWO} we can conclude that our eFEDS selection preferentially isolates obscured QSO in the fastest phase of the wind (see also \citealt{costa2018a}).

The right panel of Fig. \ref{fig:edot_lbol_QWO} shows the ratio between the kinetic power and the AGN bolometric luminosity (${\dot{\textrm{E}}_\mathrm{{ion}}/\mathrm{L_{bol}}}$)  as a function of the AGN bolometric luminosity. A larger fraction of ionised outflows occupy the 1$\%\! < \! {\dot{\textrm{E}}_\mathrm{{ion}}/\mathrm{L_{bol}}}\!<$10\% as compared to \cite{fiore2017}, and $\sim$30\% of the ionised winds have ${\dot{\textrm{E}}_\mathrm{{ion}}/\mathrm{L_{bol}}}$ in the range 1$-$100\%. 
Overall, more than half of the sample has ${\dot{\textrm{E}}_\mathrm{{ion}}/\mathrm{L_{bol}}}$ within one dex of the theoretical predictions (\citealt{dimatteo2005,hopkinsandelvis2010,schaye2015}, see also Fig.  2 in \citealt{harrison2018} for more theoretical predictions as compared to observations). However, we note that taking measurements at face value, 70\% have ${\dot{\textrm{E}}_\mathrm{{ion}}/\mathrm{L_{bol}}}\!<\!1$\%.
As discussed in \cite{harrison2018}, as the wind traverses in the surrounding ISM, some part of the original nuclear wind energy is used to do work against other forces. As a consequence, the final outflow kinetic energy is expected to be considerably lower than the original ejected energy. 
Some models (e.g. \cite{hopkinsandelvis2010} and as reviewed in \citealt{harrison2018} ) predict that even with ${\dot{\textrm{E}}_\mathrm{{ion}}/\mathrm{L_{bol}}}\!<\!$1\% radiatively driven winds are capable of having a significant effect on the host galaxy.

\begin{table}[!t]
    \caption{Results from AGN and outflow properties correlations.
}
    \begin{tabular}{ |p{2cm}|p{1.5cm}|p{1.5cm}|p{0.8cm}|p{1cm}|}
 \hline
Correlation & Slope & Constant & R$-$value & P$-$value \\
 \hline
$\mathrm{L_{bol}}$ Vs. $\mathrm{V_{max}}$  & 1.78$\pm$0.2  & 40.7$\pm$0.8 & 0.27 & $<10^{-5} $ \\
${\dot{M}}$ Vs. $\mathrm{L_{bol}}$  & 1.16$\pm$0.07 & -51.7$\pm$3.3 & 0.67 & $<10^{-5} $\\
${\dot{E}}$ Vs. $\mathrm{L_{bol}}$   & 1.47$\pm$0.09 & -24.3$\pm$4.2 & 0.66 & $<10^{-5} $\\
\hline
\end{tabular}
    \label{tab:qwofits}
\end{table}

\begin{figure}[!ht]
\includegraphics[width=0.45\textwidth]{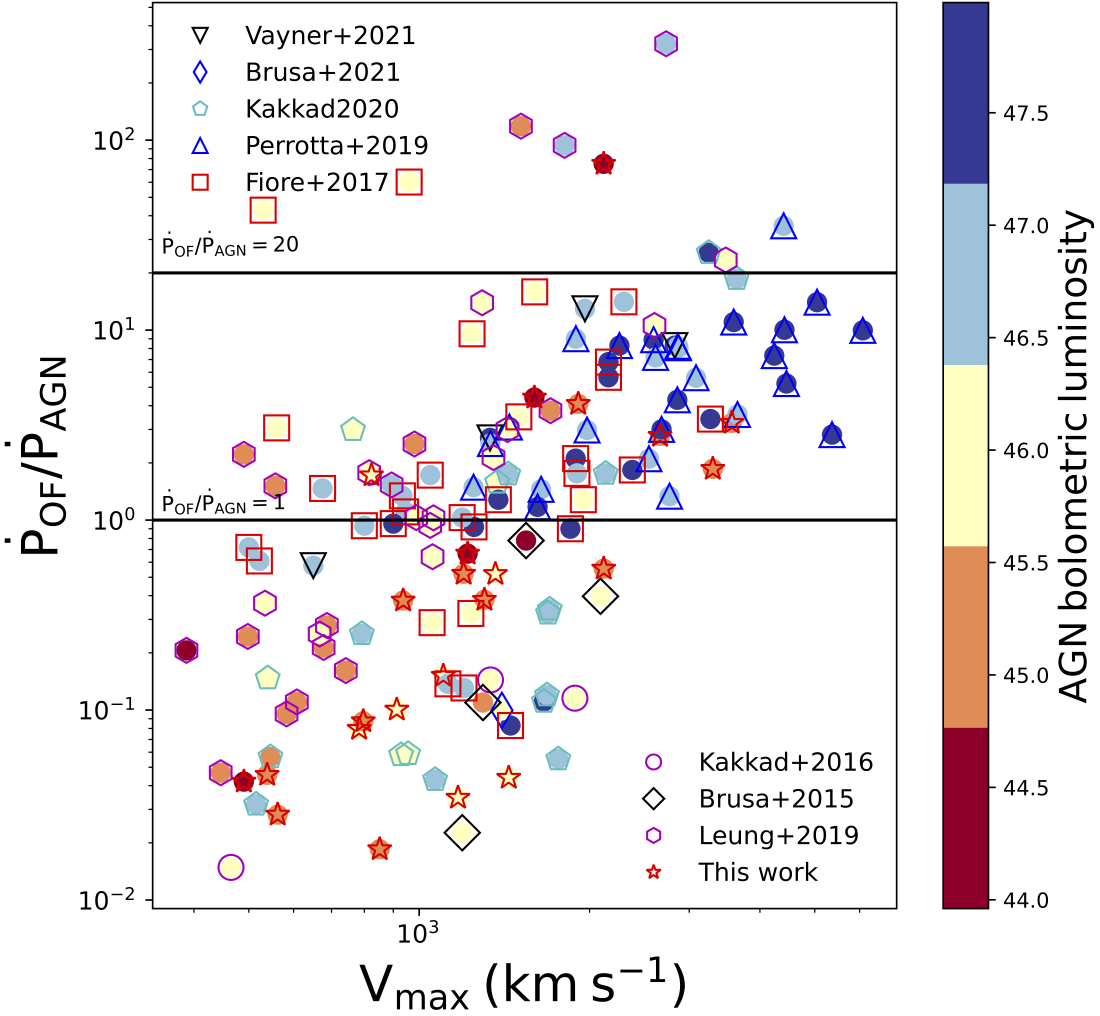} 
\caption{We plot the outflow momentum rate (momentum flux of the outflow divided by the radiation momentum flux from the central black hole) against the maximum outflow velocity and colour-coded with the AGN bolometric luminosity. The black solid lines show the ${\mathrm{{\dot{P}_{OF}/\dot{P}_{AGN} }}}\!=\!20$ and =1 as indicated in the plot. The majority of our sources have momentum flux ratios of less than 20. }
\label{fig:mom_flux}
\end{figure}

The comparison between outflow momentum rates ($\dot{\textrm{P}}_\mathrm{{OF}}$) and the radiative momentum flux from the central black hole (L$_\mathrm{{bol}/c}$) has been proposed to be key to understanding whether the detected outflows are energy or momentum conserving. 
Indeed, models predict that energy-driven winds on large-scale outflows have momentum fluxes  
${\dot{\textrm{P}}_\mathrm{{OF}}\!\sim \!\mathrm{20L_{bol}/c}}$ \citep[e.g.][]{zubovasandking2012}. On the contrary, momentum-driven winds would show  ${\dot{\textrm{P}}_\mathrm{{OF}}\!\sim\!1 \mathrm{L_{bol}/c}}$. 
Figure \ref{fig:mom_flux} shows the momentum flux versus the outflow velocity for our QWO+eFEDS homogenised sample. 
For the outflows confirmed for the first time in this work (red stars in Fig. \ref{fig:mom_flux}), a bigger fraction have momentum flux ratios lower than 20 which in principle rules out their energy-conserving nature. 
This indicates that our probed outflows could be either momentum-driven or energy-driven with momentum being carried in other gas phases. These results are consistent with the momentum flux ratios found in the previous studies for ionised outflows, for example \cite{perna2015a,leung2019} and \cite{tozzi2021} for outflow momentum rates $< 10 \rm{L_{bol}/c }$ and in \cite{carniani2015,kakkad2016} and \cite{marasco2020} for ionised outflows below the momentum flux ratio of 1. Considering the full new compilation (QWO and eFEDS targets) shown in Fig. \ref{fig:mom_flux}, we obtain median outflow momentum rates $\sim$1 times the momentum flux from the AGN.

\section{Conclusions}
\label{conclusions}
This paper focused on the selection and characterisation of z$>0.5$ AGNs in the feedback phase. 
We applied a combination of various selection methods previously proposed in the literature to the eFEDS main AGN sample (Sect. \ref{sampleselection}), to maximise the completeness (of the selection) and reduce the chances of a biased selection. We then investigated the presence of ionised gas outflows in the 0.5$<$z$<$1 subsample population via spectral fitting of the [OIII] emission line profile (Sect. \ref{efeds_spectra}). Finally, we explored the scaling relations between AGN luminosity and outflow properties of a large sample of 141 sources at z$>\!0.5$ (Sect. \ref{correlations}). 

We summarise our results from this study as follows:
\begin{itemize}
    \item By applying a combination of selection methods to eFEDS, we isolated 853 candidates AGNs in the outflow phase from flux ratios and colour selection diagnostics. Similarly, we isolated 528 candidates applying the N$_{\rm H}$ versus Eddington ratio diagnostic, to the subsample of sources with BH mass measured in SDSS-IV data (see Sect. ~\ref{opticalspectralproperties}). In total, we isolated a sample of $\sim$1400 obscured and/or red quasars expected to be in the feedback phase, which corresponds to $\sim$12\% of the z$>0.5$ AGN population ($\sim$9\% of the 0.5$<$z$<$1) in eFEDS, with a sky number density of $\sim$10/deg$^2$. 
      \item  
      We analysed the [OIII] line fitting of 50 candidates in our study. These candidates were selected based on having good quality SDSS spectra and being at 0.5$<$z$<$1. Among these candidates, we identified 23 sources that displayed evidence of a broad component with FWHM $\sim$600--2800 $\mathrm{km~s^{-1}}$, signalling the presence of unsettled gas motions and confirming their nature as outflowing QSOs. This corresponds to $\sim$45\% of the spectroscopic sample and this may increase to 80\% if we consider an additional 17 sources (see Sect.~\ref{efeds_spectra}) for which a broad component can be accommodated in the fit, but at a lower significance. 
      It is worth noting that our sample may be biased towards sources with high bolometric luminosity ($\mathrm{logL_{bol}}$ between  $\mathrm{\sim~44-46.2~erg~s^{-1}}$ ) and column densities ($\mathrm{N_{H}}$ $\mathrm{\sim10^{22}~cm^{-2}}$), due to the selection effects affecting our sample of candidates with good quality spectra, which is dominated by sources from sample B. The average [O III] luminosity, $\mathrm{L[OIII]}$ is $\sim ~10^{42}~\mathrm{erg s^{-1}}$ and by assuming the outflow radius as the half-light radius of the galaxy as measured from deep HSC data (from $\sim 2 - 10$ kpc), we obtain mass outflow rates in the range of $\mathrm{\sim 0.2 - 23~  M_{\odot}~yr^{-1}} $ and ${\log(\dot{\textrm{E}})~\sim~40-44~erg~s^{-1}}$.
    \item We complemented these 23 sources with an additional 118 sources at z$>0.5$ reported in the literature, for which we recomputed in a standardised and homogenised way the outflow properties. This constitutes the largest sample of AGNs with detected ionised outflows. From this compilation, we find a correlation between the maximum velocity of the outflow and the AGN bolometric luminosity ($\mathrm{L_{bol}\propto{V_{max}}^{ 1.78\pm0.2}}$) with a slope considerably flatter than the scaling relation presented in previous studies and a much larger scatter (see Sect. \ref{correlations}) 
     \item The mass outflow rate and the kinetic power of the outflow instead correlate well with the AGN bolometric luminosity (${\dot{\textrm{M}}_\mathrm{{ion}}\propto\rm{{L_{bol}}}^{1.16\pm0.07}}$ and ${\dot{\textrm{E}}_\mathrm{{ion}}\propto\rm{{L_{bol}}}^{1.47\pm0.09}}$). Both trends are in agreement with the relations derived in previous studies (see Sect. \ref{correlations}), despite the considerable difference in the maximum velocity and AGN bolometric luminosity correlation. This can be explained if our X--ray based, eFEDS selection preferentially isolates obscured QSO in the fastest phase of the wind.
    \item More than half of our sample have $\dot{\textrm{E}}/\rm{L_{bol}}$ close to the theoretical predictions (see discussion in Sect. \ref{correlations}). 
    About 30\% of ionised outflows have the $1\%\!< \!{\dot{\textrm{E}}_\mathrm{{ion}}}/\rm{L_{bol}} \!<\!10$\%. This is an indication that the outflows present in these sources could have a significant impact on their host galaxies. We are comparing outflows in a single phase (ionised) with the theoretical predictions that consider multi-phase outflows.
    \item 
    The majority of the sources show outflows with momentum flux ratios lower than 20, which rules out an energy-conserving nature since models predict that energy-driven winds on large-scale outflows have momentum rates that are 20 times the momentum rates from the central black hole, ${\dot{\textrm{P}}_\mathrm{{OF}}\!\sim \!\rm{20L_{bol}/c}}$. 
\end{itemize}

Overall, this study provides an improved approach to isolate quasars in the feedback phase, suggesting that the best way to select AGNs with strong winds is by applying a combination of selection methods, minimising the selection biases that result from using incomplete and single selection methods. Of the 50 objects with good quality optical spectra for which we performed spectral fitting, we have significantly detected the presence of ionised outflows in $\sim$45\%. This is a strong confirmation of the reliability of our strategy. Even though 23 sources ($\sim$45\%) may be a relatively small sample, we will be able to validate our selection techniques on larger scales with eROSITA.

From our methods, we selected $\sim$1400 sources ($\sim$12\% of the AGN subsample at z$>0.5$) from the eFEDS area of $\sim$140 deg$^2$. We, therefore, expect $\sim$140000 such candidates will exist in the eROSITA all-sky catalogue (area factor of $\sim$100 considering the best extra-galactic sky). 
We predict that following a similar strategy and with the extended spectroscopic coverage provided by dedicated and sensitive AGN surveys within 4MOST \citep{merloni2019} and SDSS-V \citep{kollmeier2019} 
we will be able to fully uncover and characterise this rare population bringing the total number of confirmed objects from a single X--ray survey to $>2500$. 

This study finally provides an observational benchmark for the investigation of the correlations between AGN and outflow properties, being the largest compiled sample (141 objects, including the eFEDS sources) of ionised outflows available at z$>$0.5 and with properties derived with standardised assumptions. This study can be extended by also including the molecular outflows and follow-up studies for the candidates with higher redshift.

\begin{acknowledgements}
This work is based on data from eROSITA, the soft X-ray instrument aboard SRG, a joint Russian-German science mission supported by the Russian Space Agency (Roskosmos), in the interests of the Russian Academy of Sciences represented by its Space Research Institute (IKI), and the Deutsches Zentrum für Luft- und Raumfahrt (DLR). The SRG spacecraft was built by Lavochkin Association (NPOL) and its subcontractors and is operated by NPOL with support from the Max Planck Institute for Extraterrestrial Physics (MPE). The development and construction of the eROSITA X-ray instrument was led by MPE, with contributions from the Dr. Karl Remeis Observatory Bamberg \& ECAP (FAU Erlangen-Nuernberg), the University of Hamburg Observatory, the Leibniz Institute for Astrophysics Potsdam (AIP), and the Institute for Astronomy and Astrophysics of the University of Tübingen, with the support of DLR and the Max Planck Society. The Argelander Institute for Astronomy of the University of Bonn and the Ludwig Maximilians Universität Munich also participated in the science preparation for eROSITA. The eROSITA data shown here were processed using the eSASS or NRTA software system developed by the German eROSITA consortium.

The Hyper Suprime-Cam (HSC) collaboration includes the astronomical communities of Japan and Taiwan, and Princeton University.  The HSC instrumentation and software were developed by the National Astronomical Observatory of Japan(NAOJ), the Kavli Institute for the Physics and Mathematics of the Universe (Kavli IPMU), the University of Tokyo, the High Energy Accelerator Research Organisation (KEK), the Academia Sinica Institute for Astronomy and Astrophysics in Taiwan (ASIAA), and Princeton University.  Funding was contributed by the FIRST program from Japanese Cabinet Office, the Ministry of Education, Culture, Sports, Science and Technology (MEXT), the Japan Society for the Promotion of Science (JSPS), Japan Science and Technology Agency (JST),the Toray Science Foundation, NAOJ, Kavli IPMU, KEK,ASIAA, and Princeton University.

Funding for the Sloan Digital Sky Survey IV has been provided by the Alfred P. Sloan Foundation, the U.S. Department of Energy Office of Science, and the Participating Institutions. SDSS acknowledges support and resources from the Center for High-Performance Computing at the University of Utah. The SDSS web site is \texttt{www.sdss.org} SDSS is managed by the Astrophysical Research Consortium for the Participating Institutions of the SDSS Collaboration including the Brazilian Participation Group, the Carnegie Institution for Science, Carnegie Mellon University, Center for Astrophysics | Harvard \& Smithsonian (CfA), the Chilean Participation Group, the French Participation Group, Instituto de Astrofísica de Canarias, The Johns Hopkins University, Kavli Institute for the Physics and Mathematics of the Universe (IPMU) or University of Tokyo, the Korean Participation Group, Lawrence Berkeley National Laboratory, Leibniz Institut für Astrophysik Potsdam (AIP), Max-Planck-Institut für Astronomie (MPIA Heidelberg), Max-Planck-Institut für Astrophysik (MPA Garching), Max-Planck-Institut für Extraterrestrische Physik (MPE), National Astronomical Observatories of China, New Mexico State University, New York University, University of Notre Dame, Observatório Nacional or MCTI, The Ohio State University, Pennsylvania State University, Shanghai Astronomical Observatory, United Kingdom Participation Group, Universidad Nacional Aut\'onoma de M\'exico, University of Arizona, University of Colorado Boulder, University of Oxford, University of Portsmouth, University of Utah, University of Virginia, University of Washington, University of Wisconsin, Vanderbilt University, and Yale University.

BM is supported by the European Union's Innovative Training  Network (ITN) funded by the  Marie  Sklodowska-Curie Actions in Horizon 2020 No 860744 (BiD4BEST). MB and CRA acknowledge support from BiD4BEST. MB acknowledges support from PRINMIUR 2017PH3WAT  ('Black hole winds and the baryon life cycle of galaxies'). BM and MB warmly thank Tiago Costa for the insightful discussion.

CRA acknowledges support from the projects 'Feeding and
feedback in active galaxies', with reference PID2019-106027GB-C42,
funded by MICINN-AEI/10.13039/501100011033, and 'Quantifying the
impact of quasar feedback on galaxy evolution', with reference
EUR2020-112266, funded by MICINN-AEI/10.13039/501100011033 and the
European Union NextGenerationEU/PRTR.
\end{acknowledgements}

\bibliographystyle{aa}
\bibliography{aanda}

\begin{appendix}
\section{More of the eFEDS selected samples represented in same selection planes}
\label{sectionAPX}
\begin{figure*}[!ht]
\centering
\begin{tabular}{cc}
    \includegraphics[width=0.45\textwidth]{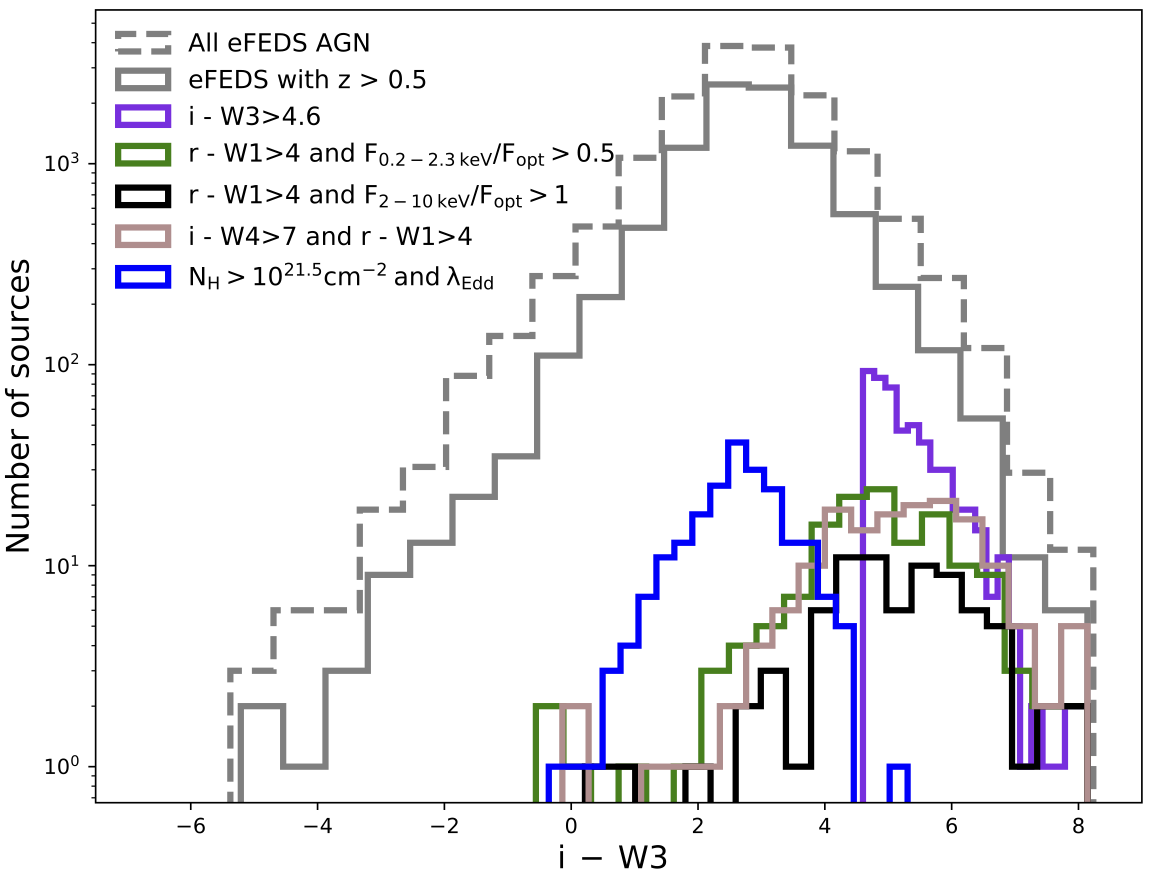} &
    \includegraphics[width=0.45\textwidth]{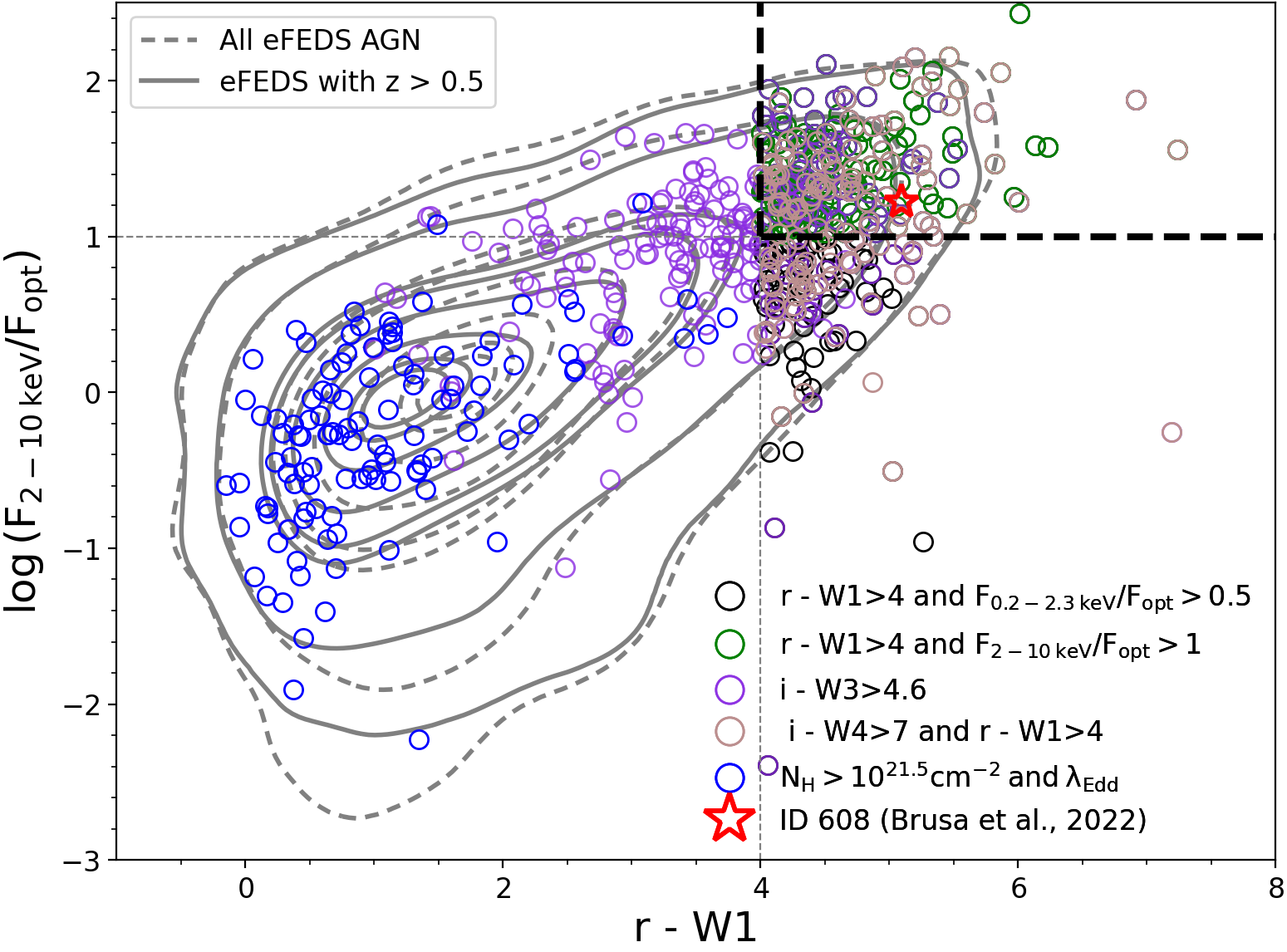}\\
    \includegraphics[width=0.45\textwidth]{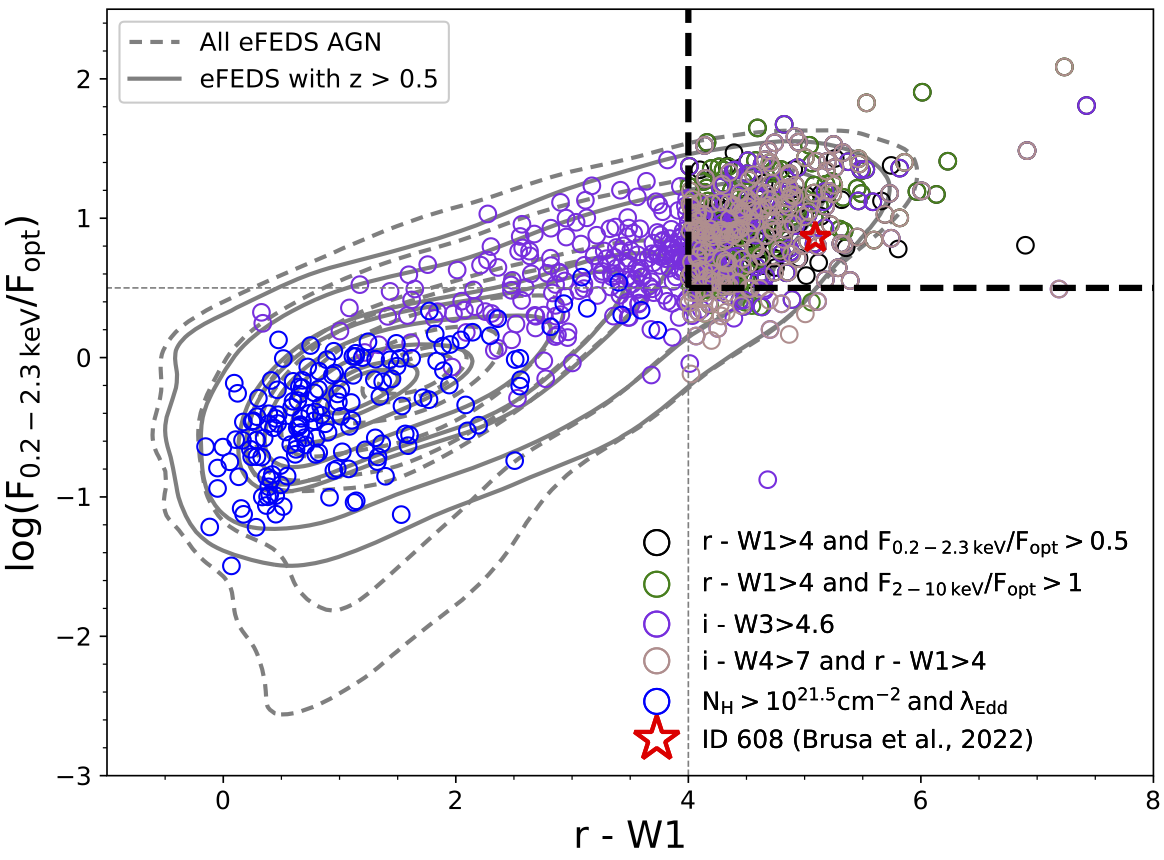} &
    \includegraphics[width=0.45\textwidth]{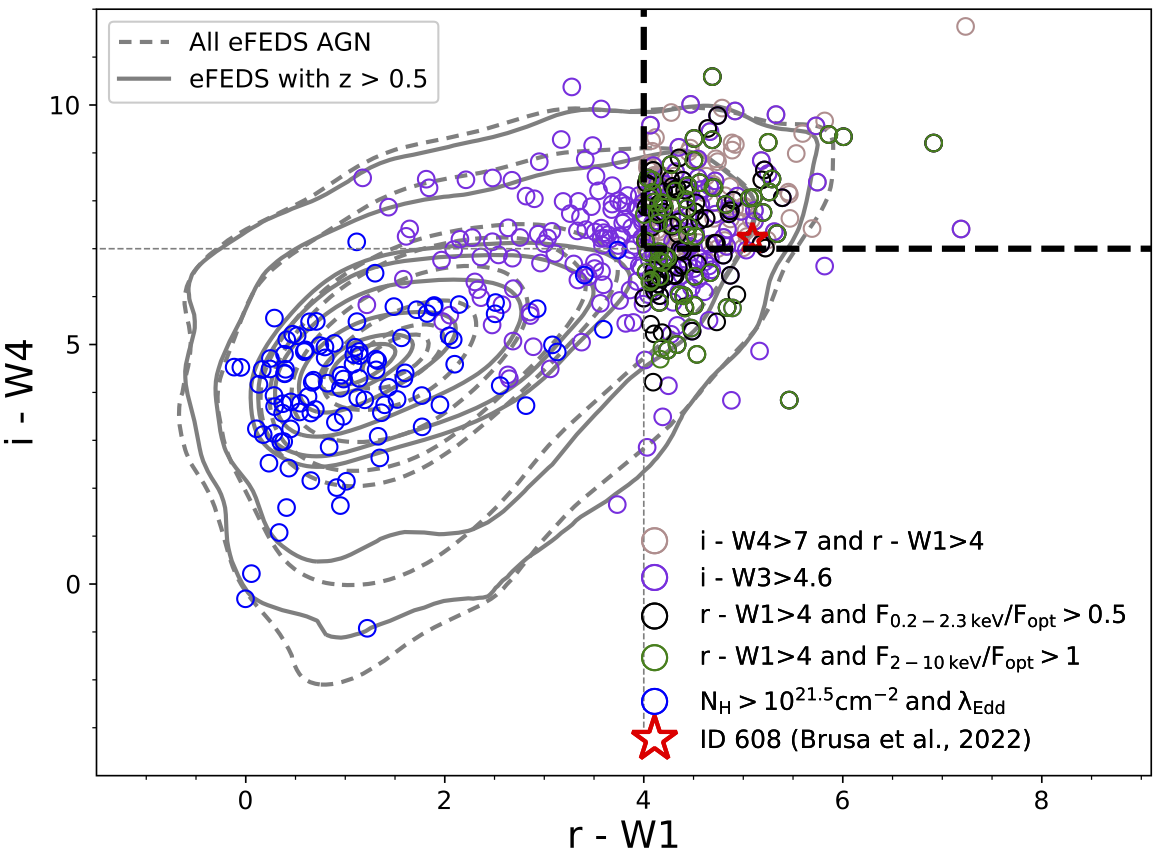}\\
\end{tabular}
\caption{ All selected candidates plotted in different planes. Top left panel: i-W3 distribution of candidates selected by different methods. Top right panel: Candidates selected by different methods plotted in $2-10$ keV to optical flux ratio vs. r-W1 colour plane. Bottom left panel:  $0.2-2.3$ keV to optical flux ratio vs. r-W1 colour. The sources from different selection methods are indicated with different colours or shapes as shown in the legend. Bottom right panel: All selected candidates plotted in i-W4 vs. r-W1. The sources from different selection methods are indicated with different colours as shown in the legend. }
    \label{fig:eFEDSAllColorSelectionMethods}
\end{figure*}

In this appendix, for each selection method, we represent our samples isolated from Sect. \ref{sampleselection}. As seen in Fig. \ref{fig:eFEDSAllColorSelectionMethods}, our isolated samples populate all the regions of the plane. This shows the complexity of isolating AGNs in the feedback phase since they do not occupy a specific region but also shows how our strategy can be used to minimise the chances of leaving out potential candidates.   

The sample selected by $\mathrm{N_H}$ and $\mathrm{\lambda_{Edd}}$ appear in the blue region of the colour selection planes. This means that, with this method, we are selecting obscured AGNs with blue colours. This can also be attributed to the blow-out phase being short. Therefore, these sources have bluer colours because they are being observed in the growing phase (immediately before the blow-out phase) or due to contamination from star formation or their host galaxy. The other possibility could be that the blue colours are due to the obscuration being inhomogeneous.

\section{The rest of the spectra.}
\label{rest of the spectra}
\begin{figure*}[h!]
\centering
\begin{tabular}{cc}
    \includegraphics[width=0.45\textwidth]{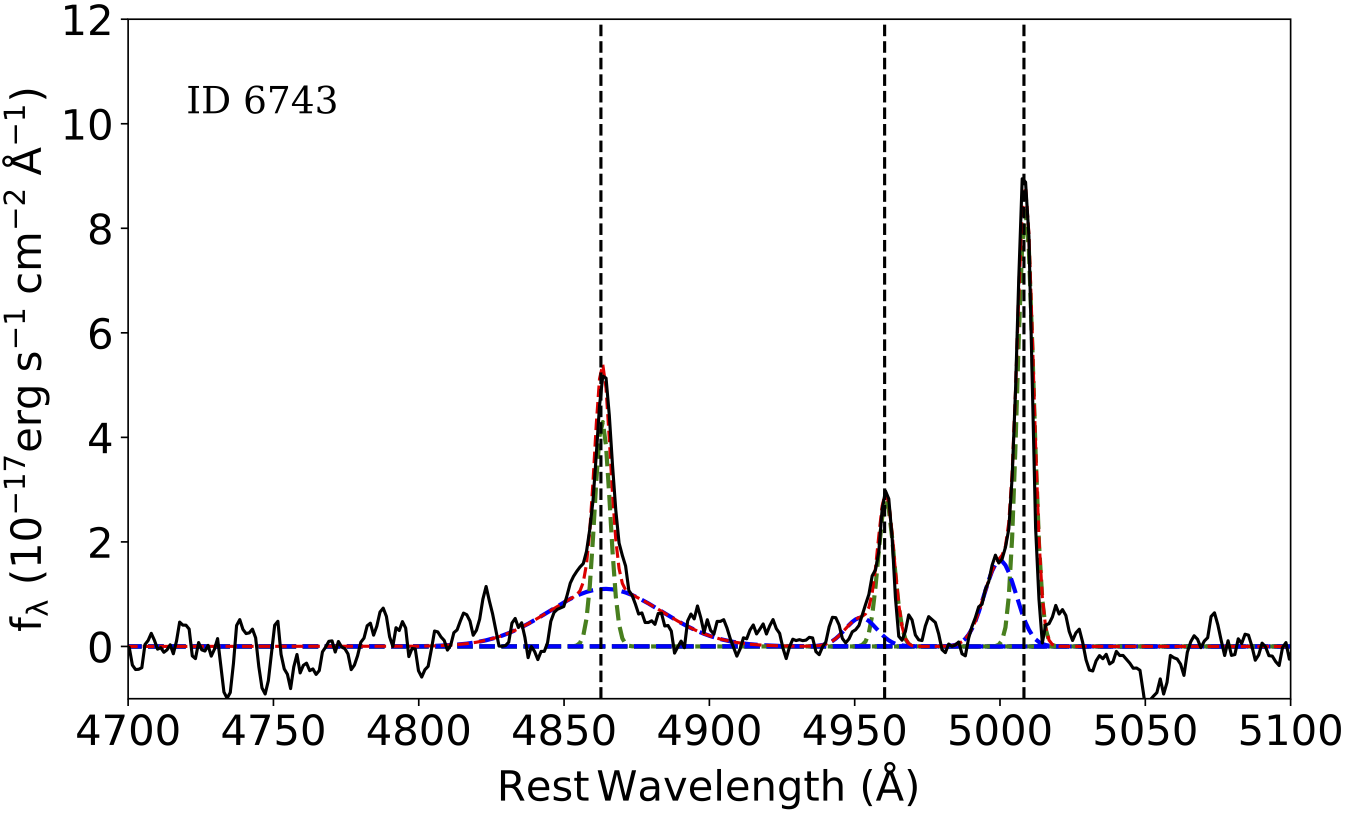} &
    \includegraphics[width=0.45\textwidth]{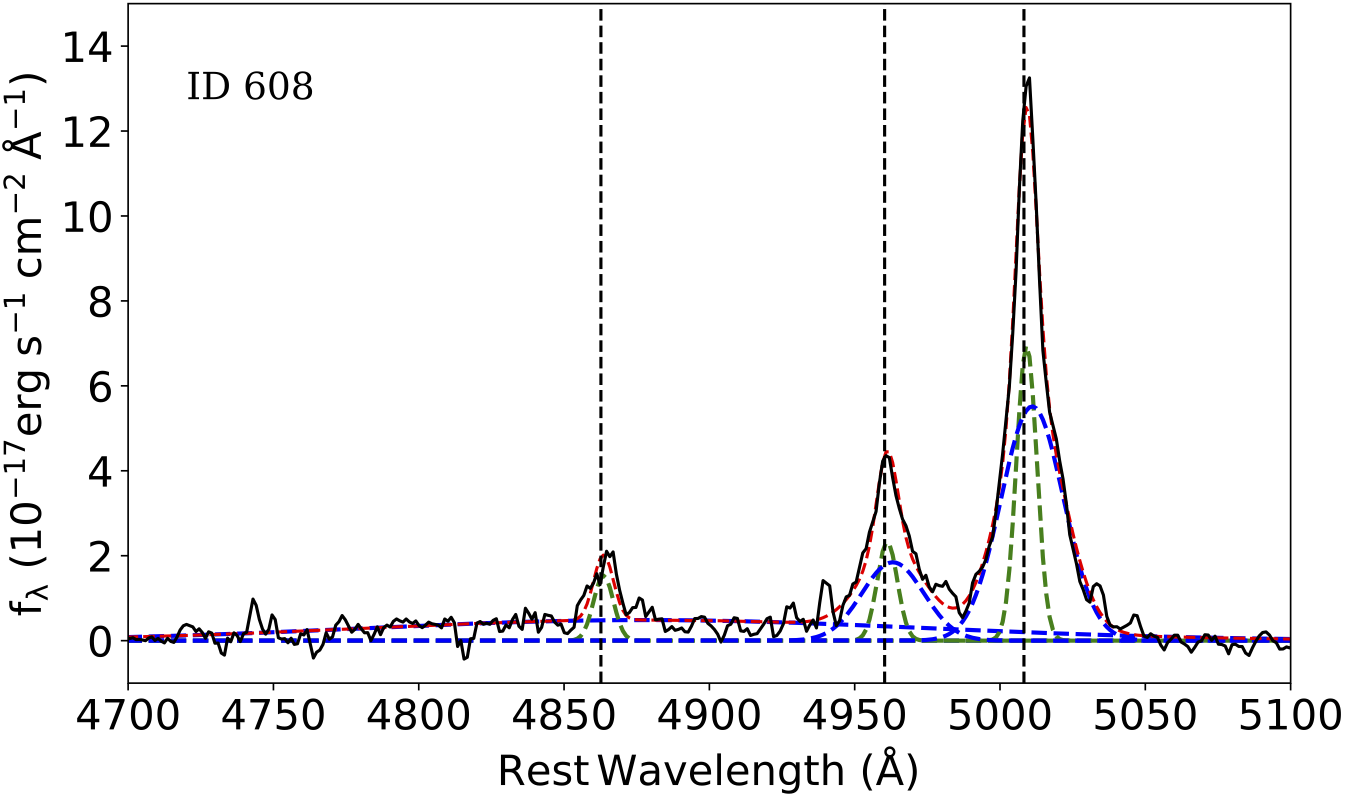}\\
    \includegraphics[width=0.45\textwidth]{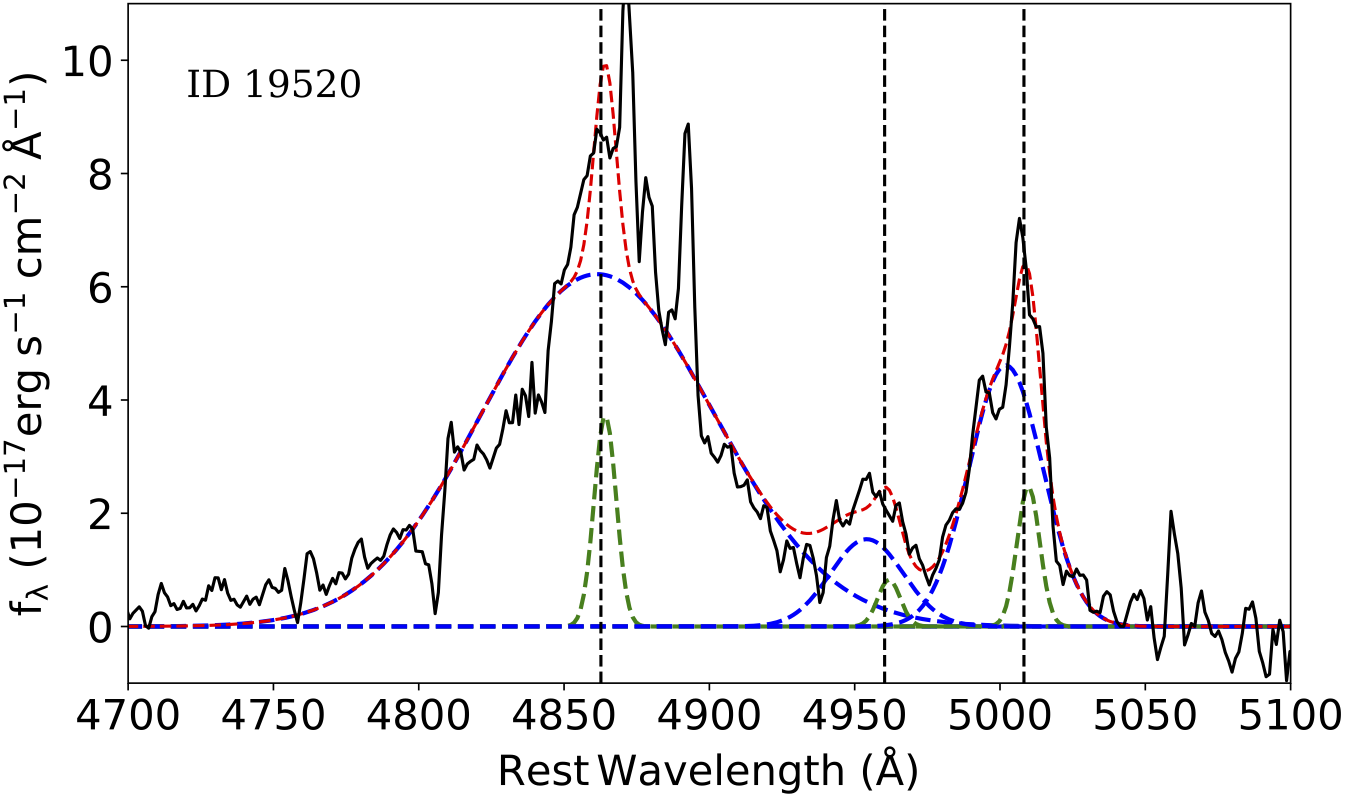} &
    \includegraphics[width=0.45\textwidth]{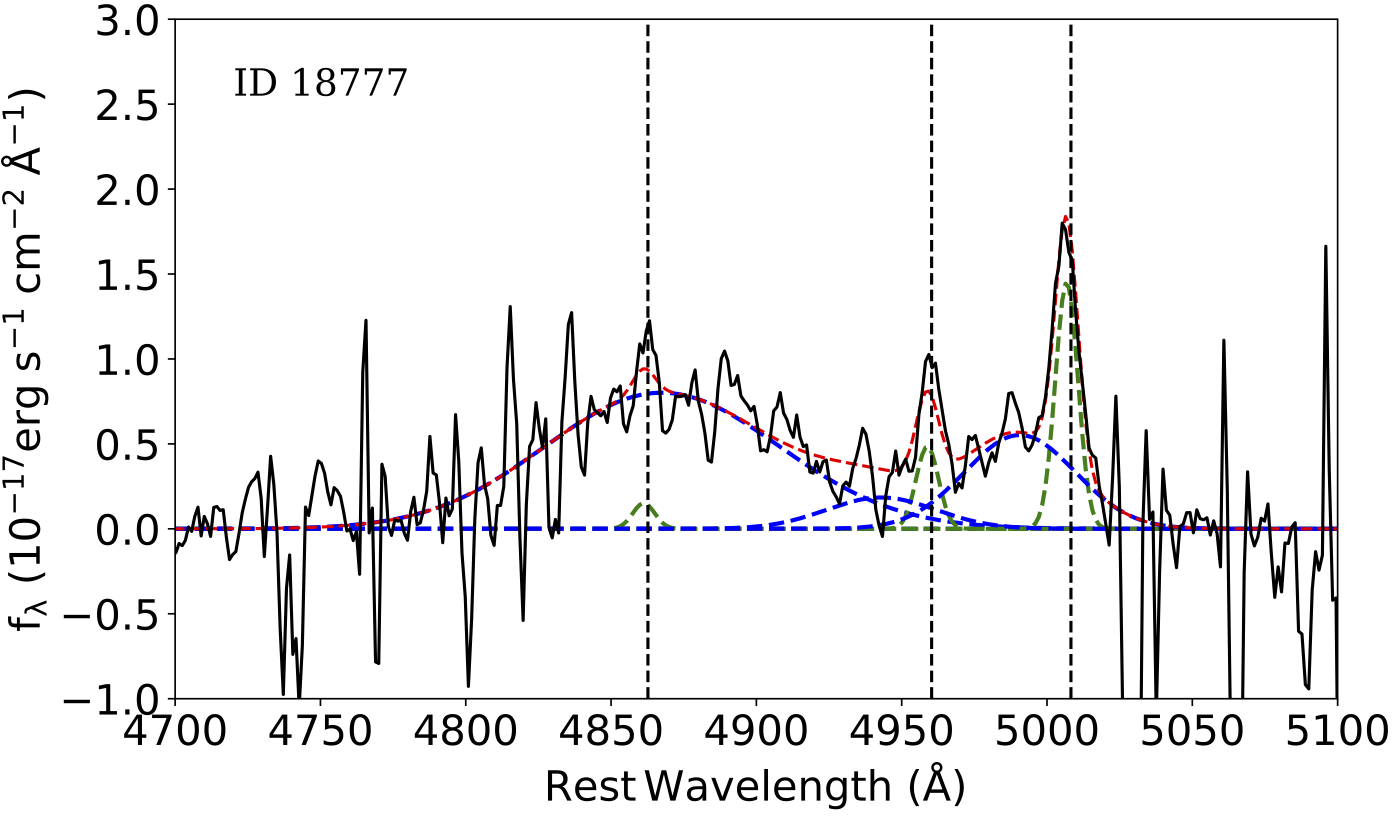}\\
    \includegraphics[width=0.45\textwidth]{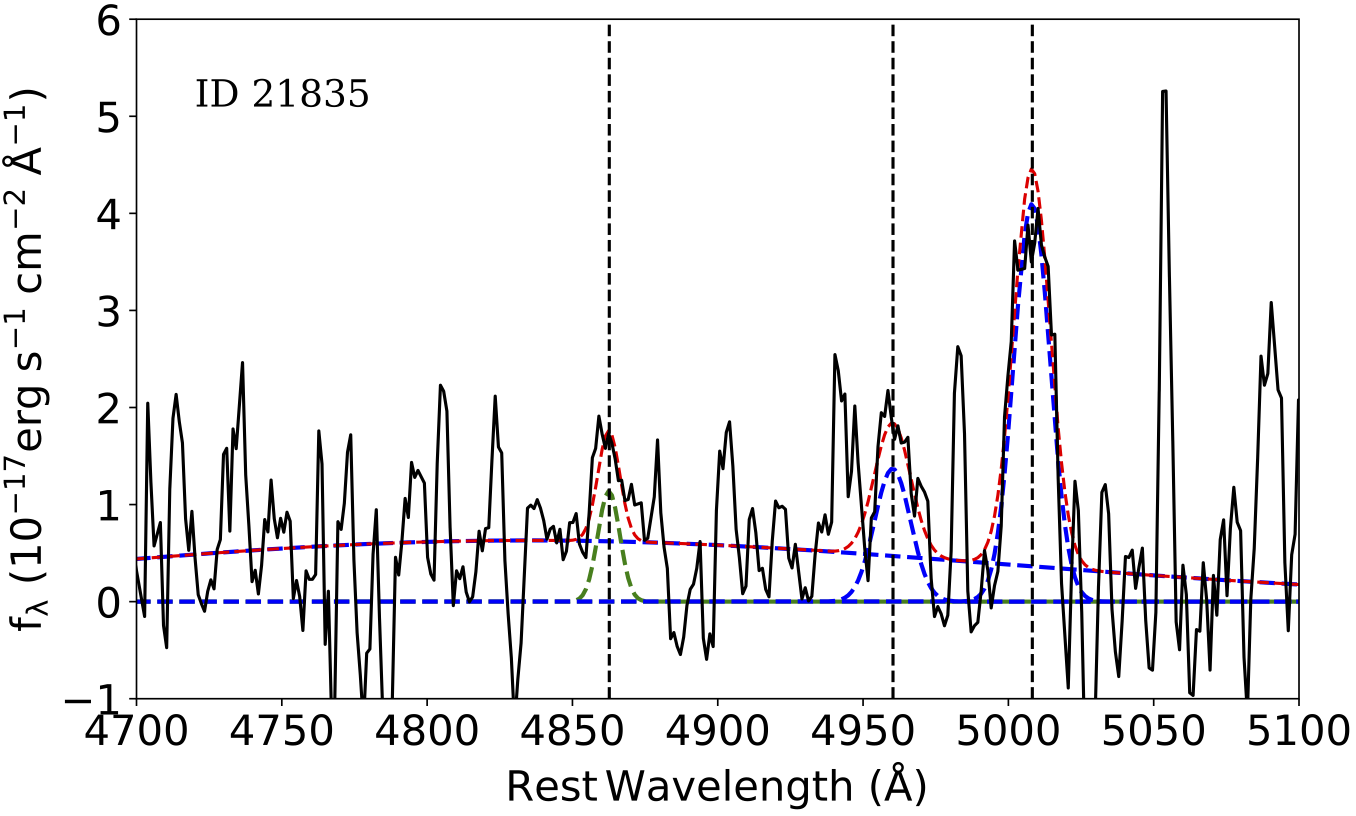} &
    \includegraphics[width=0.45\textwidth]{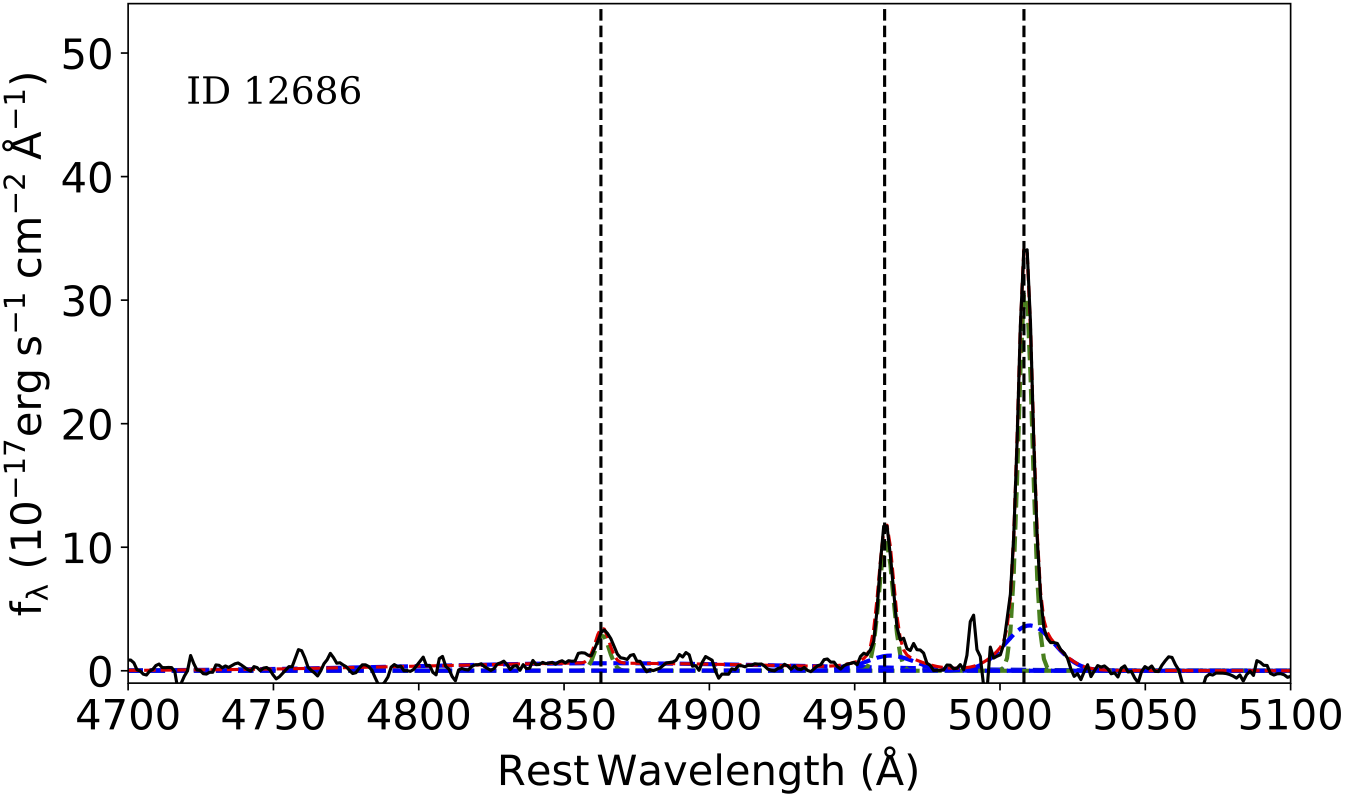}\\
   \includegraphics[width=0.45\textwidth]{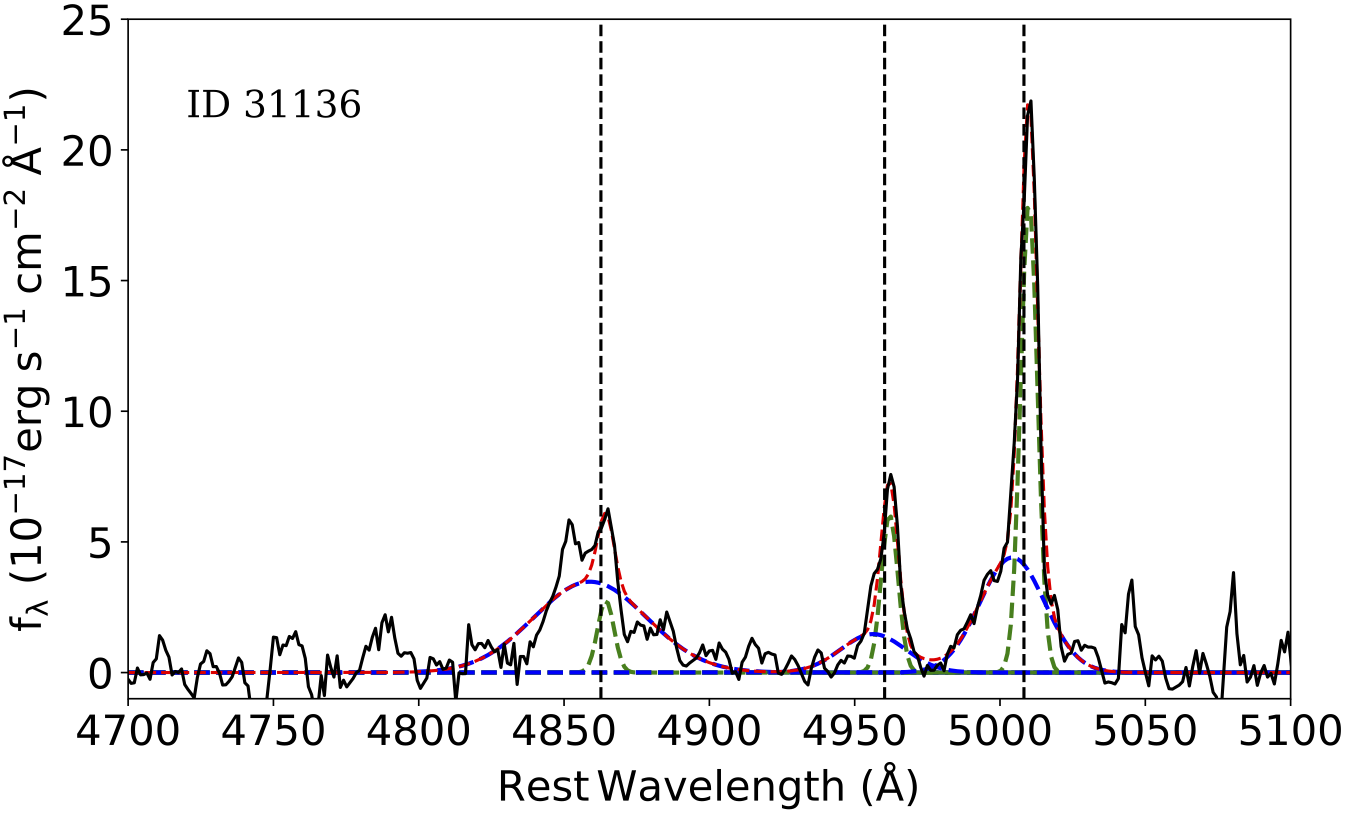} &
    \includegraphics[width=0.45\textwidth]{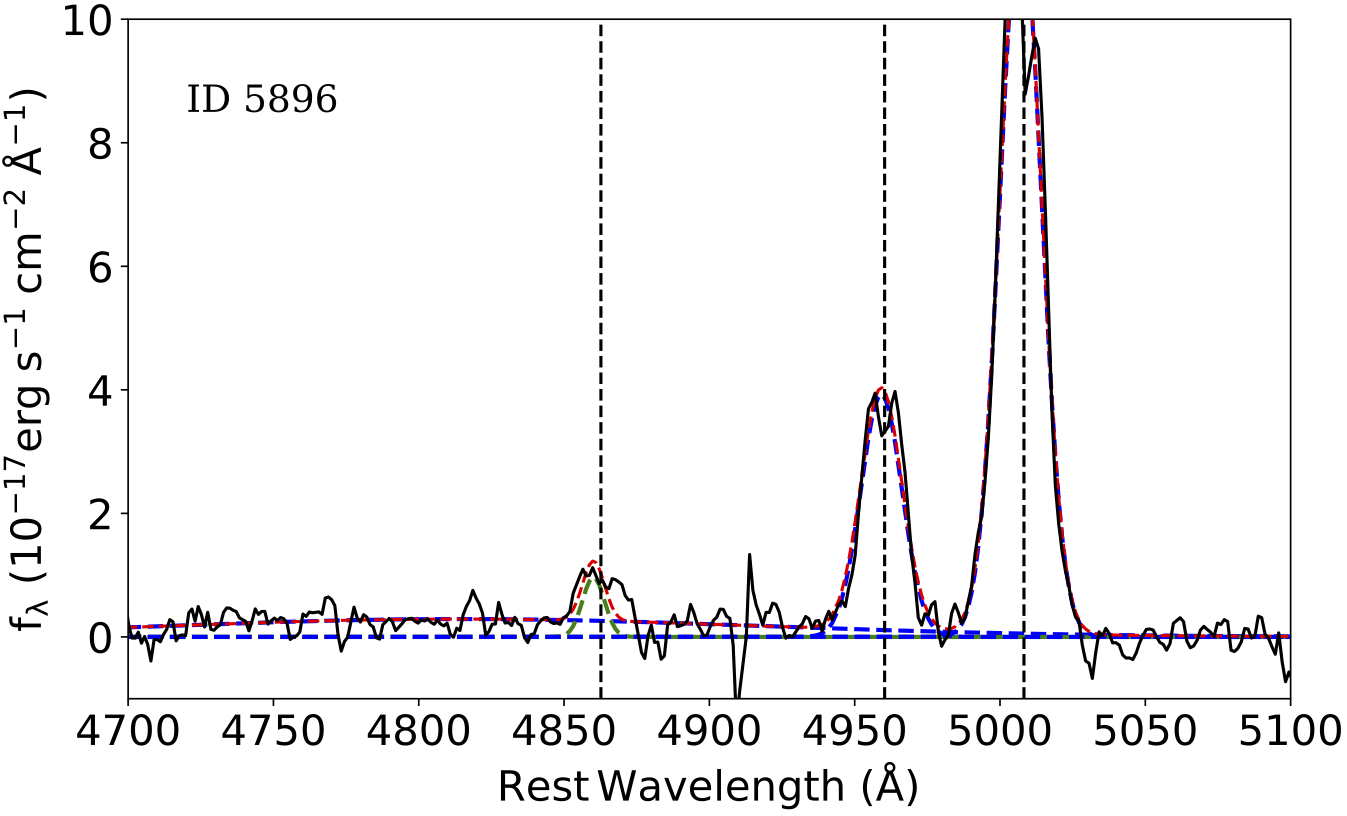}\\
\end{tabular}
\caption{SDSS emission line profiles fit for our candidates with outflows.
The blue line indicates the broad component and the green fit indicates the narrow component. The red line indicates the total fit. The vertical dotted lines indicate the peaks at 4862.68, 4960.30, and 5008.24 for $H_{\beta}$ and [OIII] rest-frame wavelength.}
    \label{fig:efedsSDSSspectra2}
\end{figure*}

Here we present the rest of the spectra as obtained in Sect. \ref{efeds_spectra} for more 22 quasars with ionised outflows (in Fig. \ref{fig:efedsSDSSspectra2}). One example of the 10 single components fits with FWHM$\rm{>800 kms^{-1}}$ (in the left panel of Fig. \ref{fig:efedsSDSSspectra_1comp_2compLowSNR}) and an example of the 17 two-component fits with flux to flux error ratio less than 2.5 (in the right panel of Fig. \ref{fig:efedsSDSSspectra_1comp_2compLowSNR}). In Fig. \ref{fig:efedsSDSSspectra_bad}, we show an example of the spectra that we classified as bad due to a very noisy continuum around the [OIII] region or bad residuals after the continuum and host galaxy subtraction.
\begin{figure*}[h!]
\centering
\begin{tabular}{cc}   
     
    \includegraphics[width=0.45\textwidth]{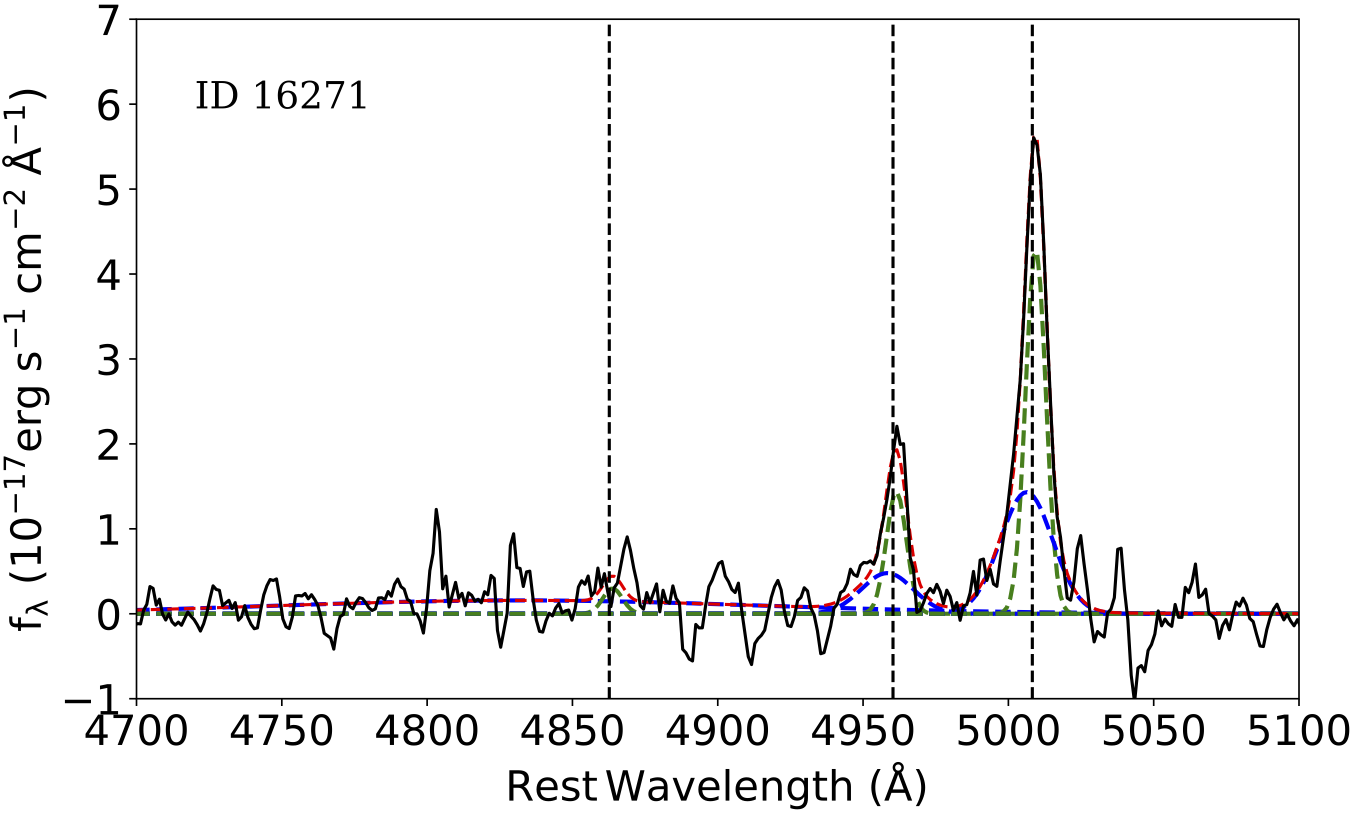} &
    \includegraphics[width=0.45\textwidth]{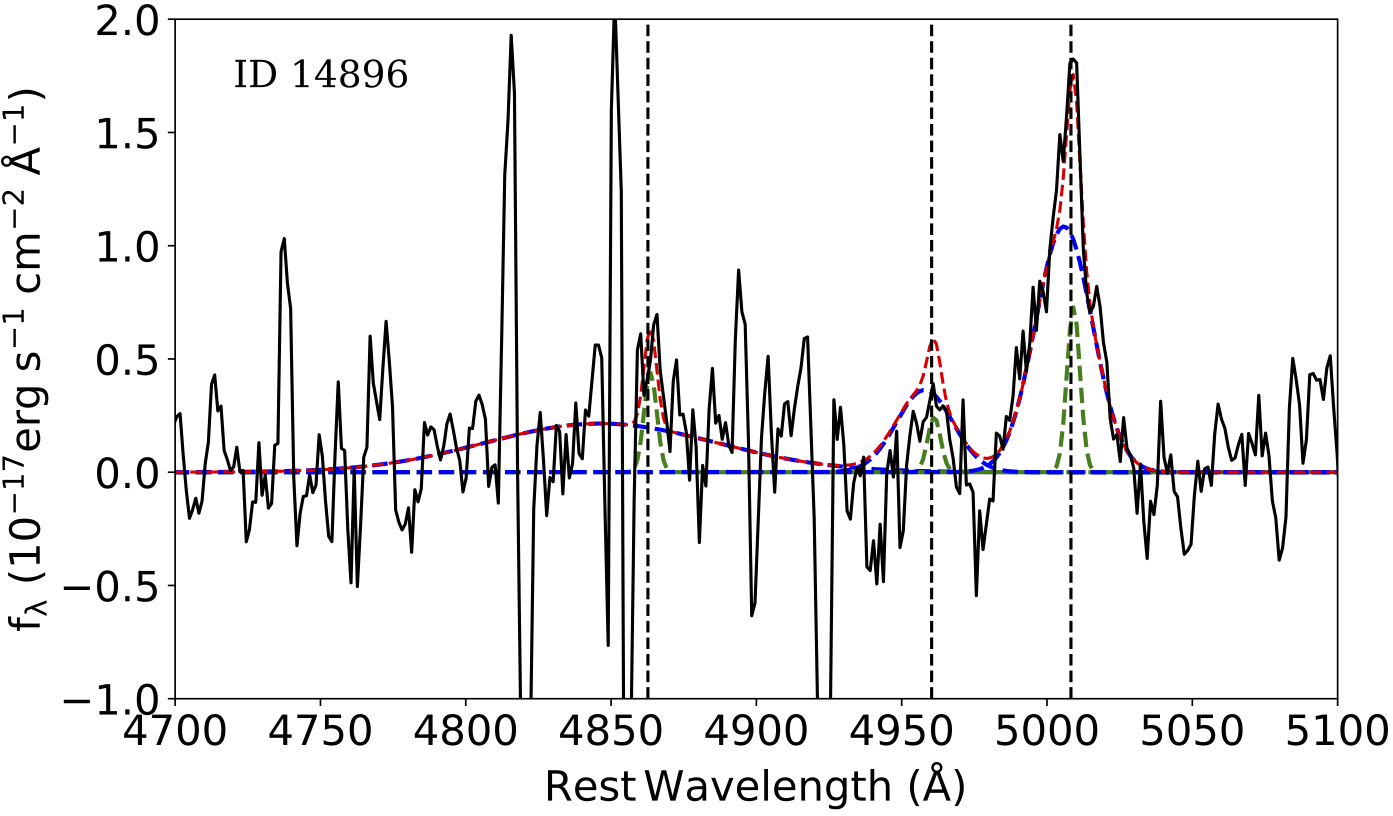} \\
    \includegraphics[width=0.45\textwidth]{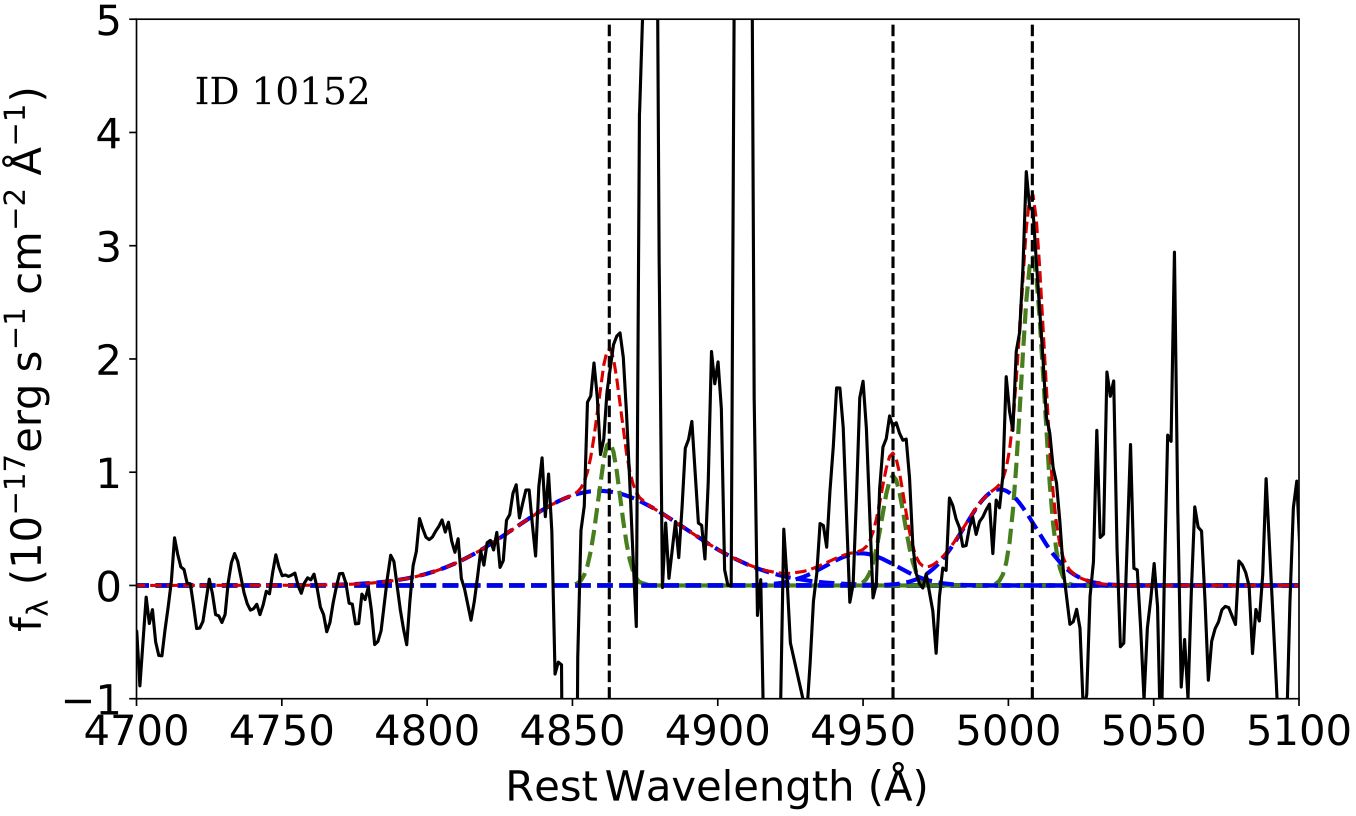} &
    \includegraphics[width=0.45\textwidth]{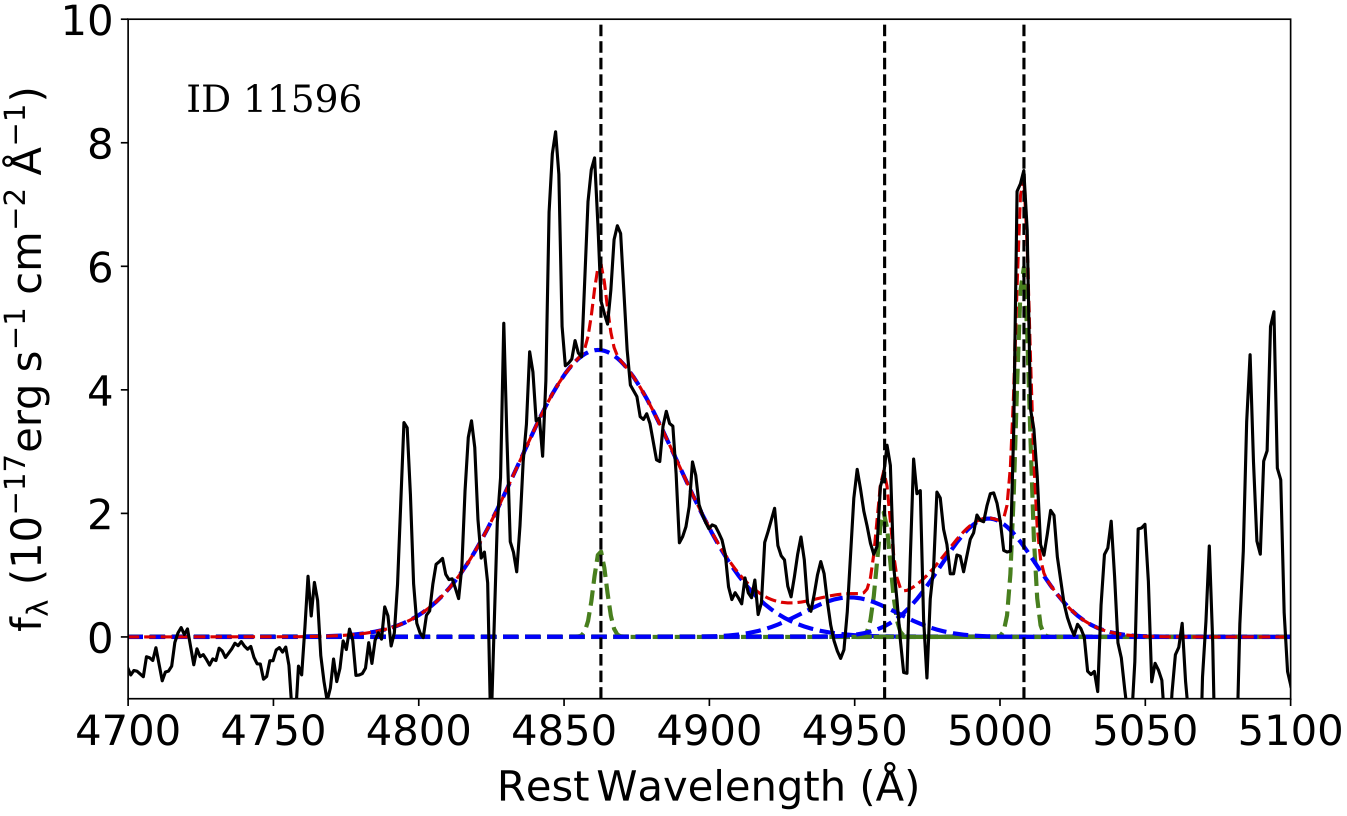} \\
    \includegraphics[width=0.45\textwidth]{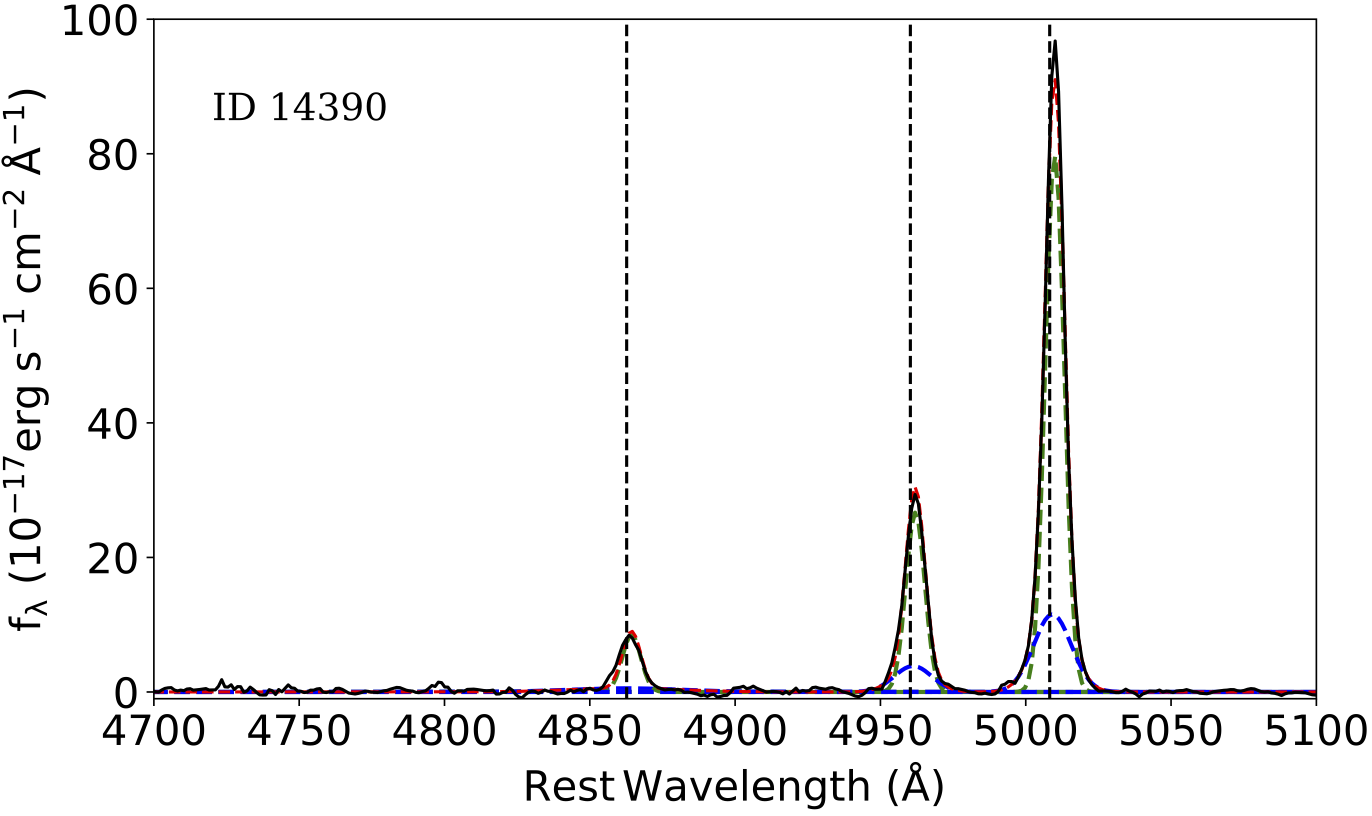} &
    \includegraphics[width=0.45\textwidth]{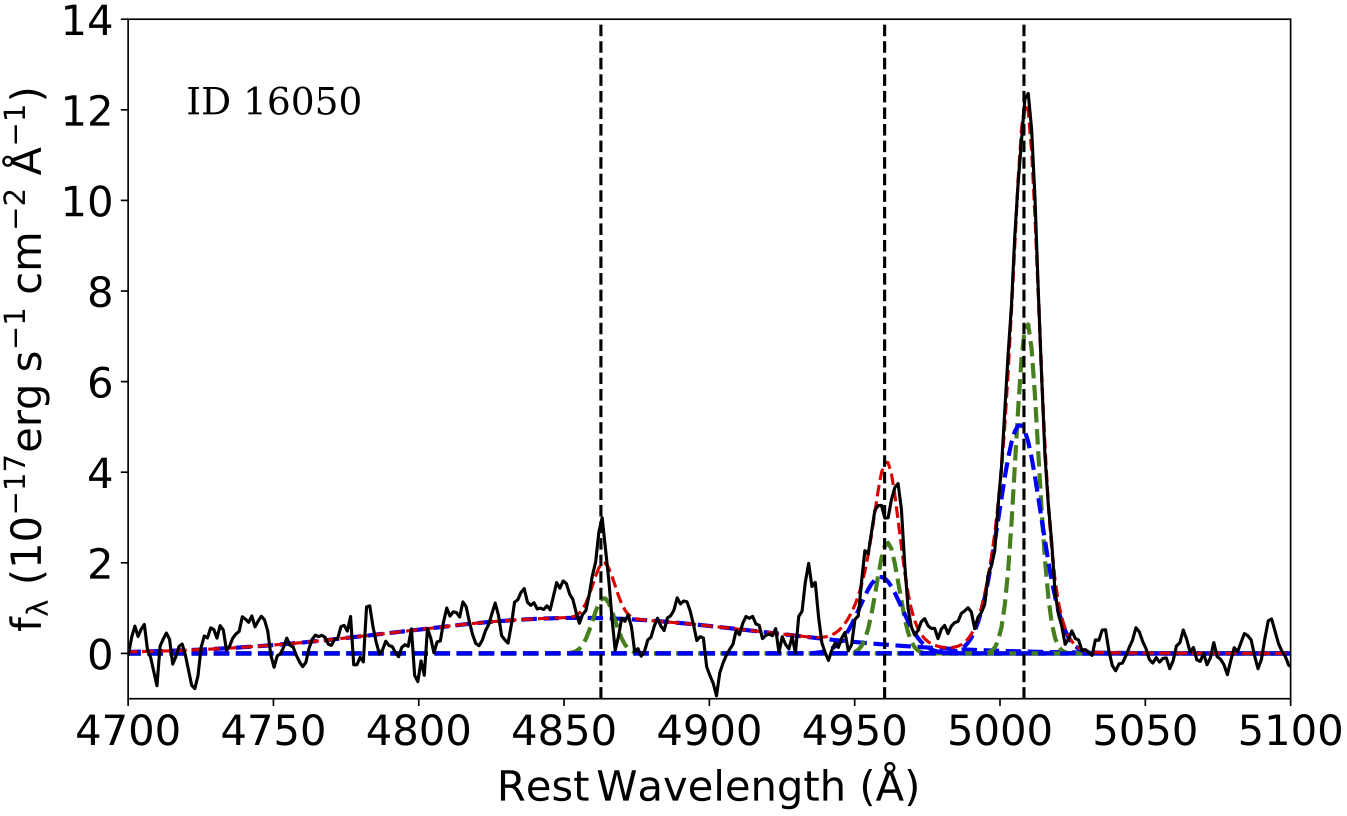} \\
    \includegraphics[width=0.45\textwidth]{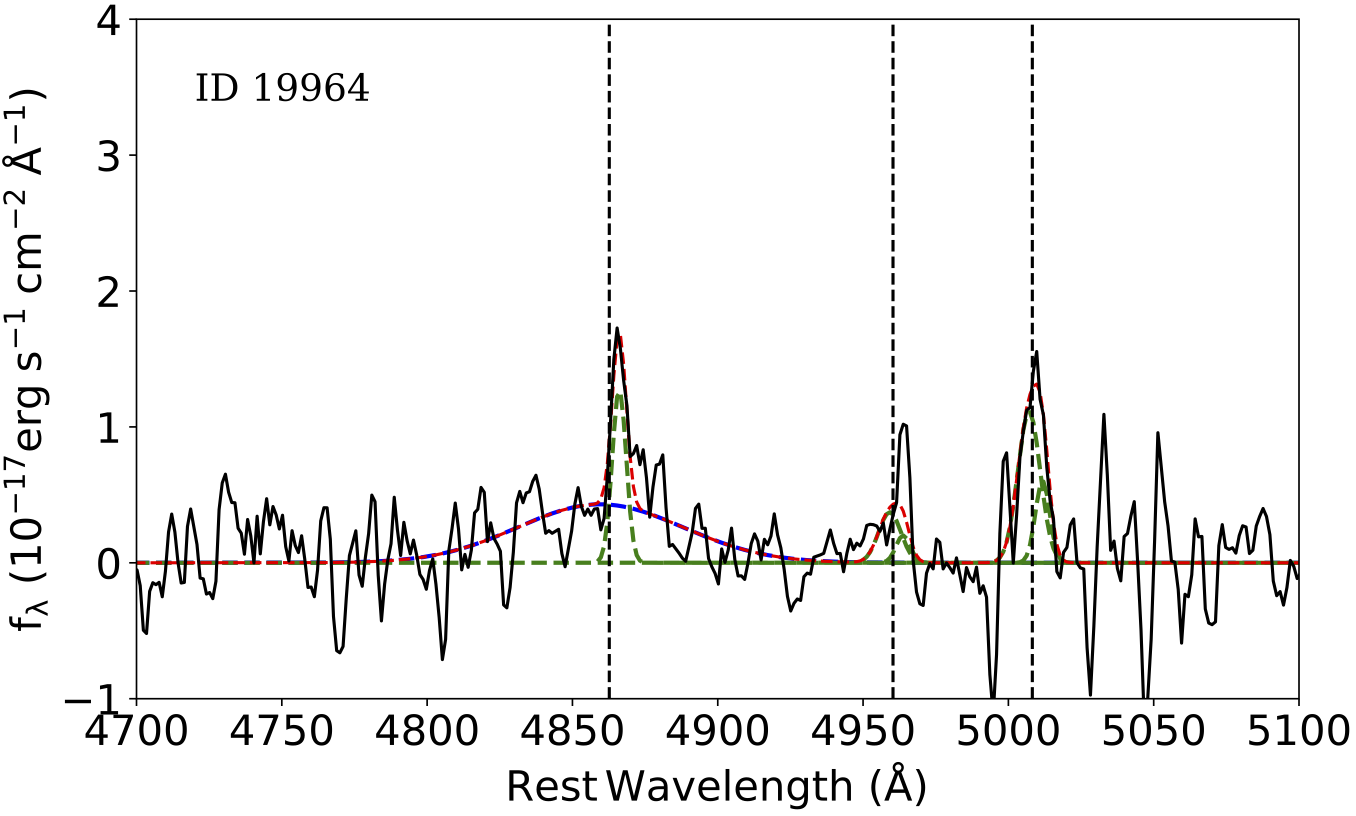}&
    \includegraphics[width=0.45\textwidth]{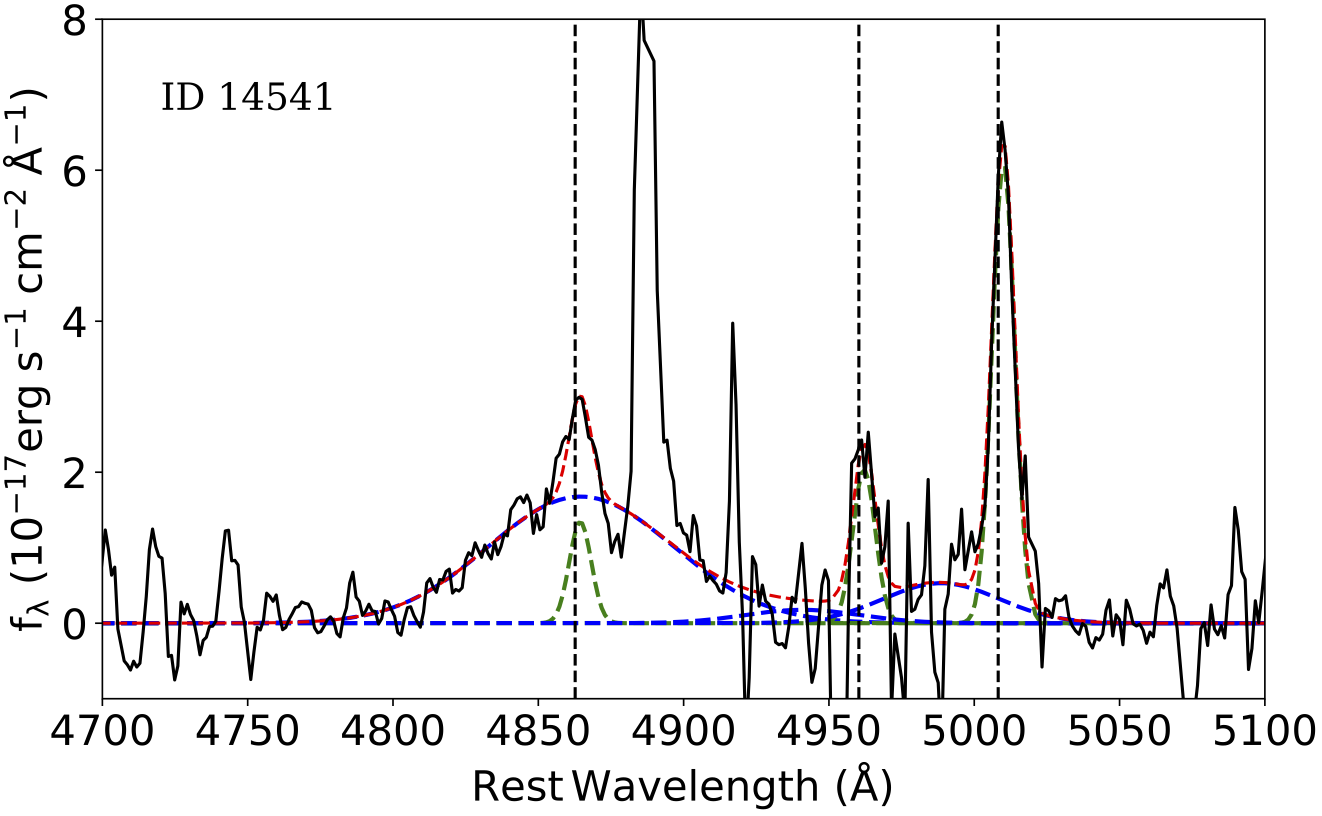} \\
    
\end{tabular}
\caption*{Fig. \ref{fig:efedsSDSSspectra2} continued}
   \label{fig:efedsSDSSspectra3}
\end{figure*}
\begin{figure*}[h!]
\centering
\begin{tabular}{cc} 
    \includegraphics[width=0.45\textwidth]{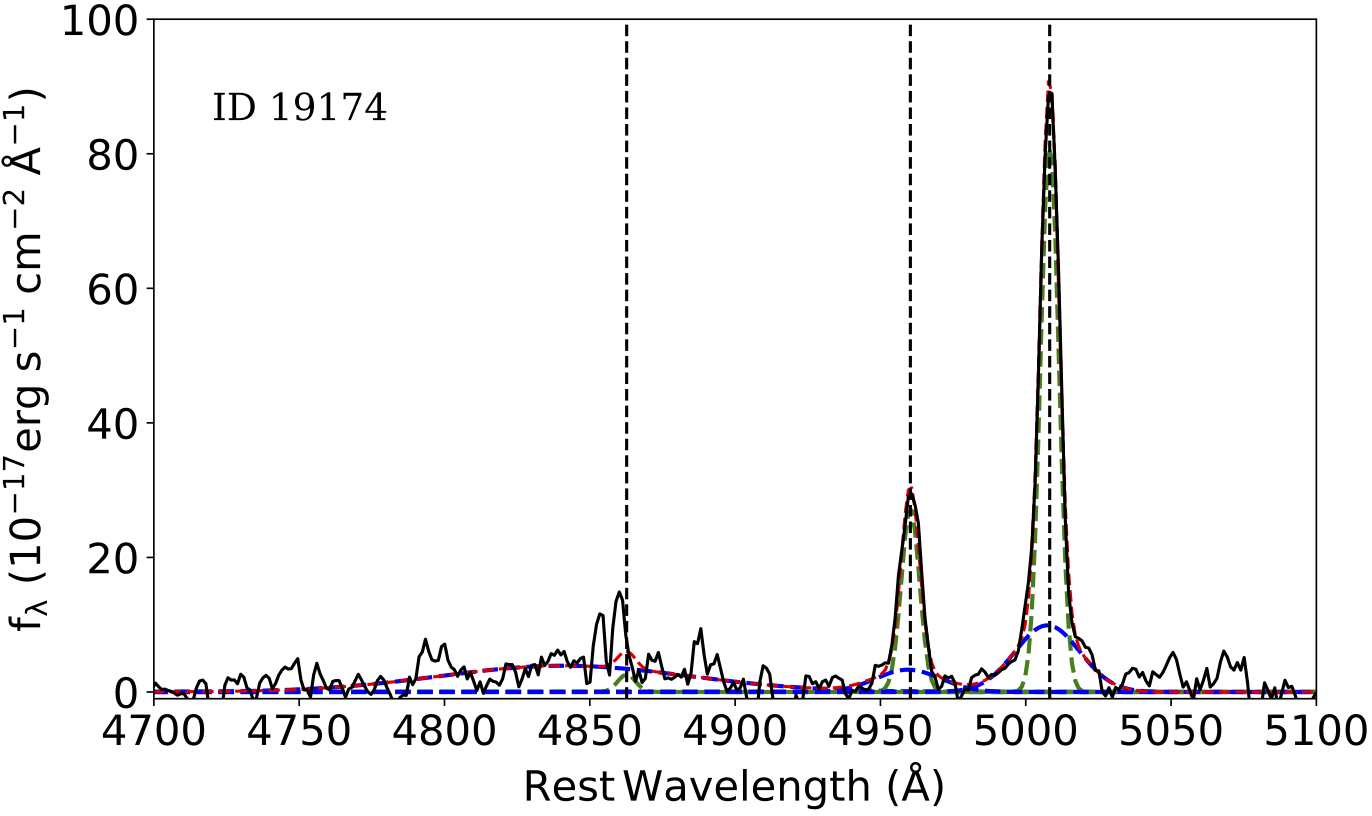} &
    \includegraphics[width=0.45\textwidth]{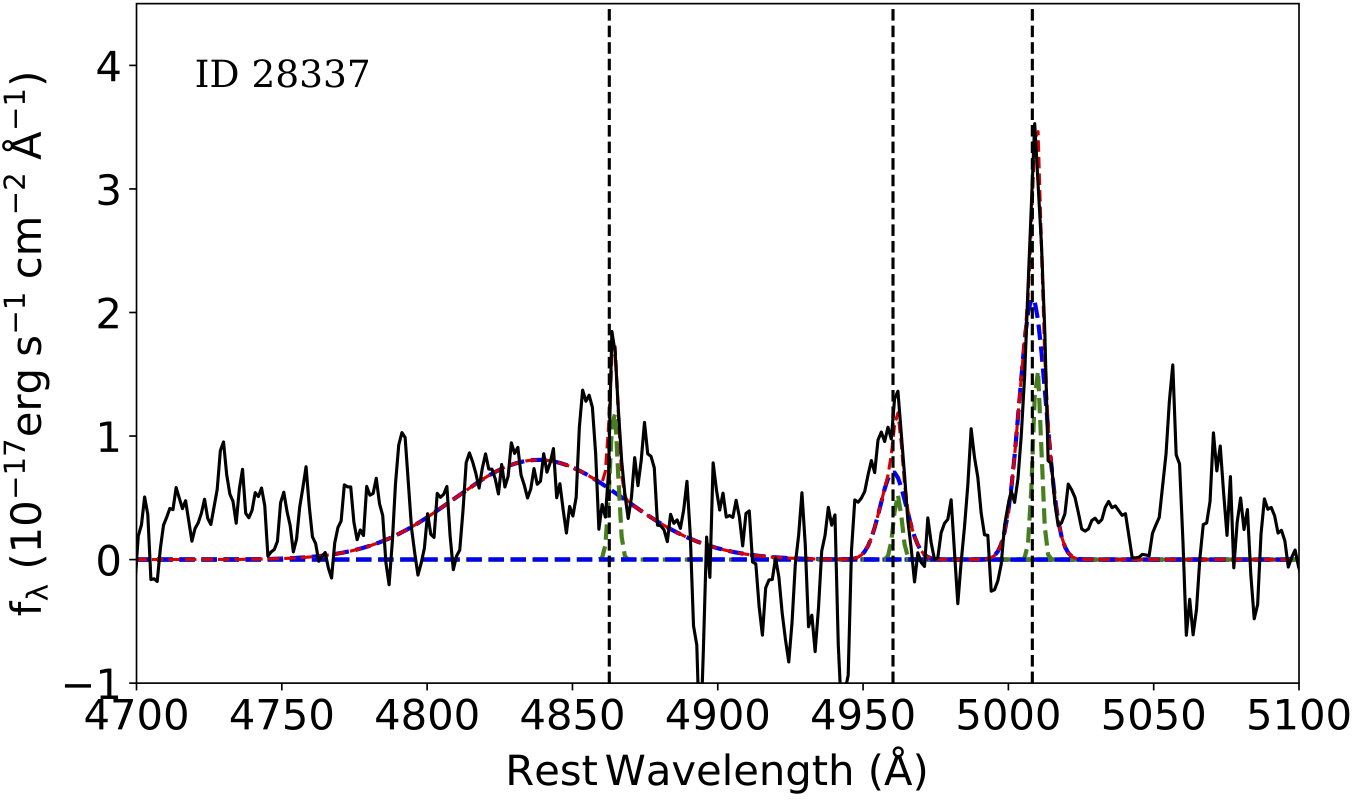} \\
    \includegraphics[width=0.45\textwidth]{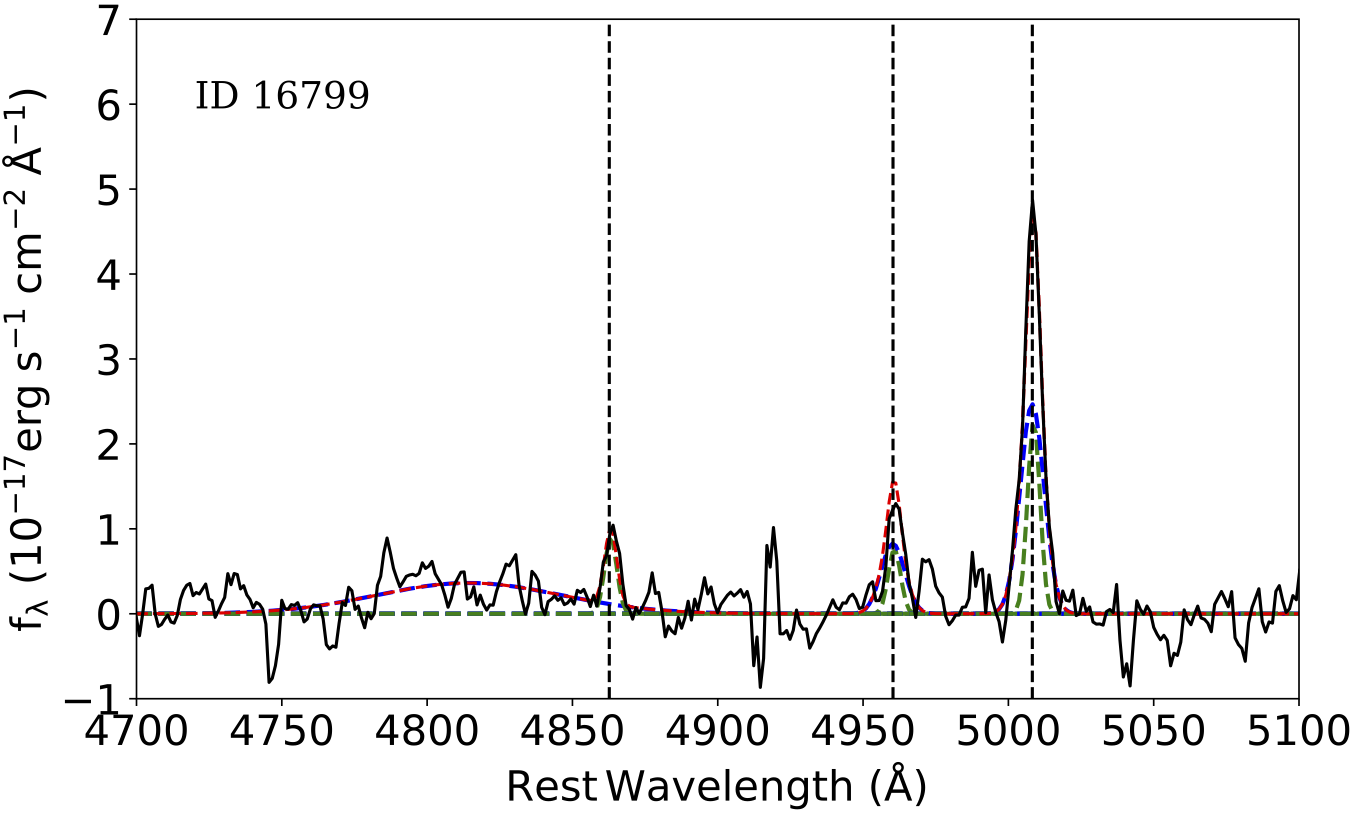} &
    \includegraphics[width=0.45\textwidth]{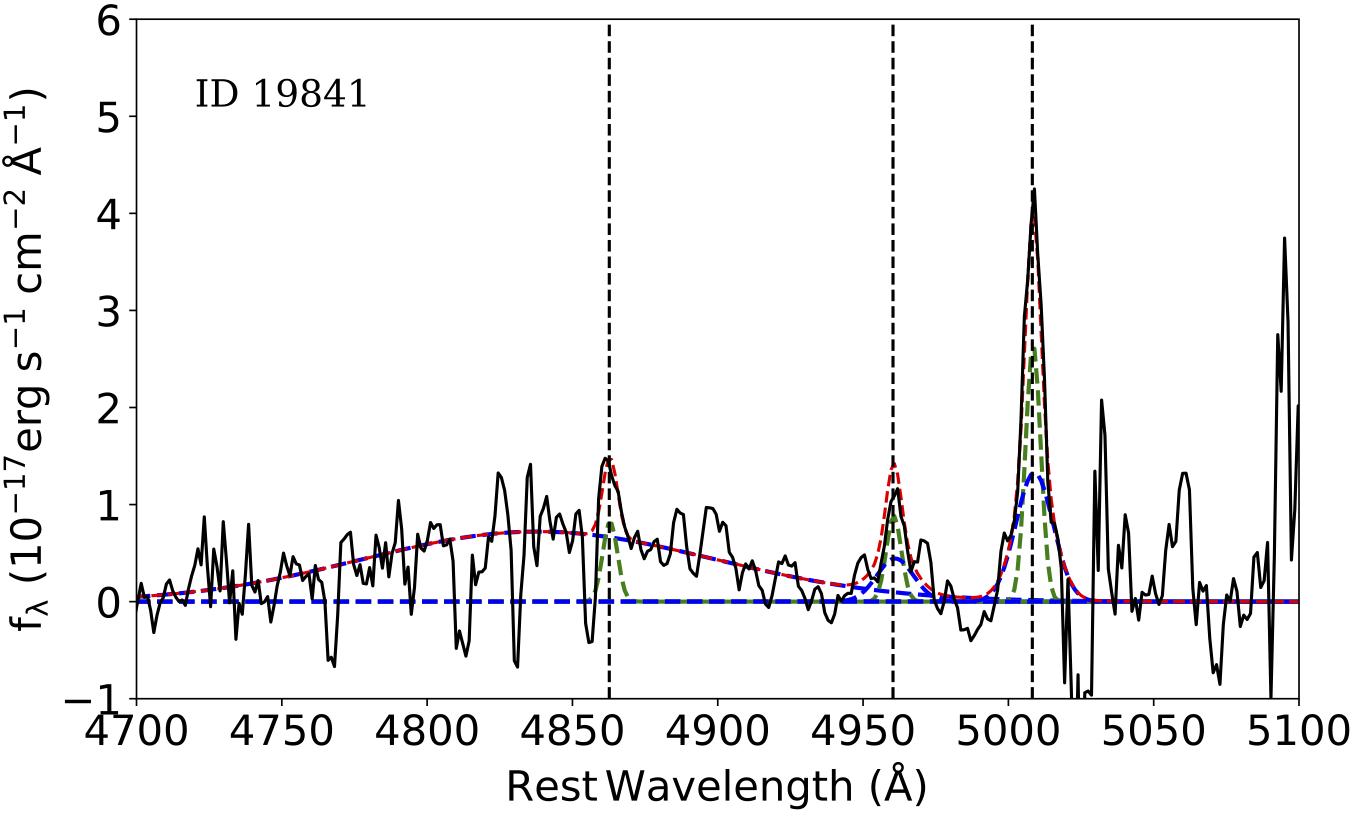} \\
    \includegraphics[width=0.45\textwidth]{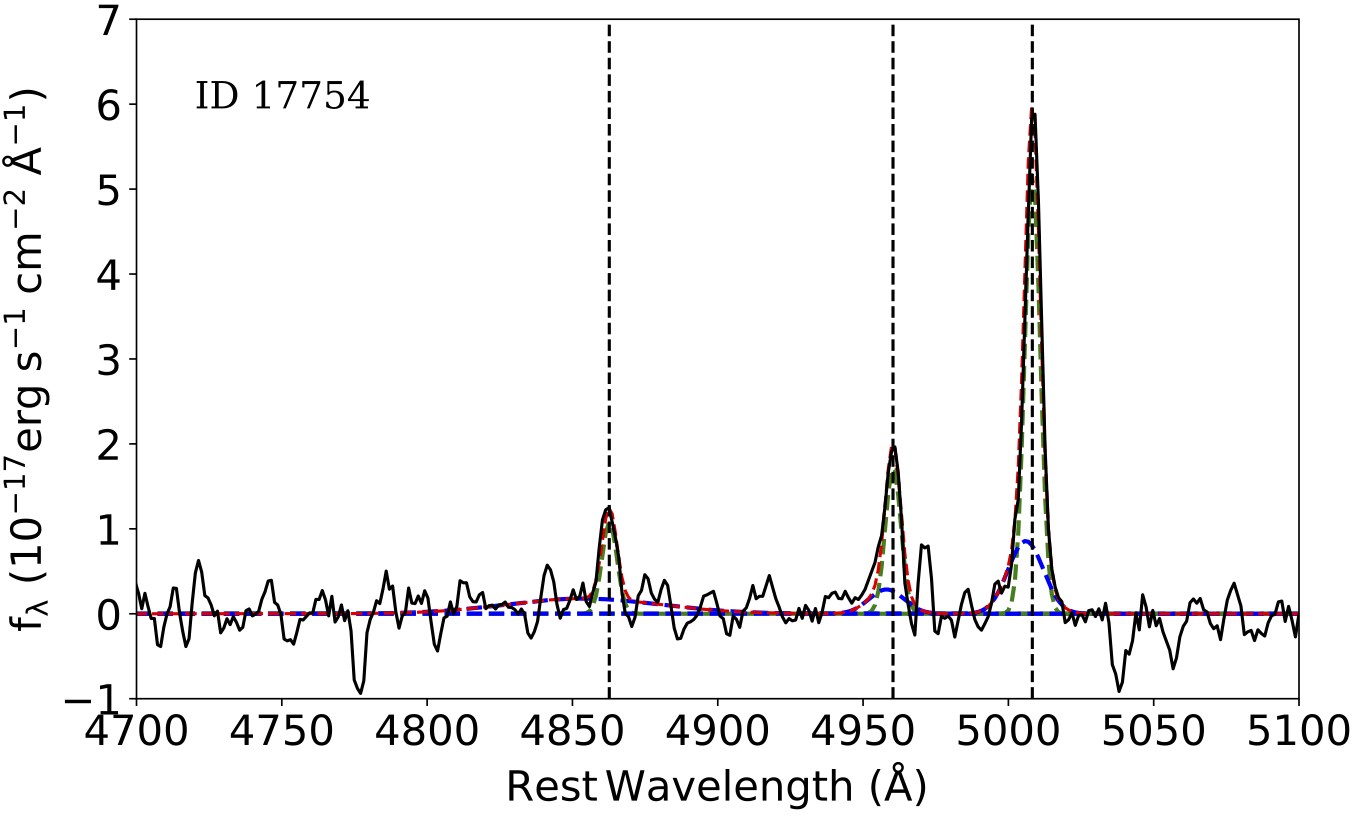} &
    \includegraphics[width=0.45\textwidth]{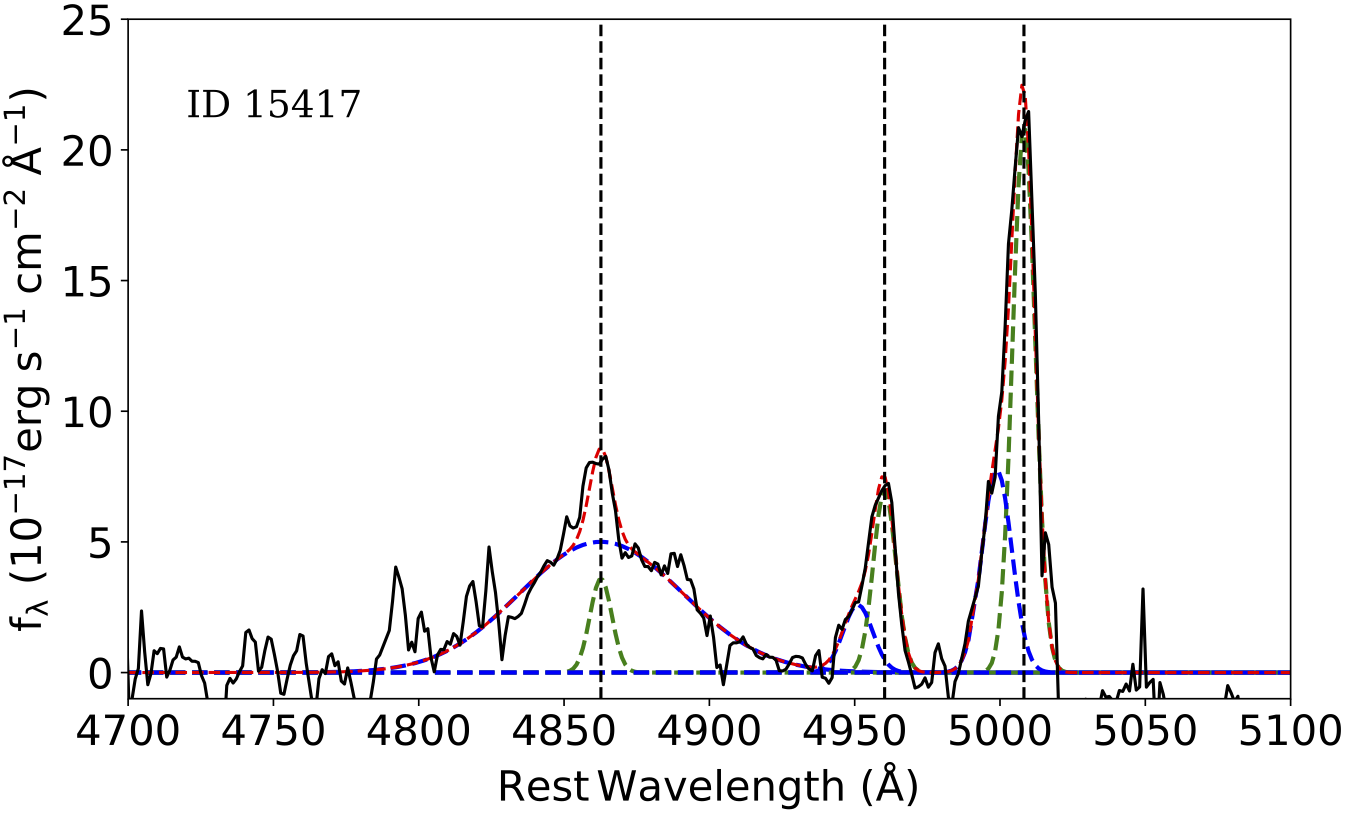} \\  
\end{tabular}
\caption*{Fig. \ref{fig:efedsSDSSspectra2} continued}
   \label{fig:efedsSDSSspectra4}
\end{figure*}
\begin{figure*}[h!]
\centering
\begin{tabular}{cc}
    \includegraphics[width=0.45\textwidth]{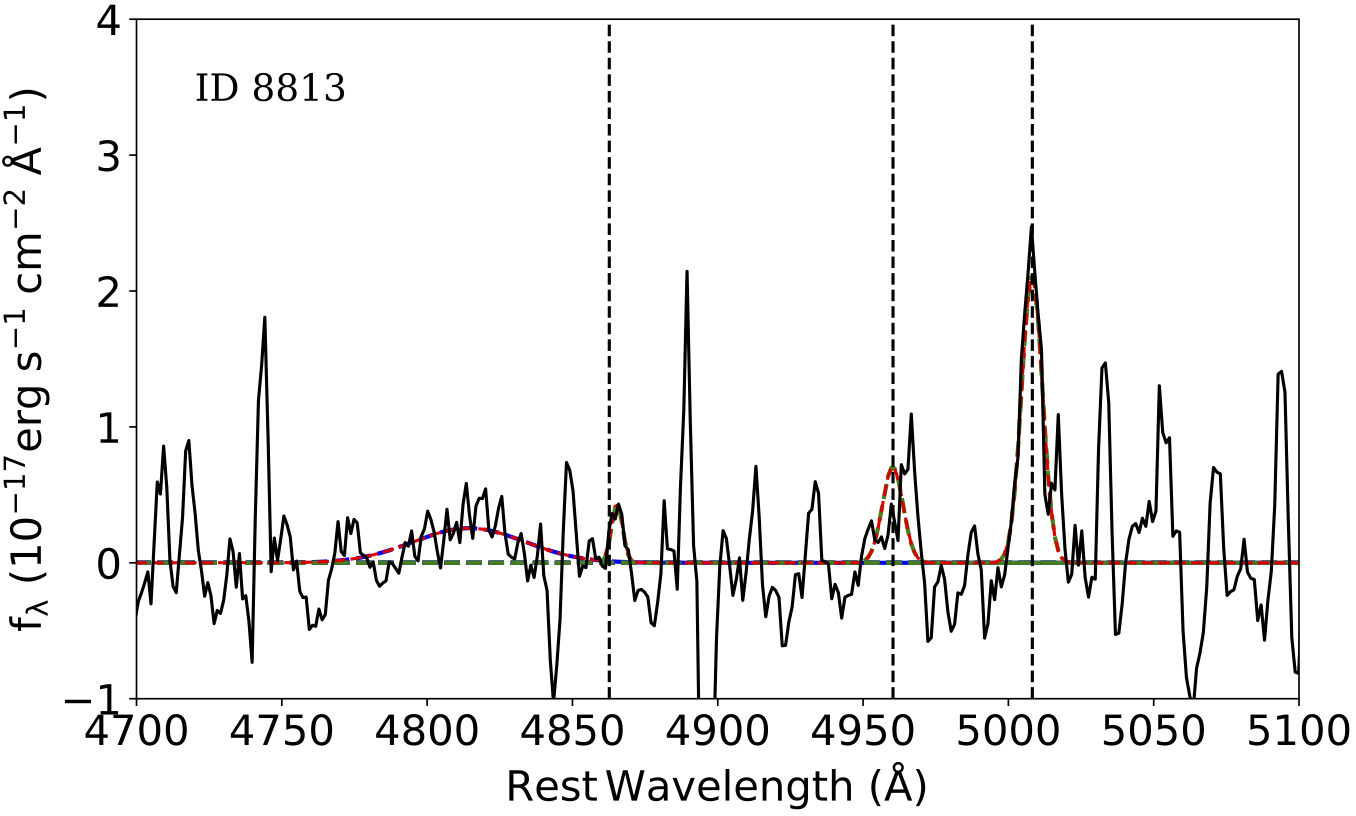} &
    \includegraphics[width=0.45\textwidth]{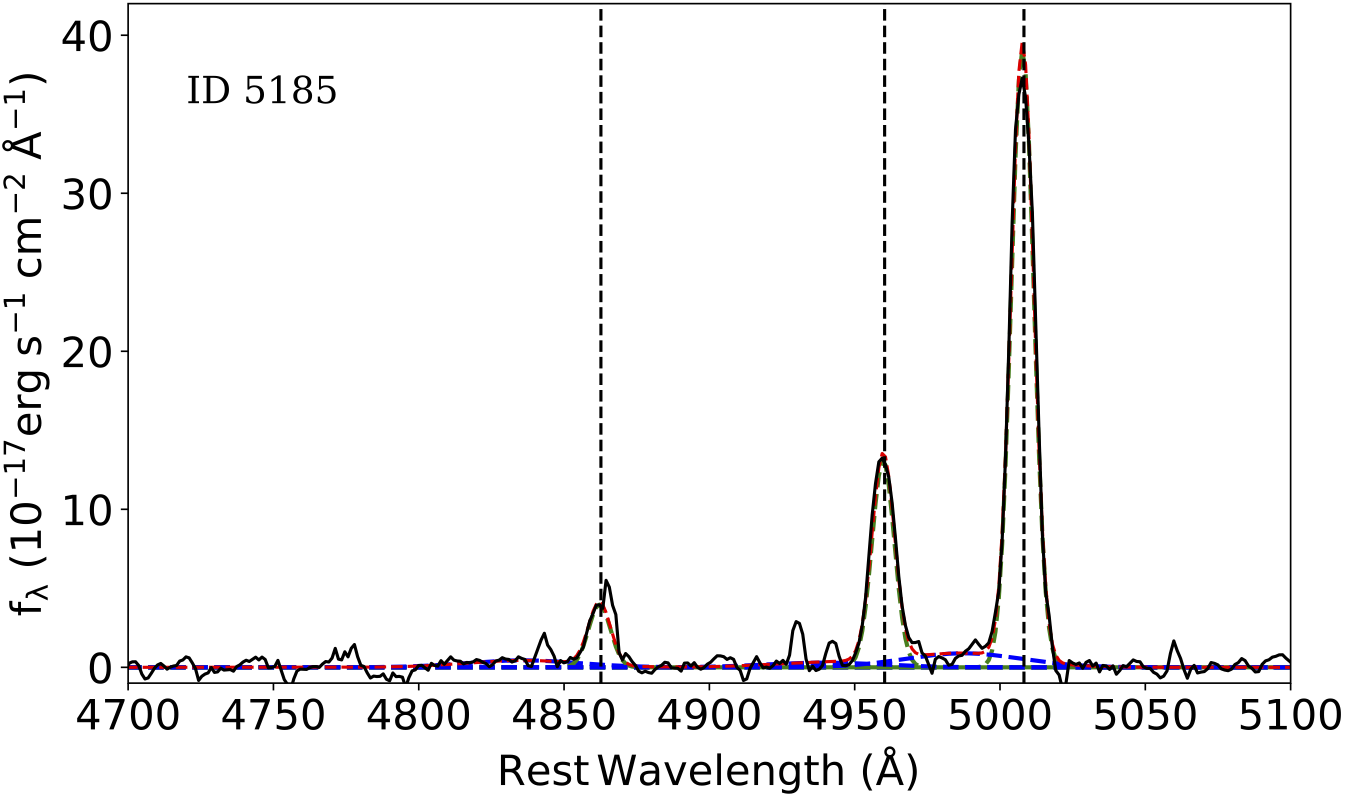}\\
\end{tabular}
\caption{Examples of spectra from two component fit but with poor spectra and single component fit. Left panel: An example of the spectra with single component fit (ID 8813) and FWHM$\rm{<800 kms^{-1}}$. Right panel: An example of the spectra with two component fit (ID 5185) but whose flux-to-flux error ratio is less than 2.5. These sources were excluded from our final candidates with outflows because the additional broad component is not significant. In both panels, the red line indicates the total fit. The vertical dotted lines indicate the peaks at 4862.68, 4960.30, and 5008.24 for $H_{\beta}$ and [OIII] rest-frame wavelength.}
    \label{fig:efedsSDSSspectra_1comp_2compLowSNR}
\end{figure*}
\begin{figure}[h!]
\centering
    \includegraphics[width=0.45\textwidth]{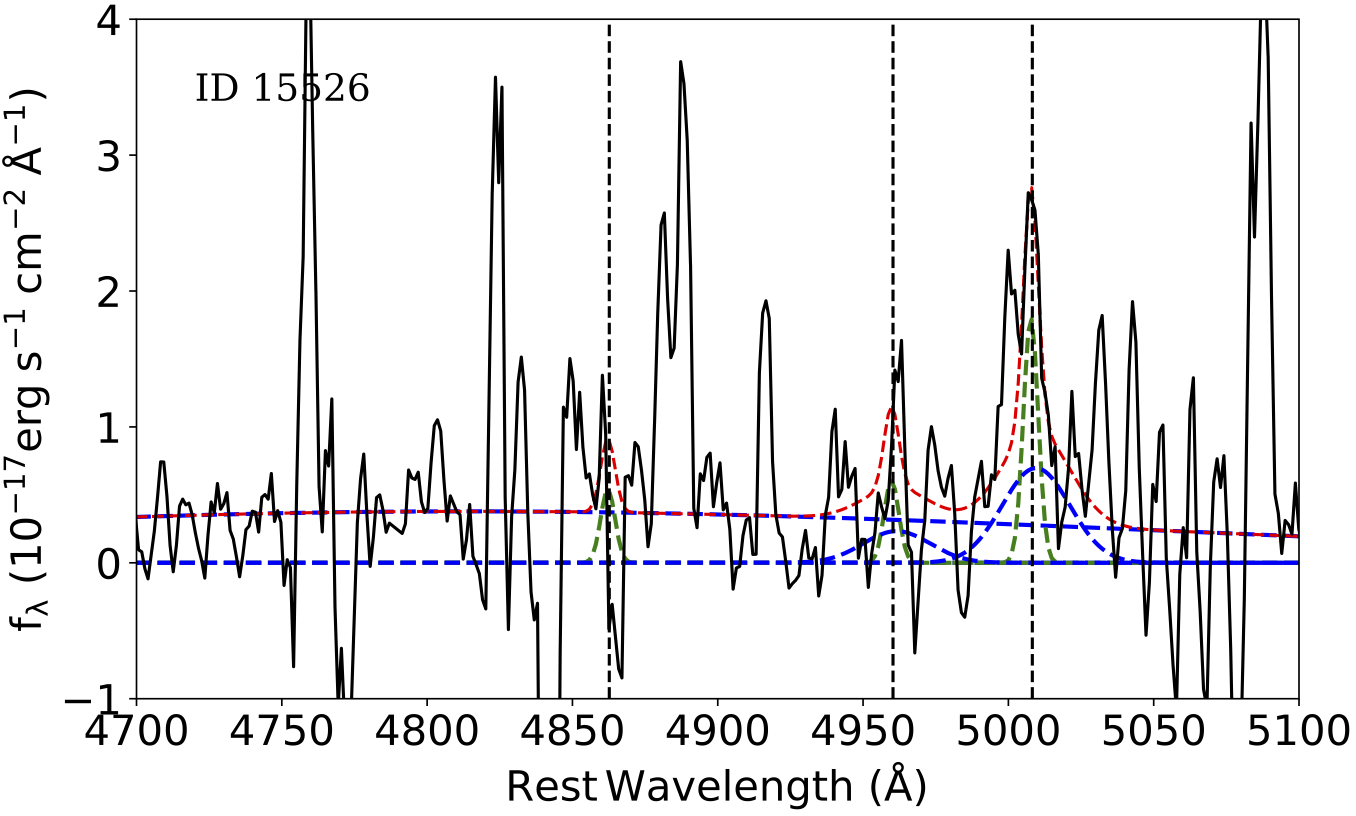} 
\caption{An example of the spectra that we classified as bad due to a very noisy continuum around the [OIII] region or bad residuals after the continuum and host galaxy subtraction. The red line indicates the total fit. The vertical dotted lines indicate the peaks at 4862.68, 4960.30, and 5008.24 for $H_{\beta}$ and [OIII] rest-frame wavelength.}
    \label{fig:efedsSDSSspectra_bad}
\end{figure}

\section{Summary of the results from each selection method. }
\label{selection_summary}
We have employed various selection techniques to identify AGNs in the feedback phase, aiming to evaluate our methods completeness and reliability through spectroscopic analysis. Unfortunately, the lack of available spectra for our candidates limits our analysis. Furthermore, the few spectra that are accessible exhibit poor quality, adding to our limitations. Consequently, in order to derive meaningful statistics from our reduced number of final subsamples relative to the original samples, we have grouped these subsamples into Sample A and Sample B instead of examining each method individually. Nevertheless, in order to demonstrate the efficacy of the individual methods, we present the performance results for different subsamples after applying various criteria in Table \ref{summary_selection}. This table effectively showcases the outcomes of our selection techniques which include; 1) r - W1 $>$ 4 and $\mathrm{F_{0.2-2.3 keV}/F_{opt}>}$1: selects 516 sources, 171 have z $<$ 1, 2 have SDSS spectra available (ID 608 and ID 4744) and only 1 source has good quality spectra and with outflows (ID 608). 2) r - W1$>$ 4 and $\mathrm{F_{2-10 keV}/F_{opt}>}$0.5: selects 274 sources, 77 have z $<$ 1, 2 have SDSS spectra available (ID 608 and ID 4744) and only 1 source has good quality spectra and with outflows (ID 608). 3)i - W4 $>$ 7 and r - W1$>$4: selects 203 sources, 66 have z $<$ 1, 1 have SDSS spectra available with good quality spectra and shows outflows (ID 608). 4) i - W3 $>$ 4.6: selects 539 sources, 231 have z $<$ 1, 6 have SDSS spectra available, 4 have good quality spectra (ID 608, ID 10152, ID 13349, ID 25592) and  3 sources show outflows (ID 608, ID 10152, ID 25592). 5) $\mathrm{log(N_{H})}>21.5$ and $\mathrm{\lambda_{Edd}}$: selects 528 sources, 78 have z $<$ 1,78 have SDSS spectra available, 47 have good quality spectra (these includes ID 608 and ID 13349 in colour selection ) and 21 sources show outflows (which include ID 608).
\begin{table*}
    \caption{Summary of the results from each selection method. To accurately convey the numerical data presented in this table, refer directly to the text. }
    \begin{tabular}{ |p{6cm}|p{1cm}|p{1.5cm}|p{1.5cm}|p{1cm}|p{1.5cm}| }
 \hline
 Selection method & Total & 0.5$<$ z $<$ 1 & available spectra& reliable spectra & confirmed outflows\\
 \hline
 r - W1 $>$ 4 and $\mathrm{F_{2-10 keV}/F_{opt}>}$1  & 516  & 171 &   2  & 1 & 1  \\ 
r - W1 $>$ 4 and $\mathrm{F_{0.2-2.3 keV}/F_{opt}>}$0.5 & 274 & 77 &  2 & 1 &  1 \\
 i - W4 $>$ 7 and r - W1$>$4 & 203  & 66 &  1 & 1 & 1 \\ 
 i - W3 $>$ 4.6 &  539  & 231 & 6 & 4 &  3\\ 
$\mathrm{log(N_{H})}>21.5$ and $\mathrm{\lambda_{Edd}}$ & 528 & 78 &  78 & 47 & 21 \\ 
\hline
\hline
Total (sample A + sample B)  & 1400 & 427 &  82 & 50 &  23 \\ 
 \hline
\end{tabular}
    \label{summary_selection}
\end{table*}
Except ID 608 which is selected by all the colour criteria, 2/3 of sources with outflows in Sample B are only selected by Eq. ~\ref{imw3}. As already discussed in \cite{hamann2016,perrotta2019}, this colour cut is reliable in isolating ERQs whose red colours are related to outflows.


\section{Additional table }
\label{additionalTable}

\begin{sidewaystable*}[!ht]
\begin{threeparttable}
    \centering
   \caption{Fit results from the broad and narrow components of [OIII] line are represented in the table below. }
    \begin{tabular}{ |c|c|c|c|c|c|c|c|c|c|c|}    
 \hline
ERO ID  & z & FWHM$\mathrm{^{n}}$ & FWHM$\mathrm{^{b}}$ & $\mathrm{\Delta}$V  & $\mathrm{\sigma}$ & $\rm{V_{max}}$ & flux & flux/flux$\rm{_{err}}$  & L[OIII]$\mathrm{^{b}}$ & sample \\
 &  & $\mathrm{km~s^{-1}}$ & $\mathrm{km~s^{-1}}$ & $\mathrm{km~s^{-1}}$ & $\mathrm{km~s^{-1}}$ & $\mathrm{km~s^{-1}}$ & $\mathrm{erg~cm^{-2}~s^{-1}}$ ($10^{-17}$)  & & $\mathrm{erg~s^{-1}}$ ($10^{42}$) &  \\
\hline
25592 & 0.660 & 554.2$\pm$12.9 & 1732.2$\pm$19.0 & 667.7$\pm$22.4 & 725.1$\pm$8.1 & 2118.0$\pm$27.6 & 284.5$\pm$4.4 & 64.1 & 5.3$\pm$0.08 & A \\
6743 & 0.652 & 394.4$\pm$29.7 & 773.1$\pm$112.0 & 526.3$\pm$60.6 & 322.3$\pm$47.6 & 1171.1$\pm$112.9 & 22.4$\pm$0.9 & 24.8 & 0.4$\pm$0.01 & B  \\
608 & 0.602 & 501.4$\pm$46.4 & 1514.0$\pm$79.1 & 171.1$\pm$9.0 & 634.2$\pm$33.6 & 1439.6$\pm$67.9 & 148.1$\pm$6.9 & 21.4 & 2.2$\pm$0.10 & A \& B  \\
19520 & 0.937 & 554.3$\pm$-99.0 & 1794.3$\pm$133.5 & 407.9$\pm$52.9 & 751.1$\pm$56.8 & 1910.2$\pm$125.4 & 146.6$\pm$10.5 & 13.8 & 6.5$\pm$0.4 & B  \\
18777 & 0.852 & 554.0$\pm$1.1 & 2684.0$\pm$60.3 & 1053.4$\pm$306.3 & 1124.4$\pm$25.6 & 3302.4$\pm$310.5 & 26.2$\pm$1.9 & 13.4 & 0.9$\pm$0.06 & B  \\
21835$^*$ & 0.981 & -99.0 & 884.6$\pm$36.8 & 44.6$\pm$90.6 & 369.2$\pm$15.7 & 783.2$\pm$95.9 & 64.0$\pm$5.1 & 12.3 & 3.1$\pm$0.2 & B  \\
12686 & 0.768 & 349.3$\pm$7.2 & 1174.1$\pm$228.0 & 119.4$\pm$77.7 & 491.8$\pm$97.0 & 1103.0$\pm$209.0 & 76.4$\pm$6.4 & 11.8 & 2.0$\pm$0.1 & B  \\
31136 & 0.655 & 412.0$\pm$21.5 & 1580.5$\pm$112.3 & 275.1$\pm$66.0 & 661.4$\pm$47.8 & 1598.0$\pm$116.2 & 123.3$\pm$11.6 & 10.5 & 2.2$\pm$0.2 & B  \\
5896$^*$ & 0.699 & -99.0 & 1009.1$\pm$97.7 & 67.4$\pm$62.4 & 422.6$\pm$41.6 & 912.8$\pm$104.0 & 209.9$\pm$21.4 & 9.7 & 4.5$\pm$0.4 & B  \\
16271 & 0.508 & 496.4$\pm$25.1 & 1270.0$\pm$257.8 & 155.4$\pm$48.6 & 531.1$\pm$109.7 & 1217.7$\pm$224.7 & 32.1$\pm$4.9 & 6.4 & 0.3$\pm$0.04 & B  \\
14896 & 0.819 & 317.8$\pm$114.9 & 1382.3$\pm$190.1 & 145.7$\pm$126.7 & 579.0$\pm$80.9 & 1303.9$\pm$205.5 & 26.5$\pm$4.2 & 6.1 & 0.8$\pm$0.1 & B  \\
10152 & 0.923 & 554.1$\pm$178.2 & 1727.0$\pm$921.4 & 671.7$\pm$492.4 & 723.5$\pm$392.1 & 2118.7$\pm$926.0 & 26.0$\pm$4.5 & 5.6 & 1.1$\pm$0.1 & A  \\
11596 & 0.956 & 323.9$\pm$84.6 & 2315.4$\pm$147.4 & 723.9$\pm$174.4 & 970.0$\pm$62.7 & 2664.1$\pm$214.9 & 78.6$\pm$14.4 & 5.4 & 3.6$\pm$0.6 & B  \\
14390 & 0.608 & 462.5$\pm$8.7 & 916.5$\pm$48.9 & 55.8$\pm$24.6 & 383.9$\pm$20.8 & 823.6$\pm$48.4 & 187.3$\pm$36.7 & 5.0 & 2.8$\pm$0.5 & B  \\
16050 & 0.615 & 554.2$\pm$117.0 & 1019.2$\pm$413.8 & 83.0$\pm$51.6 & 426.9$\pm$176.1 & 936.8$\pm$355.9 & 91.2$\pm$20.2 & 4.5 & 1.4$\pm$0.3 & B  \\
19964 & 0.670 & 358.4$\pm$189.6 & 551.4$\pm$55.2 & 104.6$\pm$110.1 & 228.8$\pm$23.5 & 562.3$\pm$119.7 & 10.7$\pm$3.1 & 3.4 & 0.2$\pm$0.06 & B  \\
14541 & 0.906 & 554.3$\pm$73.2 & 2813.6$\pm$782.9 & 1210.2$\pm$909.3 & 1178.4$\pm$333.1 & 3567.1$\pm$1127.3 & 26.4$\pm$7.8 & 3.3 & 1.0$\pm$0.3 & B  \\
19174 & 0.707 & 450.4$\pm$49.3 & 1570.7$\pm$1032.9 & 47.4$\pm$158.6 & 658.0$\pm$439.5 & 1363.5$\pm$893.3 & 275.9$\pm$88.6 & 3.1 & 6.1$\pm$1.9 & B  \\
28337 & 0.706 & 199.8$\pm$116.7 & 639.1$\pm$234.1 & 3.5$\pm$243.3 & 267.6$\pm$99.6 & 538.7$\pm$314.4 & 23.9$\pm$7.8 & 3.0 & 0.5$\pm$0.1 & B  \\
16799 & 0.546 & 306.2$\pm$117.6 & 581.5$\pm$894.5 & 2.9$\pm$268.8 & 243.5$\pm$380.6 & 489.9$\pm$807.4 & 25.4$\pm$8.4 & 3.0 & 0.3$\pm$0.1 & B  \\
19841 & 0.854 & 368.9$\pm$79.1 & 898.6$\pm$102.8 & 43.7$\pm$52.4 & 376.3$\pm$43.7 & 796.5$\pm$102.0 & 21.1$\pm$7.3 & 2.8 & 0.7$\pm$0.2 & B  \\
17754 & 0.667 & 374.7$\pm$19.3 & 854.7$\pm$219.6 & 135.5$\pm$131.3 & 358.0$\pm$93.4 & 851.5$\pm$228.5 & 12.9$\pm$4.5 & 2.8 & 0.2$\pm$0.08 & B  \\
15417 & 0.983 & 554.1$\pm$6.2 & 725.8$\pm$138.9 & 593.2$\pm$119.5 & 302.4$\pm$59.1 & 1198.0$\pm$168.1 & 99.2$\pm$35.3 & 2.8 & 4.9$\pm$1.7 & B  \\
\hline
\end{tabular}

\label{tab:fitresults1}
\begin{tablenotes}       
        \item Notes: The superscript b and n stand for broad and narrow components, respectively. The columns (left-right) are eROSITA ID (ERO ID ), redshift (z), the narrow and broad FWHM with their respective errors, change in velocity ($\mathrm{\Delta}$V), velocity dispersion ($\mathrm{\sigma}$), maximum velocity ($\rm{V_{max}}$), [OIII] flux, [OIII] luminosity (L[OIII]$\mathrm{^{b}}$ ) and the sample in which the source was selected. -99.0 is used to indicate missing values. Sources marked with $^*$ have single Gaussian components. 
      \end{tablenotes}
    \end{threeparttable}
\end{sidewaystable*}

\section{Assumptions on different parameters in the computation of outflow properties}
\label{assumptions}

The computations of outflow properties are based on several assumptions on different parameters which include; geometry, electron density, temperature, metallicity, in some cases, the velocity of the outflowing gas and radius. These assumptions introduce several uncertainties in the properties of the final values estimated. Below, we summarise some of the assumptions made in computing the outflow properties of QWO from the literature that is \cite{Brusa2015,kakkad2016,fiore2017,perrotta2019,leung2019,kakkad2020,vayner2021,brusa2022}. 

Different proxies of velocities are used to compute AGN outflow rates. In the literature, non parametric line width that contains 80\% of the total flux (W80) \citep{kakkad2016,kakkad2020}, V10 \citep{kakkad2016,vayner2021}, V98 \citep{perrotta2019}, mean radial velocity (Vr) \citep{perrotta2019}, V50 and maximum velocity ($V_{max}$ ) \citep{Brusa2015,fiore2017,leung2019,Brusa2015} have been used. Maximum velocity has been calculated as; 
\begin{equation}
    \label{vmax2}
    \rm{V_{max}= |\Delta{V}|+2\sigma_{[OIII]}^{broad}},
\end{equation}
where $\Delta{V}$ is the velocity shift between the velocity peak of the broad emission line and the systemic velocity and $\sigma$ is the velocity dispersion. Another definition of maximum velocity is used in \cite{rupke2005a,fluetch2019} ($\rm{V_{max}= |\Delta{V}|+FWHM/2}$).  W80 is close to the FWHM of the line and relates to the typical velocity of the outflow gas for a Gaussian profile. V10 gives an estimate of the velocity the gas is moving towards us at the high-velocity end. This implies that W80 gives higher velocity values than v10 \citep{kakkad2016}. W80 is highly correlated with $V_{max}$ as W80 $\sim0.9~\times~ V_{max}$ \citep{fiore2017,leung2019}. W80 can be a factor of 1.3-1.8 smaller than $V_{max}$ \citep{bischetti2017}. 

Since the mass of the outflow is inversely related to the gas electron density, the assumptions made whenever $n_{e}$ is not measured have direct uncertainties in the outflow mass which indirectly contributes to the mass outflow rate value. As compared to \cite{fiore2017},
\cite{vayner2021,kakkad2020,brusa2022} will have outflow energies lower by a factor of $\sim$2.5 while \cite{Brusa2015} and \cite{kakkad2016} will be higher by a factor of $\sim$2.  
Although $n_{e}$ has been assumed in most of the previous studies, it has been measured in a few such as \citet[][among others]{nesvadba2006,perna2015a,Brusa2015,brusa2016} from the line flux ratios of [SII]$\lambda\lambda$6717,6731 line doublet using the formula;
\begin{equation}
    n_{e}=10^2T^{0.5}\left(\frac{r-1.49}{5.62-12.8r}\right),
\end{equation}
where r is the [SII] flux ratio.
Some of the measured values of $n_{e}$ include among others; $780\pm300~cm^{-3}$ \citep{brusa2016}, $1000-3000~cm^{-3}$ \citep{Brusa2015}, $120~cm^{-3}$ \citep{perna2015a}  , $240-570~cm^{-3}$ \citep{nesvadba2006}. In addition to the commonly used method of using [SII] doublet flux ratios to estimate electron densities, there are alternative approaches in the field. These include the auroral and transauroral line method and the ionisation parameter method. A comprehensive comparison of these various methods can be found in the study conducted by \cite{davies2020}. The aforementioned study by \cite{davies2020} highlights certain limitations associated with the [SII] doublet method, particularly for electron densities $\sim10^3$ cm$^{-3}$. The researchers caution against relying solely on this method as it tends to predominantly trace low electron density regions, potentially leading to a biased perspective of the ionised gas. Similar caution is also echoed in other studies \citep[e.g.][]{baronandnetzer2019,davies2020}. Furthermore, ongoing research by Speranza et al. (in preparation) emphasises similar concerns. Collectively, they have found electron densities estimated by [SII] doublet method to be lower as compared to other methods and hence giving much higher values of mass outflow rates and even much higher values if you consider electron density of 200 cm$^{-3}$ as assumed in \cite{fiore2017} and our study.

Outflow rates computed by assuming shell$-$like geometry give higher values by a factor of $\sim$3 than biconical or spherical geometry \citep{maiolino2012,cicone2014,kakkad2016}. The use of different electron temperatures (e.g. 10000 or 25000 K) would change the normalisation factor in Eq. \ref{moutrate2} by a certain factor since the emissivity depends on the temperature \cite{kakkad2020}.

The radius of the outflow can be measured but it is always assumed if the outflow is unresolved. Mass outflow rates and outflow energies computed by assuming higher values of the radius will be times lower than those computed with lower values of radius.

Using different lines to derive the outflow properties, different assumptions, not taking into account the multi-phase nature of outflows, etc., makes it challenging to characterise AGNs in the feedback phase and comparison of these outflow properties to theoretical prediction. The assumption made in previous studies is summarised in Table \ref{outflowassumptions}.
\begin{table*}
\begin{threeparttable}
    \caption{Assumed parameters in computing mass outflow rates and energetics.  }
    \begin{tabular}{ |p{3cm}||p{3cm}|p{2cm}|p{2cm}|p{2cm}|p{2cm}| }
 \hline
 \multicolumn{6}{|c|}{Outflow properties assumptions} \\
 \hline
 Reference & Geometry & Temperature $T_{e}$ & Electron density ($n_{e}$) (cm$^-3$) & Radius (Kpc)& Velocity\\
 \hline
 1  & spherical   & - &   500 & 10 &  Eq. \ref{vmax}, V90  \\ 
 2 & conical  & - &  500 & - & V10 \\
 3 &  biconical \& thin shell  & 10$^4$ & 500 & 2 &  W80\\ 
4  & biconical & - &  200 & 1 &  V98 \&Vr \\ 
5 & thin shell & 10$^4$ &  150 & r10 &  Eq. \ref{vmax} \\ 
5 & spherical  & 10$^4$ & 200 & - & Eq. \ref{vmax}  \\   
6 &   conical  & 10$^4$& 100 & - & v10 \& W80 \\  
7 & spherical & - & 100 & 5 &  Eq. \ref{vmax} \\ 
 \hline
\end{tabular}
\label{outflowassumptions}

\begin{tablenotes}       
        \item Notes: We show the reference(1.\cite{brusa2022}, 2.\cite{vayner2021}, 3.\cite{kakkad2020}, 4.\cite{perrotta2019}, 5.\cite{leung2019}, 6.\cite{kakkad2016}, 7.\cite{Brusa2015}),  geometry is the assumed geometry whether conical, biconical, spherical or thin shell, Te is the electron temperature, $n_{e}$ is the gas electron density,  the radius at which the outflow was computed (assumed or measured), the outflow velocity used.
      \end{tablenotes}
    \end{threeparttable}
\end{table*}
By carefully considering the uncertainties that arise from the different assumptions made, we re-computed the outflow properties of the QWO sample with the best-homogenised assumptions. We have computed mass outflow, mass outflow rate and kinetic power for \cite{kakkad2020,perrotta2019,leung2019,brusa2022}  assuming a spherical outflow geometry and electron density of 200 cm$^{-3}$. We used radius as provided in the literature. For the case of velocity, we used V$\rm{max}$ or outflow velocity when provided, otherwise, we computed using $\rm{V_{max}= |\Delta{V}|+2\times{FWHM/2.355}}$. The final values will have 2-3 orders of magnitude uncertainties due to the different assumptions applied.

\section{The X--ray properties of our eFEDS candidates with outflows. }
\label{x-raypropertiesofoutflows}
\begin{figure*}[h!] 
    \centering
    \begin{tabular}{c c}
    \includegraphics[width=0.45\textwidth]{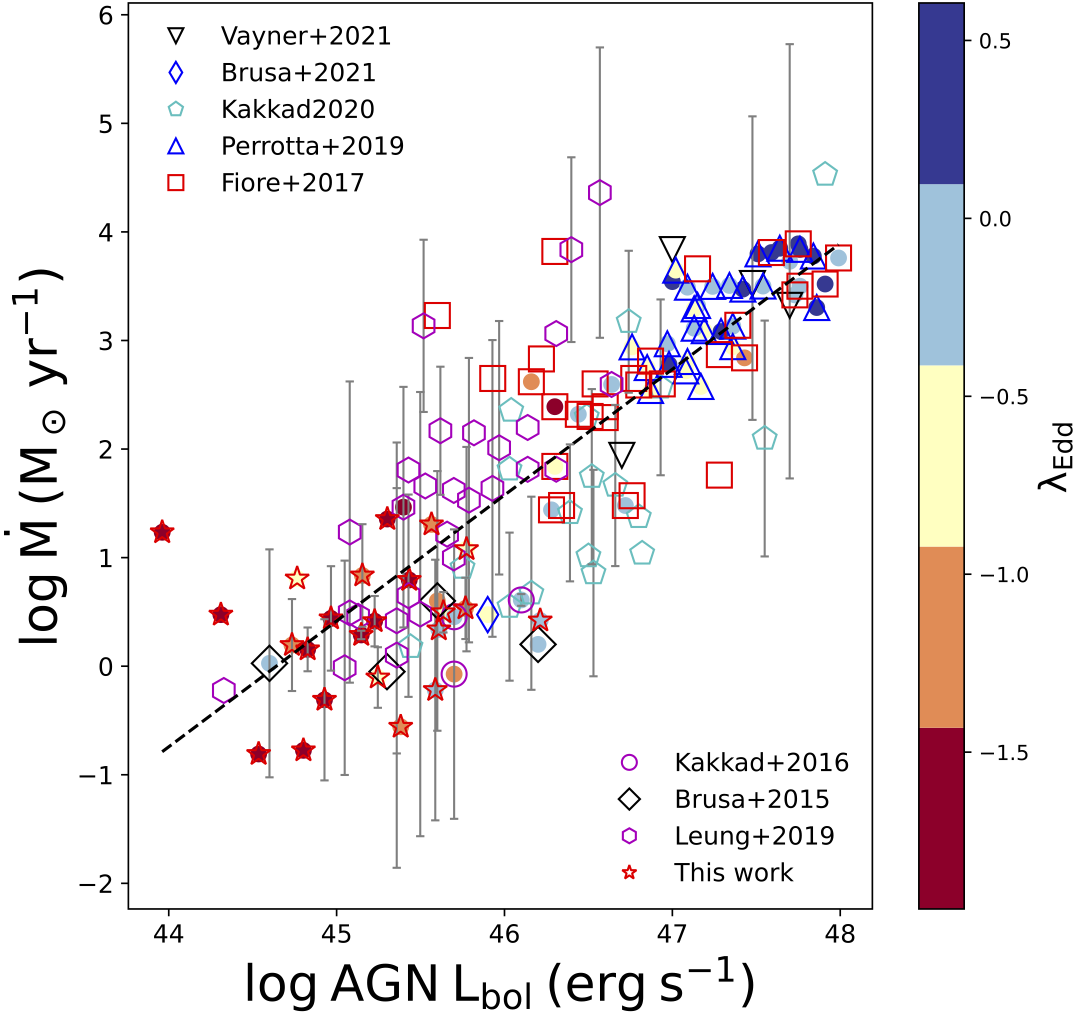} 
    \includegraphics[width=0.45\textwidth]{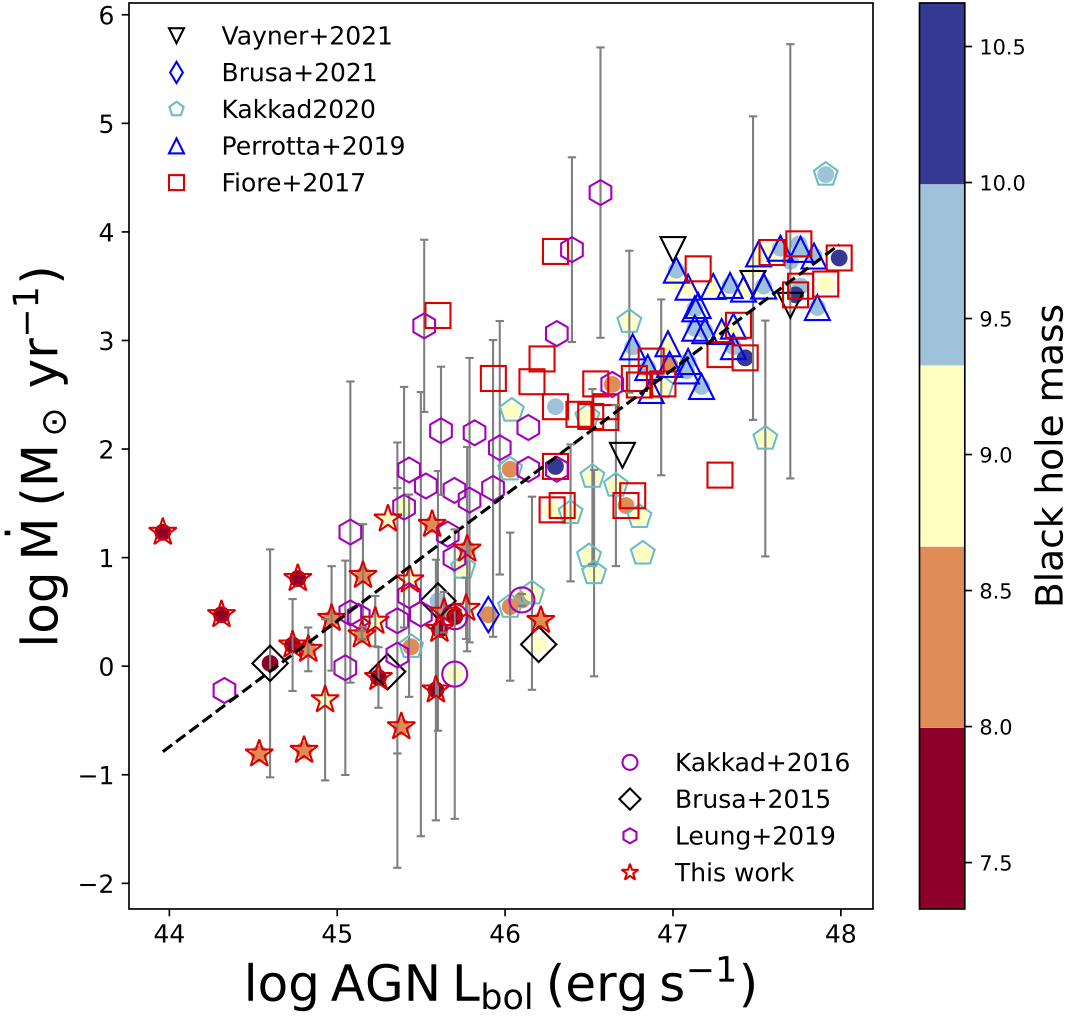} 
    \end{tabular}
    \caption{Ionised mass outflow rate as a function of AGN bolometric luminosity. Different shapes represent the literature of the sources. The black dashed line is the scaling relation obtained in Fig. \ref{fig:mdot_lbol_qwo_z}. In the left panel of  Fig. \ref{fig:mdot_lbol_qwo}, the sources are colour coded with the Eddington ratio. 
In the left panel, the sources are colour-coded
with the black hole mass.
Among the sources with black hole masses, we observe a correlation trend in the sense that AGNs with high black hole masses also appear to have high bolometric luminosity and higher mass outflow rates. In the right panel,  the sources are colour-coded with the Eddington ratio. AGNs with high Eddington ratio values appear to have high mass outflow rates and bolometric luminosity. The error bars on mass outflow rate values were obtained by extrapolating the errors on FWHM or $\rm{\delta{V}}$ whenever available. }
    \label{fig:mdot_lbol_qwo}
\end{figure*}

In this section, we present the correlations between the mass outflow rate and AGN bolometric luminosity with the best-fit relation obtained in Sect. \ref{correlations}. In Fig. \ref{fig:mdot_lbol_qwo}, left panel, the data is colour-coded with the Eddington ratio while in the right panel, the data is colour-coded with the black hole mass. We see a trend between these quantities as discussed already in Sect. \ref{correlations}. 

\end{appendix}
\end{document}